\documentclass[pre,aps,showpacs,twocolumn,reprint,longbibliography]{revtex4-1}

\usepackage{graphicx}
\usepackage{amsfonts}
\usepackage{amsmath}
\usepackage{mathtools}
\usepackage{float}
\usepackage{epsfig}
\usepackage{color}
\usepackage{gensymb}
\usepackage{multirow}
\usepackage[mathscr]{euscript}
\usepackage{pifont}
\usepackage[cm]{fullpage}

\newcommand{\ee}[1]{\ensuremath{10^{#1}}}

\begin{document}

\title{Realizable Hyperuniform and Nonhyperuniform Particle Configurations with Targeted Spectral Functions via Effective Pair Interactions}

\author{Ge Zhang}

\affiliation{Department of Physics and Astronomy, University of Pennsylvania,  Philadelphia,  Pennsylvania 19104, USA}
\author{Salvatore Torquato}%
\email{torquato@princeton.edu}
\homepage{http://chemlabs.princeton.edu}
\affiliation{Department of Chemistry, Princeton University, Princeton, New Jersey 08544, USA}
\affiliation{Department of Physics, Princeton University, Princeton, New Jersey 08544, USA}
\affiliation{Princeton Institute for the Science and Technology of Materials, Princeton University, Princeton, New Jersey 08544, USA}
\affiliation{Program in Applied and Computational Mathematics, Princeton University, Princeton, New Jersey 08544, USA}

\begin{abstract}

The
capacity to identify realizable many-body configurations associated with targeted
functional forms for the pair correlation function $g_2(r)$ or its corresponding structure factor $S(k)$ is of great fundamental and practical
importance. While there are obvious necessary conditions that a prescribed
structure factor at number density $\rho$ must satisfy to be configurationally realizable,
sufficient conditions are generally not known due to the infinite degeneracy
of configurations with different higher-order correlation functions. 
A major aim of this paper is to
expand our theoretical knowledge of the class of pair correlation
functions or structure factors that are realizable by classical disordered
ensembles of particle configurations, including exotic ``hyperuniform''
varieties. We first introduce a theoretical formalism that provides a
means to draw classical particle configurations from canonical ensembles with certain
pairwise-additive potentials that could correspond to targeted analytical
functional forms for the structure factor. This formulation enables us
to devise an improved algorithm to construct systematically canonical-ensemble
particle configurations with such targeted pair statistics, whenever
realizable. As a proof-of-concept, we test the algorithm  by targeting several different structure factors across dimensions that are known to be realizable and one hyperuniform target that is known to be nontrivially unrealizable.
Our algorithm succeeds for all realizable targets and appropriately fails for the unrealizable target,  demonstrating the accuracy and power of the method to numerically investigate the realizability problem.
Subsequently, we also target  several families of
structure-factor functions that meet the known necessary realizability conditions but were heretofore
not known to be realizable by disordered hyperuniform point configurations, including $d$-dimensional Gaussian structure factors, $d$-dimensional generalizations of the 2D one-component 
plasma (OCP), the $d$-dimensional Fourier duals of the previous OCP cases.
Moreover, we also explore unusual nonhyperuniform targets, including ``hyposurficial'' and ``anti-hyperuniform'' examples.
In all of these instances, the targeted structure
factors were achieved with high accuracy, suggesting that they are indeed realizable
by equilibrium configurations with pairwise interactions at positive temperatures.
Remarkably, we also show that the structure factor of nonequilibrium ``perfect glass'' specified by two-, three-, and four-body interactions, can also be realized by equilibrium pair interactions at positive temperatures.
Our findings lead us to the conjecture that any realizable structure factor corresponding to either a translationally invariant equilibrium or nonequilibrium system can be attained by an equilibrium ensemble involving only effective pair interactions. 
Our investigation not only broadens our knowledge of analytical functional forms for
$g_2({\bf r})$ and $S({\bf k})$ associated with disordered point configurations
across dimensions but also deepens our understanding of many-body physics.
Moreover, our work can be applied to the design of materials with
desirable physical properties that can be tuned  by their pair statistics.
\end{abstract}

\maketitle

\section{Introduction}

An outstanding problem in condensed matter physics, statistical physics and materials science
is the capacity to construct, at will, many-particle configurations
with prescribed correlation functions. Solutions to this general problem are of great  importance both fundamentally and practically.
Advances in this direction will shed light on the unsolved theoretical ``realizability" problem, as described below. 
Practical implications of progress on this problem include the  design of material microstructures with novel physical properties.

A classical many-particle system in  $d$-dimensional Euclidean space $\mathbb{R}^d$
is completely specified by the $n$-particle probability density function $\rho_n({\bf r}_1,\ldots, {\bf r}_n)$ for all $n\ge 1$,
where ${\bf r}_1,\ldots, {\bf r}_n$ are the particle position vectors.
In the field of statistical mechanics, the one-particle function $\rho_1({\bf r}_1)$
and the two-particle function $\rho_2({\bf r}_1,{\bf r}_2)$ are the most important ones.
These functions play crucial roles in determining
equilibrium and nonequilibrium
properties of systems and can be ascertained experimentally from 
scattering data \cite{Han13}.
In the case of statistically homogeneous systems, which is the focus of this work, $\rho_1({\bf r}_1)=\rho$,
where $\rho$  is the number density, and the two-particle function depends
only on the pair displacement vector ${\bf r}= {\bf r}_2 -{\bf r}_1$ so that
$\rho_2({\bf r}_1,{\bf r}_2)= \rho^2 g_2({\bf r})$, where  $g_2({\bf r})$ is the pair correlation function.
Of course, these two functions 
alone cannot completely specify the
ensemble of configurations, i.e., 
there is generally a high degeneracy
of configurations with the same
$\rho$ and $g_2({\bf r})$ but different
higher-order statistics ($g_3, g_4,\ldots$) \cite{torquato2006new,jiao2010geometrical}.

This degeneracy issue naturally leads
to the following version of the realizability 
problem:
Given a prescribed $g_2({\bf r})$ with a fixed positive number density $\rho$, 
are there ensemble configurations of particles that realize
such prescribed statistics?   This realizability problem has a rich and long history \cite{yamada1961geometrical, lenard1973correlation, lenard1975states, lenard1975states2, rintoul1997reconstruction, crawford2003aspects, costin2004construction, koralov2005existence, stillinger2005realizability, uche2006realizability, torquato2006new,kuna2007realizability, kuna2011necessary}, but it is still a wide open
area for research.  There are obvious necessary conditions for a given pair correlation function to be realizable; for example, $g_2(\mathbf r)$ 
must be nonnegative function {\it i.e.,}
\begin{equation}
g_2({\bf r}) \ge 0 \qquad \mbox{for all}\, {\bf r}.
\label{g2Definition}
\end{equation}
Moreover, the corresponding ensemble-averaged structure factor 
\begin{equation}
S(\mathbf k)= 1+\rho {\tilde h}({\bf k}) \ge 0,
\label{eq:Sk_g2}
\end{equation}
must be nonnegative for all wave vectors $\bf k$, where ${\tilde h}({\bf k})$ is the Fourier
transform  of the total correlation function $h(\mathbf r)\equiv g_2({\bf r})-1$.
Another simple realizability condition is that the number variance $\sigma^2(R)$ associated   
with a randomly placed spherical window of radius $R$, which is entirely determined
by $\rho$ and $g_2(\mathbf r)$ [or $S(\mathbf k)$] \cite{torquato2003local}, must satisfy
the following lower bound \cite{yamada1961geometrical}:
\begin{equation}
\sigma^2(R)\ge \theta(1-\theta),
\label{yamada}
\end{equation}
where $\theta$ be the fractional part of $\rho v_1(R)$ and 
\begin{equation}
v_1(R)=\frac{\pi^{d/2} R^d}{\Gamma(1+d/2)}
\label{v1}
\end{equation}
is the volume of a $d$-dimensional sphere of radius $R$.
The Yamada condition (\ref{yamada}) is relevant only in very low dimensions,
often only for $d=1$ \cite{torquato2006new}. Indeed, generally speaking, it is known that the lower the space dimension, the more difficult it is to satisfy
realizability conditions \cite{torquato2006new}, a point elaborated in Sec. \ref{unknown}.

Conditions for realizability have also been found for special types of many-particle systems \cite{costin2004construction}. 
Moreover, necessary and sufficient conditions for the particular class of point configurations
with ``hard" cores have been identified \cite{kuna2011necessary, rey2015regularity},  but these conditions are difficult to check in practice.
Thus, knowledge of necessary conditions beyond inequalities (\ref{g2Definition}), (\ref{eq:Sk_g2}), and (\ref{yamada}) that can be applied to determine the realizability of general pair correlation
functions are, for the most part, lacking.

This places great importance on the need to formulate algorithms to construct
particle configurations that realize targeted hypothetical functional forms of the pair statistics 
with a certain density. Successful numerical techniques could provide theoretical
guidance on attainable pair correlations. Algorithms have been devised in ``direct space" to generate
particle realizations that correspond to hypothetical pair correlation functions \cite{rintoul1997reconstruction, crawford2003aspects},
but only up to intermediate values of the pair distance $|\bf r|$. This prevents
one from accurately constructing the large-scale structural characteristics of the systems.


Therefore, such direct-space methods are not suitable to explore
the realizability of hypothetical functional forms of pair correlation functions that could
correspond to disordered {\it hyperuniform} point configurations with high fidelity.
Disordered hyperuniform many-particle systems are exotic amorphous
states of matter that are like crystals in the manner in which their
large-scale density fluctuations are anomalously suppressed and
yet behave like typical liquids or glasses in that they are statistically
isotropic without any Bragg peaks. More precisely, hyperuniform point configurations possess
a structure factor $S({\bf k})$ that goes to zero as the wavenumber $|\bf k|$ vanishes  \cite{torquato2003local, To18a}, i.e.,
\begin{equation}
\lim_{|{\bf k}| \to 0} S({\bf k})=0.
\label{hyper}
\end{equation}
For a large class of ordered and disordered systems, the number variance $\sigma^2(R)$ has the following large-$R$
asymptotic behavior~\cite{torquato2003local, To18a}:
\begin{equation}
  \sigma^2(R)=2^d\phi\left[A\left(\frac{R}{D}\right)^d+B\left(\frac{R}{D}\right)^{d-1}+\ell\left(\frac{R}{D}\right)^{d-1}\right] 
  \label{eq:sigma}
\end{equation}
where $\phi=\rho v_1(D/2)$ is a dimensionless density,
$D$ is a characteristic length, 
$A$ and $B$ are ``volume'' and ``surface-area''
coefficients, respectively, while $\ell\left(R/D\right)^{d-1}$ represents terms of lower order
than $\left(R/D\right)^{d-1}$.
The coefficients $A$ and $B$ can be expressed as:
\begin{equation}
  A=\lim_{|\mathbf{k}|\rightarrow 0}S(\mathbf{k})=1+d2^d\phi\left<x^{d-1}\right>
  \label{eq:A}
\end{equation}
and
\begin{equation}
  B=-\frac{d^22^{d-1}\Gamma\left(\frac{d}{2}\right)}{\Gamma\left(\frac{d+1}{2}\right)\Gamma\left(\frac{1}{2}\right)}\phi\left<x^d\right>
  \label{eq:B}
\end{equation}
where $\Gamma(x)$ is the gamma function, $x=r/D$, and $<x^d>=\int_{0}^{\infty}x^dh(x)dx$ is the $d$-th moment of $h(x)$
In a perfectly hyperuniform system~\cite{torquato2003local},
the non-negative volume coefficient vanishes, i.e., $A=0$.
On the other hand,
when $A>0$ and $B=0$, the system is \textit{hyposurficial}; examples include
homogeneous Poisson point
patterns and hypothetical hard-sphere systems~\cite{torquato2003local}.
Finally, in {\it anti-hyperuniform} systems
\begin{equation}
\lim_{|{\bf k}| \to 0} S({\bf k})=+\infty
\label{antihyper}
\end{equation}
and $A$ becomes unbounded \cite{To18a}. Anti-hyperuniform systems include fractals as well as systems at
thermal critical points.

When the structure factor goes to zero in the limit $|{\bf k}| \rightarrow 0$ with the power-law  form 
\begin{equation} 
S({\bf k}) \sim |{\bf k}|^\alpha, 
\label{power} 
\end{equation} 
where $\alpha >0$, hyperuniform systems can be categorized 
into three different classes according to the large-$R$ 
asymptotic scaling of the number variance  \cite{To18a}: 
\begin{eqnarray} 
\sigma^2(R) \sim \left\{ 
\begin{array}{lr} 
R^{d-1}, \quad \alpha >1 \quad \mbox{(CLASS I)}\\ 
R^{d-1} \ln R, \quad \alpha = 1 \quad \mbox{(CLASS II)}\\ 
R^{d-\alpha}, \quad 0 < \alpha < 1 \quad \mbox{(CLASS III)} 
\end{array}\right. 
\label{sigma-asy} 
\end{eqnarray} 
Class I is the strongest form of hyperuniformity in the sense 
that it is the scaling that provides the greatest suppression of large-scale density fluctuations.

Disordered hyperuniform systems have attracted great attention because of their
deep connections  to problems that arise in physics, materials science, mathematics
and biology \cite{chremos2017particle, To18a, ghosh2018generalized,brauchart2018hyperuniform, crowley2018quantum, lei2019nonequilibrium, lacroix2019intermediate} as well as for their emerging technological
importance, including disordered cellular solids that have complete isotropic photonic band gaps \cite{fl09b, Ri19}, surface-enhanced Raman spectroscopy \cite{Zi15}, transparent materials \cite{Le16}, terahertz quantum cascade lasers \cite{ma20163d}, and certain Smith-Purcell radiation patterns \cite{saavedra2016smith}. While a variety of equilibrium and nonequilibrium hyperuniform systems
have been generated via computer simulations \cite{To18a}, current numerical techniques (with the exception of the
``collective-coordinate" approach \cite{uche2006collective, batten2008classical})
cannot guarantee perfect hyperuniformity  \cite{To18a}. 

Remarkably, very little
is known about analytical forms of two-body and higher-order
correlation functions that are
exactly realizable by disordered hyperuniform systems. An exception to this dearth of knowledge is the special class of determinantal
point processes \cite{So00,mehta2004random,costin2004construction,torquato2008point,hough09,Ab17}, examples of which are considered in Sec. \ref{known}.
Furthermore,  no one to date
has shown  the rigorous existence of  hyposurficial point configurations, even if they have been shown
to arise in computer simulation study of phase transitions involving amorphous ices  \cite{martelli2017large}.

The purpose of the present investigation is to expand our theoretical knowledge
of the class of pair correlation functions or, equivalently, structure-factor functions
that are realizable by disordered hyperuniform ensembles of statistically homogeneous classical particle configurations
at some number density $\rho$, including $d$-dimensional Gaussian structure factors, $d$-dimensional generalizations of the 2D one-component 
plasma (OCP), the $d$-dimensional Fourier duals of the previous OCP cases.
We also demonstrate the realizability of unusual {\it nonhyperuniform} point configurations, including ``hyposurficial'' and ``anti-hyperuniform'' examples.
Our findings lead us to the conjecture that any
realizable structure factor corresponding to either an equilibrium or nonequilibrium homogeneous system can be
attained by an equilibrium ensemble involving only effective pair interactions
in the thermodynamic limit.

We begin by introducing a theoretical formalism that provides a means to
draw equilibrium particle configurations from canonical ensembles with certain pairwise-additive potentials that could correspond to 
targeted  analytical functional forms for the structure factor (Sec. \ref{theory}). 
Using this theoretical foundation, we then 
devise a new algorithm to construct systematically canonical-ensemble particle configurations with such targeted pair statistics
whenever realizable (Sec. \ref{newAlgorithm}).
We demonstrate the efficacy of  our targeting method in two ways. 
First, as a proof-of-concept, we test it to target several different structure factor functions 
across dimensions that are known to be realizable by determinantal hyperuniform point processes (Sec. \ref{known}).
We verify that all of these considered targets are indeed realizable. 
As another proof-of-concept, we also show that this methodology indeed fails on a nontrivial target that is known to be unrealizable, even though the target meets all explicitly known necessary realizability conditions (Sec. \ref{impossible}). 
Taken together, these benchmark tests  demonstrate
the accuracy and power of the method to numerically investigate the realizability problem.
Finally, we apply the methodology to target several families of structure-factor functions that meet the known necessary realizability conditions
but  were heretofore not known to be realizable by disordered hyperuniform and nonhyperuniform (hyposurficial and anti-hyperuniform) systems (Sec. \ref{unknown}-\ref{unknown-nonhu}). In all of these instances, we are able 
to achieve the targeted structure factor with high accuracy, suggesting that
these targets are indeed truly realizable by disordered hyperuniform many-particle systems in equilibrium
with effective pairwise interactions at positive temperatures.
Our results leads to a conjecture that any realizable
structure factor can be attained by an equilibrium ensemble involving only effective pair interactions, which is presented in Sec. \ref{sec:conjecture}. 
We further demonstrate the validity of this conjecture by showing that a previously numerically found structure factor of a nonequilibrium state of a two-, three-, and four-body interaction can also be realized by equilibrium pair interactions.
Concluding remarks and discussion of our results are presented 
in Sec. \ref{conclusions}.

\section{Theoretical Analysis and Novel Algorithm}

\subsection{Motivation}

At first glance, the aforementioned Fourier-based collective-coordinate optimization procedure may
seem to be ideally suited  to construct possibly realizable hyperuniform configurations,  since it enables one to
obtain configurations with desired structure factors for wave vectors around the origin  with very high accuracy \cite{uche2006collective, batten2008classical, zachary2011anomalous,zh16a}.
One begins with a single classical configuration of $N$ particles with positions $\mathbf r^N \equiv {\bf r}_1,\ldots, {\bf r}_N$ in a fundamental cell under periodic boundary conditions.
Here, the structure factor of a single configuration, $\mathcal S(\mathbf k)$, is constrained to be equal to a target function $\mathcal S_0(\mathbf k)$ for $\mathbf k$ in a certain {\it finite} set $\mathbb K$.
These constraints are enforced by minimizing a fictitious potential energy $\Phi(\mathbf r^N)$,
defined to be  the square of  the difference between $\mathcal S({\bf k})$ and $\mathcal S_0({\bf k})$:
\begin{equation}
\Phi(\mathbf r^N)=\sum_{\mathbf k \in \mathbb K} [\mathcal S(\mathbf k)-\mathcal S_0(\mathbf k)]^2,
\label{eq:targetFunction}
\end{equation}
where, for a single configuration, the structure factor at a non-zero $\mathbf k$ vector is given by 
\begin{equation}
\mathcal S(\mathbf k)=\frac{1}{N}\left|{\tilde \rho}(\mathbf k)\right|^2 \mbox{ ($\mathbf k \neq \mathbf 0$)},
\label{eq:Sk_direct}
\end{equation}
and
\begin{equation}
{\tilde \rho}(\mathbf k)=\sum_{j=1}^{N} \exp(-i\mathbf k \cdot \mathbf r_j)
\label{eq:collectiveDensity}
\end{equation}
is the complex collective density variable.
Throughout the paper, we will use $\mathcal S$ to denote single-configuration structure factors and  $S$ to denote ensemble-averaged structure factors.
It was shown that the potential energy given by (\ref{eq:targetFunction}) is equivalent to a certain
combination of a long-ranged two-, three-, and four-body interactions \cite{uche2006collective}. Therefore, constraining $\mathcal S(\bf k)$ to a target function $\mathcal S_0(\bf k)$ using this method is equivalent to finding a single ground-state configuration with these interactions.
One calculates $\mathcal S(\mathbf k)$ from (\ref{eq:Sk_direct}) rather than (\ref{eq:Sk_g2}) not only because it applies only for
a single finite-size configuration (not an ensemble), but  (\ref{eq:Sk_direct}) allows $\mathbb K$ to include $\mathbf k$ vectors very close to the origin.

However, this standard collective-coordinate method cannot be used for the realizability problem. It suffers from numerical difficulty if the cardinality of $\mathbb K$ is too large, meaning
that only a portion of wave vectors can be targeted. Indeed, if the number of independent constraints is larger than the total number of degrees of freedom $dN$, then the system ``runs out of degrees of freedom'' and the potential energy surface often becomes so complicated that one cannot find an $\Phi=0$ state, even if the target $\mathcal S_0(\mathbf k)$ is known to be realizable \citep{batten2008classical}. This method also enforces $\mathcal S(\mathbf k)= \mathcal S_0(\mathbf k)$ for $\mathbf k \in \mathbb K$ for a single configuration, while in many cases we only expect the ensemble average $S(\mathbf k)$ to be equal to $S_0(\mathbf k)$. 
These drawbacks will be overcome by our new algorithm, as detailed in Sec.~\ref{newAlgorithm}.

\subsection{Theoretical formalism}
\label{theory}

The discussion above suggests that targeting an ensemble-averaged structure factor,  which would enable
the toleration of fluctuations in individual configurations, may be a possible way to bypass the
limitations of the standard collective-coordinate procedure for the realizability problem.  We now show
on theoretical grounds how this is indeed the case. Specifically, we demonstrate
that targeting ensemble-averaged structure factors results in an enormous increase
in the number of degrees of freedom, which in turn enables one to  extend
the range of  constrained wave vectors over an infinite set, in principle.
Moreover, we show that the configurations are sampled from a canonical
ensemble with a certain pair potential.

Let us begin by imagining imposing constraints such that the average structure factor for a finite number of  configurations  $N_c$ is equal to a target
functional form for $\mathbf k \in \mathbb K$, i.e.,
\begin{equation}
\langle \mathcal S(\mathbf k) \rangle=S_0(\mathbf k)\mbox{, for any } \mathbf k \in \mathbb K,
\end{equation}
where $\langle \mathcal S(\mathbf k)\rangle $ is the average structure factor of these $N_c$ configurations. 
We will assume the $N_c \to +\infty$ limit in this theoretical subsection, so that 
\begin{equation}
\langle \mathcal S(\mathbf k)\rangle =S(\mathbf k),
\end{equation}
where $S(\mathbf k)$ is the ensemble-averaged structure factor defined in (\ref{eq:Sk_g2}). 
In this thermodynamic limit ($N_c \to +\infty$), there is an infinite number of degrees of freedom,
which enables one to extend the range of constrained wave vectors over all space.

A critical question is what is the behavior of each individual configuration when such constraints are imposed? 
Although it appears that these configurations interact with each other in some complex manner, we now 
demonstrate that these configurations follow the {\it canonical-ensemble} distribution of a pairwise additive potential energy.
To begin with, let us consider constraining the target structure factor at a single point, $\mathbf k=\mathbf q$:
\begin{equation}
S_0(\mathbf q)=\frac{\mathcal S_1(\mathbf q)+\mathcal S_2(\mathbf q)+\cdots+\mathcal S_{N_c}(\mathbf q)}{N_c},
\label{eq:constr}
\end{equation}
where $\mathcal S_i(\mathbf q)$ ($i>0$) is the structure factor of the $i$th configuration at wave vector $\mathbf k = \mathbf q$.
In Eq.~(\ref{eq:constr}), we are treating $\mathcal S_1(\mathbf q)$, $\mathcal S_2(\mathbf q)$, $\cdots,\mathcal S_{N_c}(\mathbf q)$ as $N_c$ random variables and constrain their arithmetic mean to be equal to $S_0(\mathbf q)$.
As we will show later, $\mathcal S_i(\mathbf q)$ ($i>0$) is exponentially distributed, which implies that $\mathcal S_i(\mathbf q)$ is not self-averaging \cite{kirkpatric1989, parisi2004scale}.

\begin{figure}
\begin{center}
\includegraphics[width=0.45\textwidth]{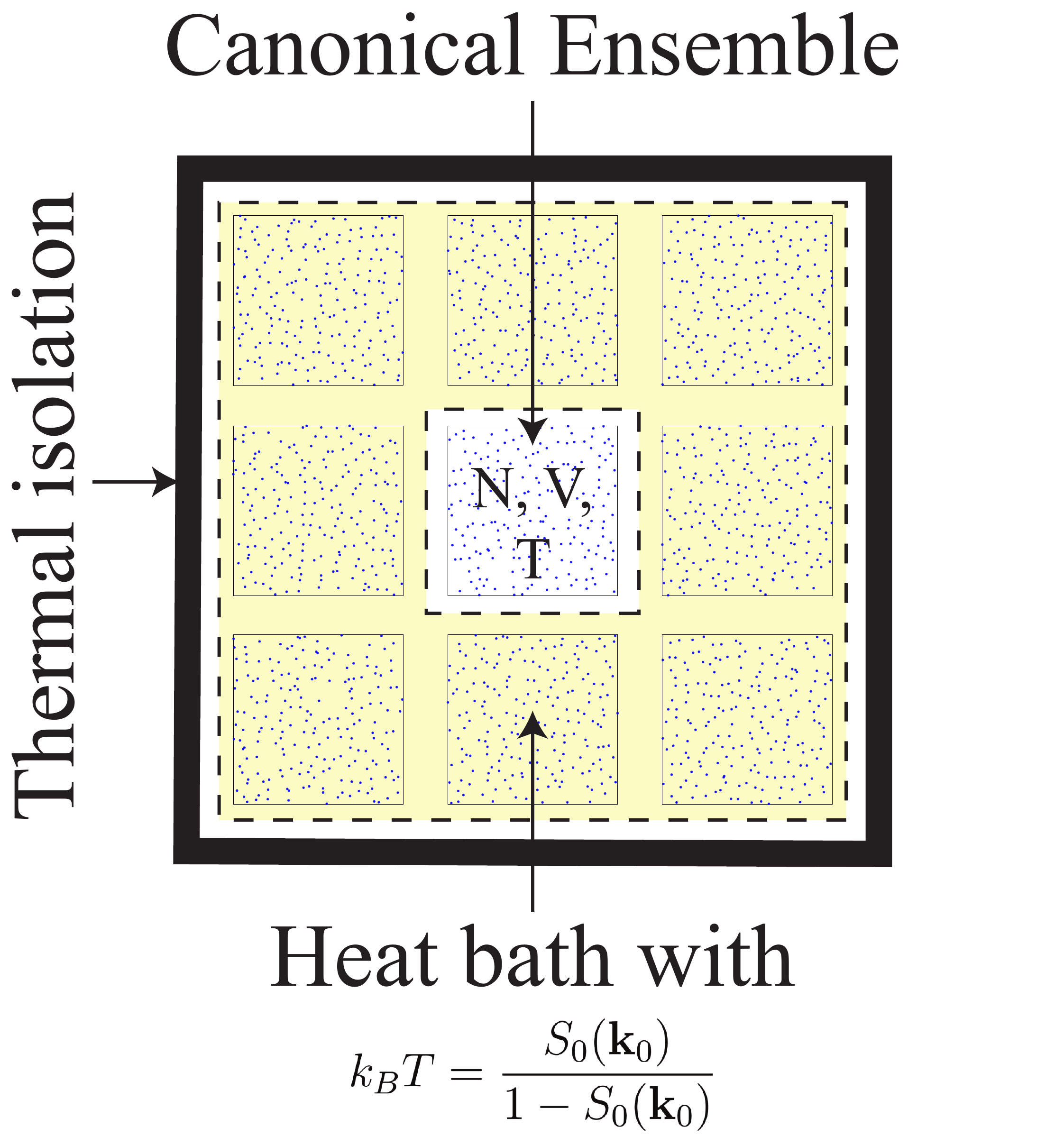}
\end{center}
\caption{When a large number of systems can exchange heat with each other but not with the environment, one can focus on one system (central one 
indicated with a white background) and treat the rest as a heat bath (systems having yellow background). The central system  
with fixed $N$, $V$ and $T$ follows the distribution function of a classical canonical ensemble with a temperature $T$ 
determined by the total energy of the systems  and degeneracy of the heat bath. For the problem at hand, we show that the temperature
is determined by the target structure factor $S_0({\bf k})$, as specified by the relation $k_BT=S_0(\mathbf q)/[1-S_0(\mathbf q)]$.}
\label{fig:canonicalEnsemble}
\end{figure}

The key idea is that because we allow an arbitrary large number of configurations $N_c$ and constrain their arithmetic mean structure factor [right-hand side of Eq. (\ref{eq:constr})], we can focus on one configuration, which we call the ``reference system,'' with
fictitious energy $E_R=\mathcal S(\mathbf q)$ and treat the rest of the
$N_c-1$ configurations as a {\it heat bath} with temperature
\begin{equation}
k_BT=\frac{S_0(\mathbf q)}{1-S_0(\mathbf q)},
\label{T}
\end{equation}
Relation (\ref{T}) for the temperature is derived immediately below.
Under these conditions, the reference system obeys the distribution function of a canonical ensemble in the limit $N_c \to \infty$.  
Figure~\ref{fig:canonicalEnsemble} schematically describes this scenario.

The probability density function (PDF) of the total energy of any system  in the canonical ensemble is given by
\begin{equation}
P(E) =\frac{g(E)\exp(-E/k_B T)}{Z},
\label{pdf}
\end{equation}
where $Z$
is the partition function, $g(E)$
is energy degeneracy or density of states, and $T$ is the temperature. 
To determine the temperature of the heat bath explicitly in terms of $S_0(\mathbf q)$, we make the simple observation that in the high-temperature or ideal-gas limit, the PDF (\ref{pdf})
is proportional to the density of states
\begin{equation}
g(E)\propto \lim_{T\to\infty}P(E).
\label{eq:proportion}
\end{equation}
Thus, to determine the heat bath's density of states and corresponding temperature
for general correlated systems, we only need to know the distribution of its total energy, $E_H$, assuming the heat bath consists of $N_c-1$ uncorrelated (infinite-temperature) systems.

 Let us first focus on one system comprising the heat bath, and denote its energy by $E_1$. For an ideal gas, each particle's location is random and independent. Thus, for any particle $j$, $\exp(-i\mathbf q \cdot \mathbf r_j)$ is a unit vector of a random orientation in the complex plane. Therefore, the collective density variable (\ref{eq:collectiveDensity})
is the sum of $N$ random unit vectors in the complex plane. From the theory of random walks, we know that for large $N$, ${\tilde \rho}(\mathbf q)$ in the complex plane follows a Gaussian distribution, and the PDF of $|{\tilde \rho}(\mathbf q)|$ is given by
\begin{equation}
P_{uncorr}(|{\tilde \rho}(\mathbf q)|)=\frac{2|{\tilde \rho}(\mathbf q)|}{N}\exp(-|{\tilde \rho}(\mathbf q)|^2/N).
\end{equation}
Therefore, the PDF of the energy $E_1=\mathcal S(\mathbf q)=|{\tilde \rho}(\mathbf q)|^2/N$ is given by
\begin{equation}
\begin{split}
P_{uncorr}(E_1)&=P_{uncorr}(|{\tilde \rho}(\mathbf q)|) \left[ \frac{d\mathcal S(\mathbf q)}{d|{\tilde \rho}(\mathbf q)|} \right]^{-1}\\ &=\exp(-E_1).
\end{split}
\end{equation}
Thus, the energy (i.e., the single-configuration structure factor) is exponentially distributed. 
This combined with (\ref{eq:proportion}) implies that the density of states of a single configuration is also an exponential function of the energy
\begin{equation}
g(E_1)\propto \exp(-E_1).
\label{eq:singleConfigDOS}
\end{equation}

For two uncorrelated configurations, the probability distribution of their total energy $E_{12}=E_1+E_2$ is
\begin{align}
P_{uncorr}(E_{12}) &=\int_0^{E_{12}} P_{uncorr}(E_1)P_{uncorr}(E_2) dE_1\\ &= \int_0^{E_{12}} \exp(-E_1)\exp\left[-(E_{12}-E_1)\right]dE_1\\ &=E_{12}\exp(-E_{12}).
\end{align}
The distribution for the total energy of three configurations is then
\begin{equation}
\begin{split}
&P_{uncorr}(E_{123}=E_{12}+E_3) \\ &=\int_0^{E_{123}} P_{uncorr}(E_3)P_{uncorr}(E_{12}) dE_3\\ &=\frac{E_{123}^2}{2}\exp(-E_{123}).
\end{split}
\end{equation}
Similarly, one can find that the distribution of the total energy of $N_c-1$ configuration is
\begin{equation}
P_{uncorr}(E=E_1+E_2+\cdots+E_{N_c-1})=\frac{E^{N_c-1}}{(N_c-1)!}\exp(-E).
\end{equation}
As we detailed before, the density of states of the heat bath, made from $N_c-1$ systems, is proportional to the probability distribution function of the total energy of $N_c-1$ uncorrelated systems:
\begin{equation}
g(E) \propto P_{uncorr}(E)=\frac{E^{N_c-1}}{(N_c-1)!}\exp(-E).
\end{equation}

The temperature of the heat bath is therefore
\begin{equation}
k_BT=\left[ \frac{\partial \ln g(E)}{\partial E} \right ]^{-1}=\left[ \frac{N_c-1}{E}-1 \right ]^{-1},
\end{equation}
where $E=E_H$ is the energy of the heat bath. On average, each configuration has an energy of $S_0(\mathbf q)$. Therefore, in the $N_c \to\infty$ limit, $E_H=(N_c-1)S_0(\mathbf q)$ and the heat-bath temperature is explicitly given by
\begin{equation}
k_BT=\frac{S_0(\mathbf q)}{1-S_0(\mathbf q)},
\end{equation}
which is what we set out to prove.

In the $N_c \to\infty$ limit, the heat bath is infinitely large, and we can determine the probability density function of the energy of the reference system, $E_R$, which is not included in the heat bath, using the canonical distribution function, {\it i.e.,}
\begin{align}
P(E_R)&=\frac{g(E_R)\exp(-E_R/k_BT)}{Z}\\
&\propto \exp(-E_R)\exp(-E_R/k_BT),
\label{eq:canonicalDistribution}
\end{align}
After normalization, one finds
\begin{equation}
P(E_R)=\frac{\exp(-E_R)\exp(-E_R/k_BT)}{S_0(\mathbf q)}=\frac{\exp[-E_R/S_0(\mathbf q)]}{S_0(\mathbf q)}.
\end{equation}
Since we previously defined $E_R=\mathcal S(\mathbf q)$,
\begin{equation}
P[\mathcal S(\mathbf q)]=\frac{\exp[-\mathcal S(\mathbf q)/S_0(\mathbf q)]}{S_0(\mathbf q)}.
\end{equation}
By symmetry, this distribution is applicable not only to the reference system, but also to the other $N_c-1$ systems as well.
This means that for any particular configuration, its structure factor at a constrained $\mathbf k$ vector is exponentially distributed. We will numerically verify this conclusion in the Appendix.
As in the ideal-gas case, the exponential distribution of $S(\mathbf q)$ implies that ${\tilde{\rho}}(\mathbf q)$ is Gaussian distributed.

As we previously showed, if one constrains the ensemble-averaged structure factor $S(\mathbf q)$ to be equal to $S_0(\mathbf q)$, the resulting configurations follow canonical-ensemble distribution of a system with energy defined as $E=\mathcal S(\mathbf q)$ at temperature $k_BT=S_0(\mathbf q)/(1-S_0(\mathbf q))$. Equivalently, one could also define a rescaled energy as 
\begin{equation}
E={\tilde v}(\mathbf q)\mathcal S(\mathbf q),
\label{eqn:canonicalEnergy1}
\end{equation}
where 
\begin{equation}
{\tilde v}(\mathbf q) = 1/S_0(\mathbf q)-1
\label{eqn:tildeV}
\end{equation}
and set $k_BT={\tilde v}(\mathbf q)S_0(\mathbf q)/(1-S_0(\mathbf q))=1$.
As detailed in our earlier papers \cite{fan1991constraints, uche2004constraints, torquato2015ensemble}, such a definition of $E$ is equivalent to a pairwise additive potential. For example, in the thermodynamic
limit, the total energy of such
a system of particles with pairwise interactions is given by
\begin{equation}
E=\frac{\rho}{2}\int_{\mathbb{R}^d} v(\mathbf r) g_2(\mathbf r)d\mathbf r,
\end{equation}
which can be represented in Fourier space using Parseval's theorem \cite{torquato2015ensemble}:
\begin{equation}
E=\frac{\rho}{2} {\tilde v}(\mathbf k=\mathbf 0) - \frac{1}{2}v(\mathbf r = \mathbf 0) + \frac{\rho}{2(2\pi)^d}\int_{\mathbb{R}^d} {\tilde v}(\mathbf k) S(\mathbf k)d\mathbf k.
\end{equation}

The derivation of (\ref{eqn:canonicalEnergy1}) and (\ref{eqn:tildeV}) applies to the cases where one constrains $S(\mathbf k)$ at a single $\mathbf k$ vector. Can one generalize it to constraining $S(\mathbf k)$ at multiple $\mathbf k$ vectors? If one constrains $S(\mathbf k)$ at up to $d$ different orthogonal wave vectors (inner product being zero), 
formulas (\ref{eqn:canonicalEnergy1}) and (\ref{eqn:tildeV})  would still apply exactly. This is because such constraints affect particle positions in different, independent directions, and can thus be treated separately.
If the $d$ wave vectors are linearly independent but not orthogonal, one could still apply a linear transformation to reduce the problem to the orthogonal case.

To treat cases where the number of $\mathbf k$ vectors is larger than $d$, we recall that in Gibbs formalism, the inverse temperature is a Lagrange multiplier of energy. For multiple constrained wave vectors, $\mathbf k_1$, $\mathbf k_2$, $\cdots,\mathbf k_{N_k}$, we can use a separate Lagrange multiplier for each constraint. Consider maximizing the Gibbs entropy of the reference configuration
\begin{equation}
\mathscr S=-k_B \int P(\mathbf r^N)\ln P(\mathbf r^N) d\mathbf r^N,
\end{equation}
where $P(\mathbf r^N)$ is the probability density function of the reference configuration, subject to constraints
\begin{equation}
C_{\mathbf k}=\int P(\mathbf r^N) \mathcal S(\mathbf k) d\mathbf r^N-S_0(\mathbf k)=0
\end{equation}
for each constrained $\mathbf k$, and
\begin{equation}
D=\int P(\mathbf r^N) d\mathbf r^N-1=0,
\end{equation}
If these constraints are satisfiable, we can find $P(\mathbf r^N)$ by
constructing the Lagrangian function
\begin{equation}
L=\mathscr S-\sum_{\mathbf k} \lambda_{\mathbf k} C_{\mathbf k} - \lambda_D D,
\end{equation}
where $\lambda_{\mathbf k}$ and $\lambda_D$ are Lagrange multipliers.
Setting $\delta L/\delta P(\mathbf r^N)=0$, we find
\begin{equation}
P(\mathbf r^N)=\frac{1}{Z} \exp\left [-\sum_{\mathbf k} \lambda_{\mathbf k} \mathcal S(\mathbf k)\right],
\end{equation}
where $Z=\int \exp\left [-\sum_{\mathbf k} \lambda_{\mathbf k} \mathcal S(\mathbf k_j)\right] d\mathbf r^N$ is the partition function of the reference system.
We see that if we define the fictitious energy as 
\begin{equation}
E=\sum_{\mathbf k} \lambda_{\mathbf k} \mathcal S(\mathbf k),
\label{eqn:canonicalEnergy2}
\end{equation}
which is still a pair potential \cite{fan1991constraints, uche2004constraints, torquato2015ensemble}, then $P(\mathbf r^N)$ follows the equilibrium distribution at $k_BT=1$.
However, we could not find an explicit expression for $\lambda_{\mathbf k}$.
Theoretically, $\lambda_{\mathbf k}$ is completely determined by the target structure factor. However, the dependency is nontrivial due to the correlation between $S(\mathbf k)$ at different $\mathbf k$ vectors.
We proved that $S(\mathbf k)$ is exponentially distributed in the single-constraint or independent-constraint cases. 
While we cannot prove that when multiple non-independent wave vectors are constrained, we do provide numerical evidence for such behavior in Appendix~\ref{appendix:justification}.

To summarize, we have proved that if one constrains the ensemble-averaged structure factor at one or multiple $\mathbf k$ vectors, and if the constraints are satisfiable, then the resulting configurations are drawn from the canonical ensemble with a pairwise additive interaction (\ref{eqn:canonicalEnergy1}) or (\ref{eqn:canonicalEnergy2}).
Thus, when we constrain $S(\mathbf k)$ for all $\mathbf k$ vectors to be equal to a structure factor realized by some $n$-body interactions, our method finds an effective pair interaction that mimics the configurations produced by such $n$-body interactions.
For the single-constraint case, we showed that the structure factor for a single configuration at the constrained $\mathbf k$ vector is exponentially distributed. 
For the multiple-constraint case, we will also show strong numerical evidence in Appendix~\ref{appendix:justification} that the exponential distribution still holds.
If the structure factor at the constrained $\mathbf k$ vectors are independent from one another, then the interaction is given by Eqs.~(\ref{eqn:canonicalEnergy1})-(\ref{eqn:tildeV}), and the temperature is $k_BT=1$. However, if the structure factor at these $\mathbf k$ vectors are correlated, then (\ref{eqn:tildeV}) is inexact.
We numerically test and verify these conclusions in Appendix~\ref{appendix:justification} for specific examples. 

\subsection{Ensemble-Average Algorithm}
\label{newAlgorithm}
Based on this theoretical formalism, we can now straightforwardly devise a new algorithm to construct 
a canonical ensemble of a finite but large number of configurations, $N_c$,  targeting a particular functional form
for the structure factor.
Specifically, we minimize the squared difference for $N_c$ configurations but {\it simultaneously}, i.e.,
\begin{equation}
\mbox{minimize } \Phi(\mathbf r^{\cal{N}})=\sum_{\mathbf k \in \mathbb K} [\langle \mathcal S(\mathbf k)\rangle-S_0(\mathbf k)]^2,
\label{Potential_Fourier2}
\end{equation}
where ${ \cal N}= N N_c$.
Thus, compared to the standard collective-coordinate procedure \cite{uche2006collective, batten2008classical, zachary2011anomalous,zh16a}, which has available $d N$ number of degrees
of freedom, the canonical-ensemble-average generalization, enables us to  substantially increase the number
of degrees of freedom to $d N N_c$. Thus, in practice, the range of wave vectors over which we can constrain the
structure factor to have a prescribed functional form can be made to be larger and larger by increasing the number of configurations.

Our algorithm involves minimizing a target function that is a sum over all $\mathbf k$ vectors within a wavenumber $K$ from the origin.  The number of such  $\mathbf k$ vectors scales as $K^dV$, where $V=N/\rho$ is the volume of the system. For each such  $\mathbf k$ vector, we need to calculate $N_c$ structure factor values, each involves a summation over $N$ particles. Thus, the computational cost scales as $K^dVNN_c=K^d\rho^{-1} N^2N_c$. As $N$ grows, the computational cost grows quadratically and can become very large. However, the calculations for different  $\mathbf k$ vectors can be carried out in parallel, and we can thus employ multiple GPUs to overcome the high computational cost. GPUs generally perform single-precision calculations faster than double-precision calculations, but we discovered that double precision is necessary for large system sizes ($N>1000$).

 For cases in which  $N >2000$,  we found that the number of iterations needed to minimize the target function becomes 
computationally costly. By inspecting the intermediate configurations during minimization, we discovered that $S(\mathbf k)$ near the origin ($\mathbf k=\mathbf 0$) converges to $S_0(\mathbf k)$ at much slower rate than that of $S(\mathbf k)$ at other wave vectors. It is reasonable to assume that this slow-up is caused by the fact that changing $S(\mathbf k)$ at $\mathbf k \approx \mathbf 0$ requires long-range particle motions. To improve the convergence speed for small $k$ when $N>2000$, 
we introduce a weight of $w(k)=1/k$ in the previous objective
function and then carry out the following minimization:
\begin{equation}
\mbox{minimize } \Phi(\mathbf r^{\cal{N}})=\sum_{\mathbf k \in \mathbb K} w(k)[\langle \mathcal S(\mathbf k)\rangle-S_0(\mathbf k)]^2,
\label{Potential_Fourier3}
\end{equation}
As a proof of concept of these modifications, we were able to generate $N_c=100$ configurations, each consisting of $N=20000$ particles, targeting the 1D fermionic target structure factor (\ref{1DFermionic_S}), after five days of computation using four NVIDIA Tesla P100 GPUs, as reported in Sec.~\ref{sec:oneComponentPlasma}. Minimizing the objective function usually requires $\sim \ee{4}$ iterations in 1D but only $\ee{2}-\ee{3}$ iterations in 2D and 3D. Thus, we can generate higher-dimensional configurations with $N=20000$ particles much faster (about 30 times faster with our hardware).

In subsequent sections, we present results  using $N_c=100$, which is large but still computationally manageable.
Unless otherwise stated, each configuration consists of $N=400$ particles in a linear (1D), square(2D), or cubic (3D) simulation boxes with periodic boundary conditions. 
As we will show in Fig.~\ref{1DFermi}, $N=400$ is large enough to produce pair statistics indistinguishable from $N=20000$ ones.
The pair statistics are averaged over $5000$ configurations to reduce statistical fluctuations.
Since the weight $w(k)=1/k$ is necessary only for sufficiently large configurations
($N > 2000$), we omit it for simplicity.
The set $\mathbb K$ contain half of all $\mathbf k$ vectors such that $0<|\mathbf k|<K$, where $K$ is a constant cutoff.
We can omit one half of the $\mathbf k$ vectors within the range due to the inversion symmetry of the structure factor: $S(-\mathbf k)=S(\mathbf k)$.
Unless otherwise stated, we use $K=30$ in 1D and $K=15$ in 2D and 3D. 
We use the low-storage BFGS algorithm \cite{nocedal1980updating, liu1989limited, nlopt} to minimize $\Phi$, starting from random initial configurations. After the minimization, $\Phi$ is on the order of $\ee{-4}-\ee{-6}$. Considering that  $\Phi$ is a sum over contributions from $N_k=\ee{3}-\ee{4}$ wave vectors, the difference between $S(\mathbf k)$ and $S_0(\mathbf k)$ at a particular wave vector is about $\sqrt{\Phi/N_k}=\ee{-3.5}-\ee{-5}$. The efficiency and accuracy of this algorithm 
is verified by applying to a variety of target structure factors
described in Secs. \ref{known}-\ref{sec:conjecture}. We present justifications of this algorithm and parameter choices in Appendix~\ref{appendix:justification}.

\section{Proof of Concept: Targeting known realizable $S(k)$}
\label{known}

As a proof-of-concept, we test our ensemble-average algorithm here  by targeting several different structure factor functions 
across dimensions that are known to be exactly realizable. All of these examples
are special cases of determinantal point processes, which are those
whose $n$-point correlation functions are completely characterized by the determinant of
some function \cite{So00,mehta2004random,costin2004construction,
torquato2008point,hough09}.

\subsection{Dyson's One-Dimensional Log Coulomb Gases}
\label{sec:oneComponentPlasma}

\begin{figure}[ht]
\begin{center}
\includegraphics[width=0.45\textwidth]{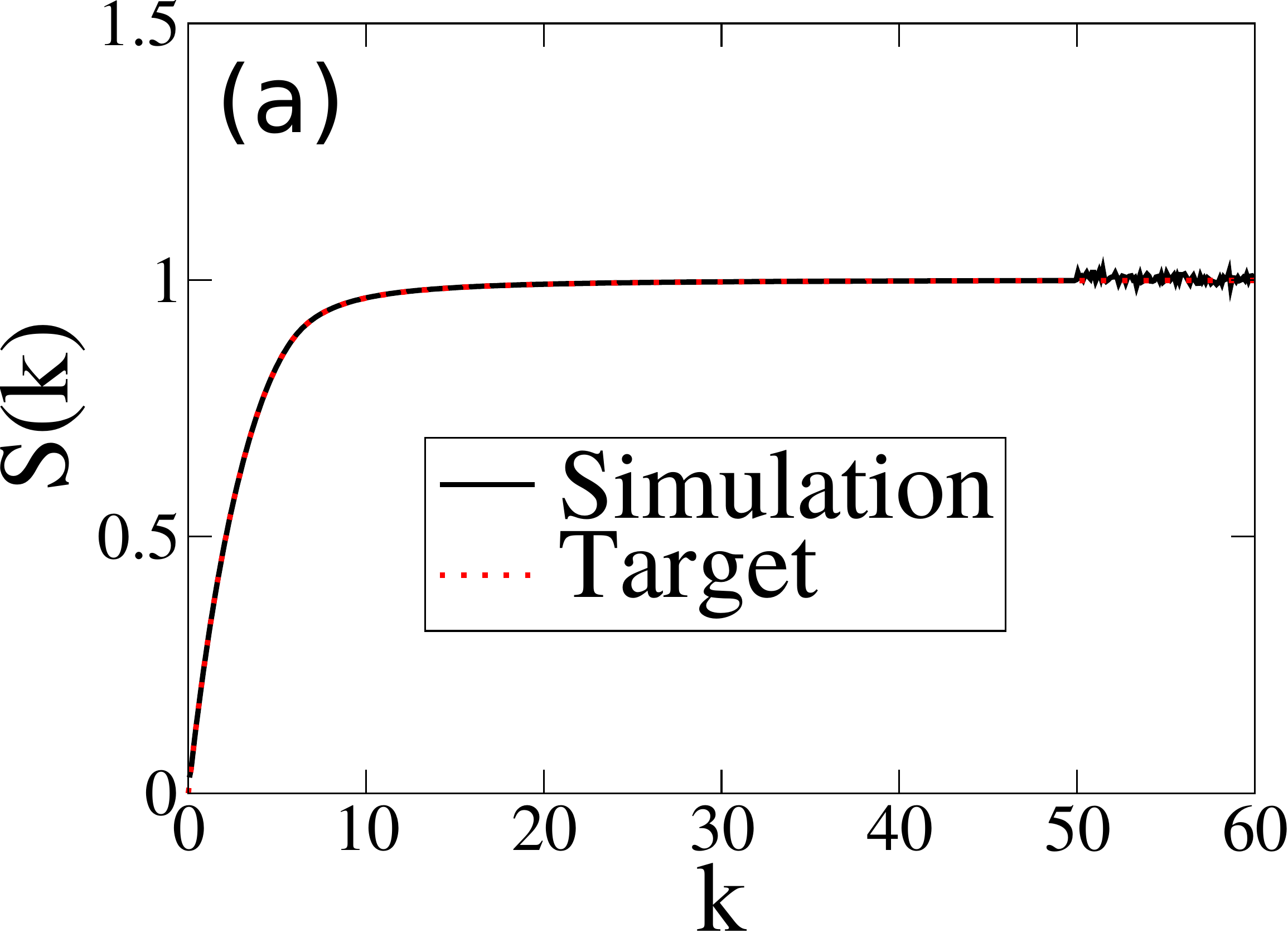}
\includegraphics[width=0.45\textwidth]{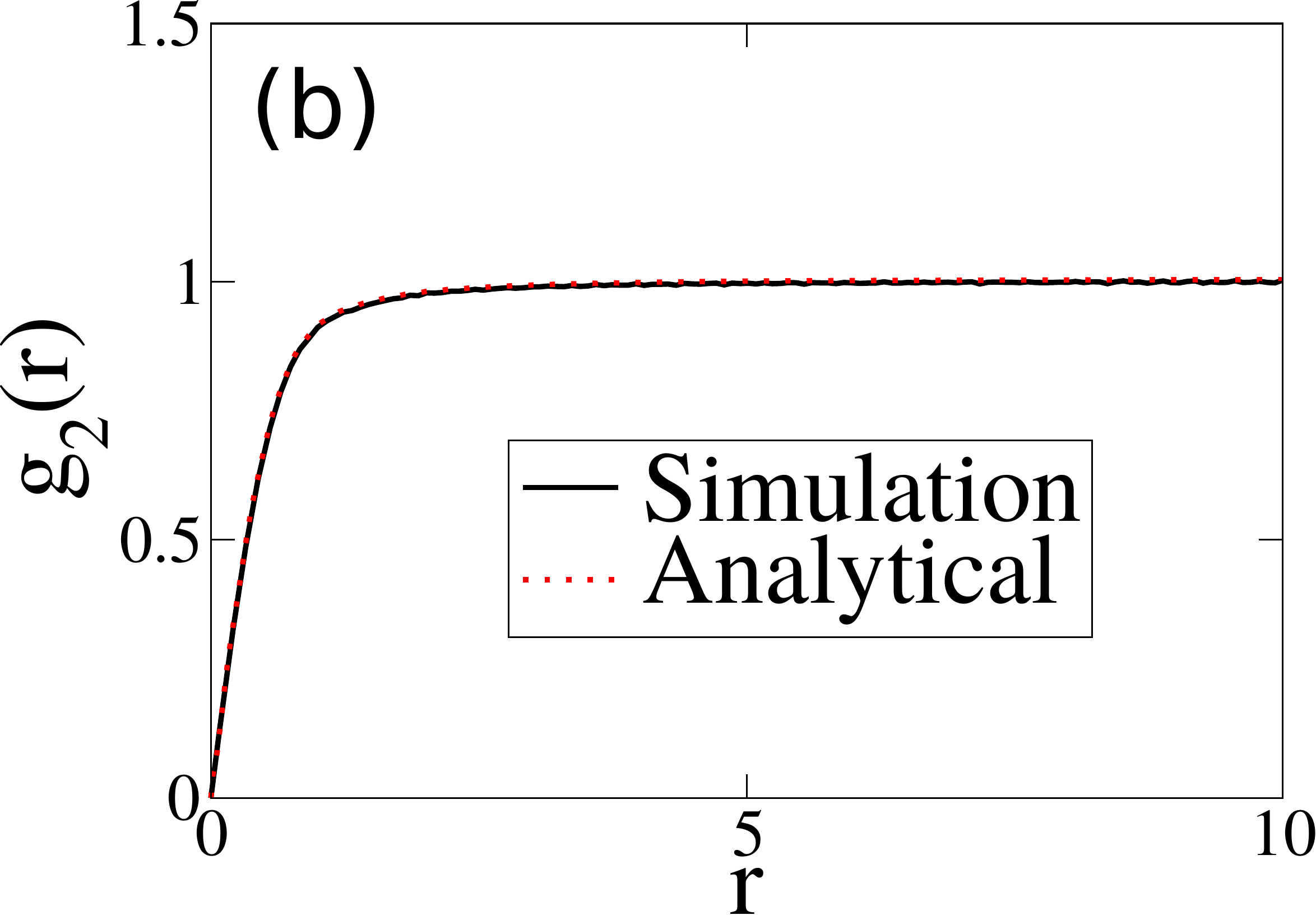}
\end{center}
\caption{(a): The structure factor obtained by sampling ensembles of 1D configurations in which
the  target function $S_0(k)$ is taken to be  Eq.~(\ref{1DSalLog1_S}) at $\rho=1$. (b): 
The corresponding pair correlation function sampled from simulations and the  analytical formula (\ref{coe-g2}).
Here it turns out that our usual reciprocal-space cutoff of $K=30$ is not large enough, 
and so we use $K=50$ instead.}
\label{1DSalLog1}
\end{figure}

\begin{figure}
\begin{center}
\includegraphics[width=0.45\textwidth,clip=]{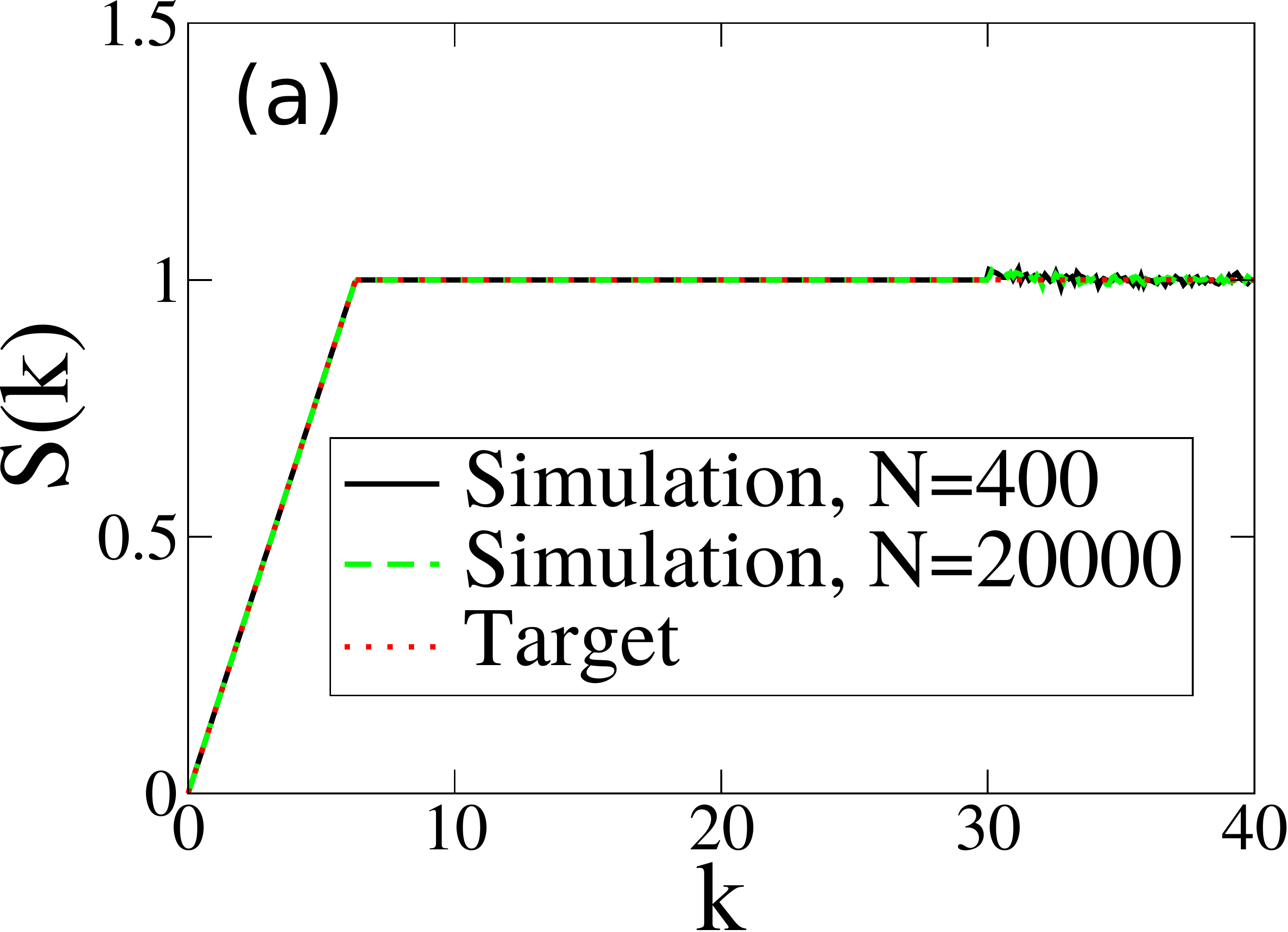}
\includegraphics[width=0.45\textwidth,clip=]{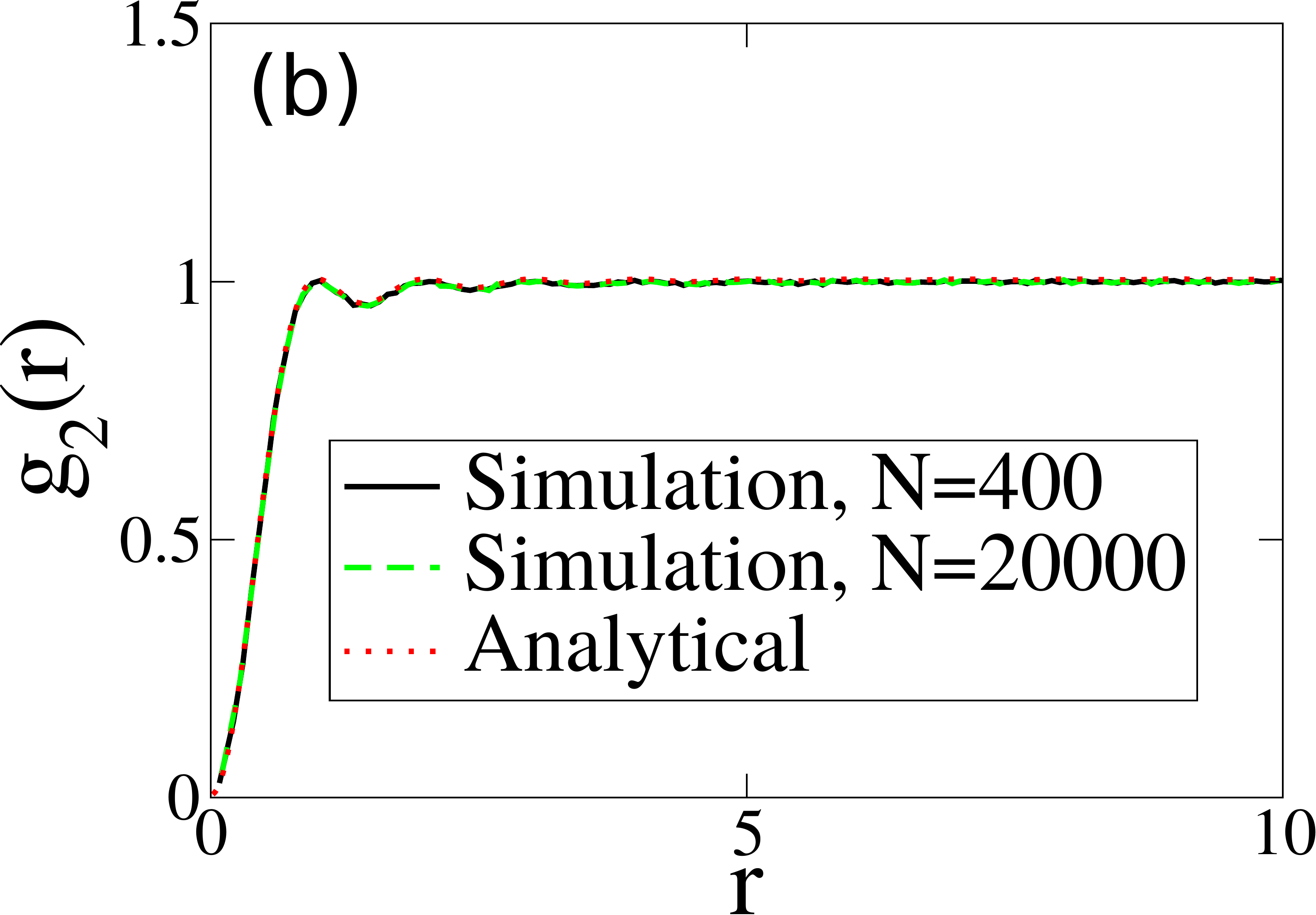}
\end{center}
\caption{(a): The structure factor obtained by sampling ensembles of 1D configurations in which
the  target function $S_0(k)$ is taken to be  Eq.~(\ref{1DFermionic_S}) at $\rho=1$. We use two different system sizes, and show here that their pair statistics are indistinguishable.
For $N=20000$, we generate 100 configurations instead of the usual 5000 configurations.
(b): 
The corresponding pair correlation function sampled from simulations and the  analytical formula (\ref{cue-g2}).}
\label{1DFermi}
\end{figure}

\begin{figure}
\begin{center}
\includegraphics[width=0.45\textwidth]{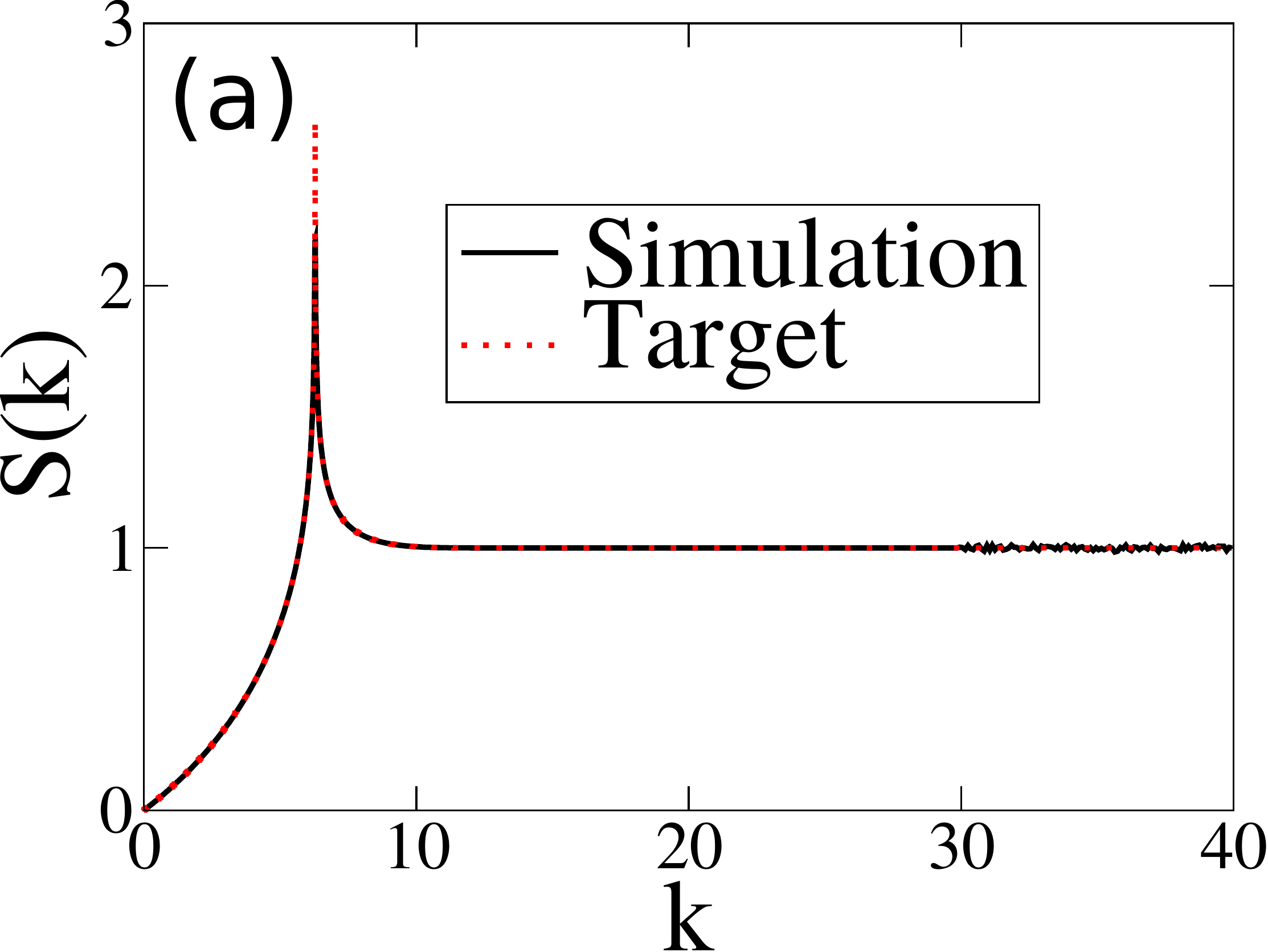}
\includegraphics[width=0.45\textwidth]{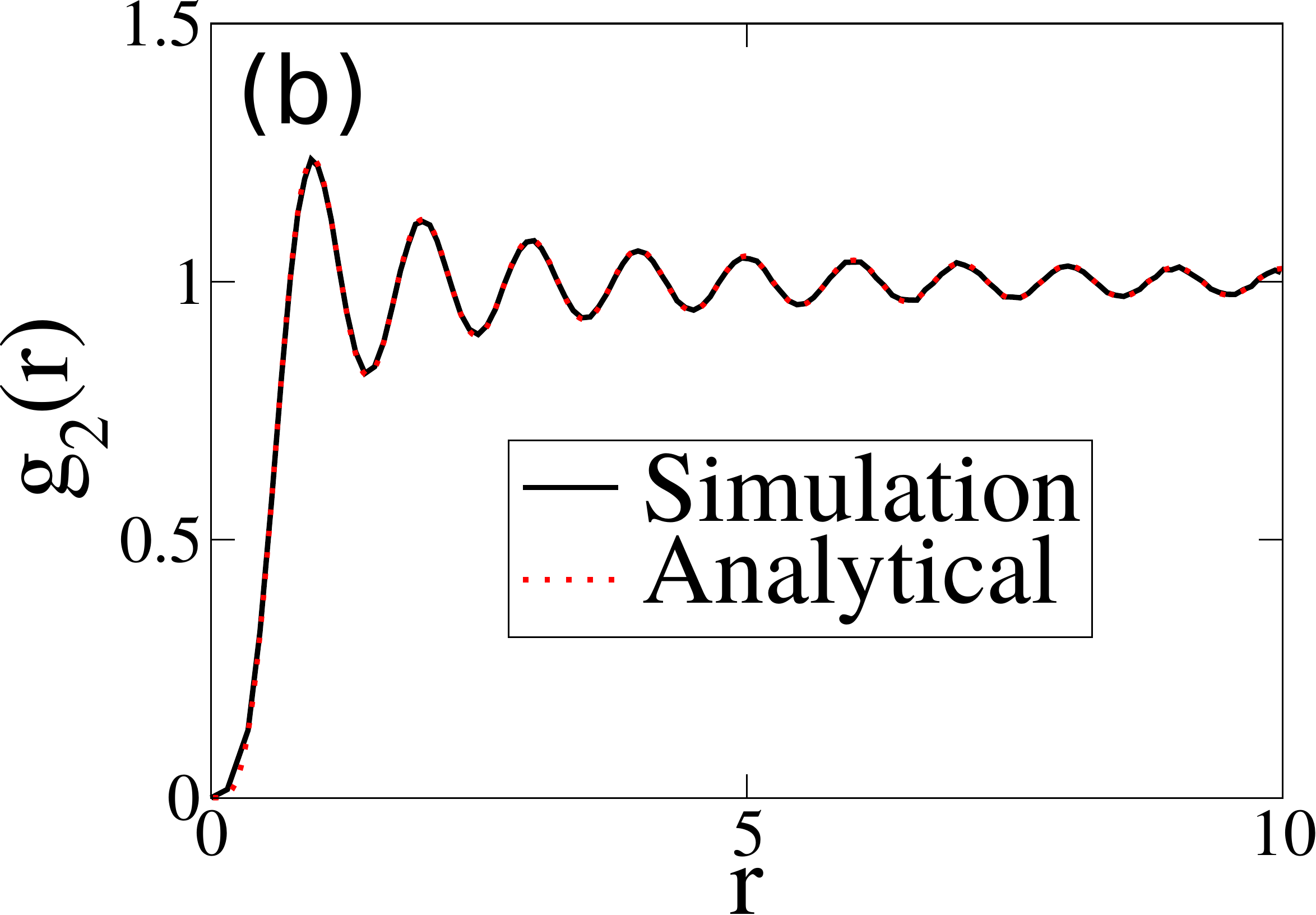}
\end{center}
\caption{(a): The structure factor obtained by sampling ensembles of 1D configurations in which
the  target function $S_0(k)$ is taken to be  Eq.~(\ref{1DSalLog2_S}) at $\rho=1$. (b): 
The corresponding pair correlation function sampled from simulations and the  analytical formula (\ref{cse-g2}).}
\label{1DSalLog2}
\end{figure}

Circular ensembles in the theory of random matrices \cite{mehta2004random}
are measures on spaces of unitary matrices introduced by Dyson as modifications of  Gaussian matrix ensembles.  
These different ensembles are equivalent to one another when the matrix size tends to infinity.
Dyson showed that the distribution of eigenvalues can be mapped to systems of  particles
on a circle interacting with a two-dimensional Coulomb potential (logarithmic function) at
positive temperatures. These systems in turn can be mapped to logarithmically interacting particles in $\mathbb{R}$
with an appropriately confining potential.

The structure factors that correspond to those of the circular orthogonal ensemble (COE), circular unitary ensemble (CUE), and circular symplectic ensemble (CSE), respectively \cite{mehta2004random, torquato2008point, scardicchio2009statistical, forrester2016analogies} at unit density ($\rho=1$) in the thermodynamic limit are:
\begin{eqnarray}
S(k) = \left\{
\begin{array}{lr}
{\displaystyle \frac{k}{\pi} -\frac{k}{2\pi}\ln(k/\pi+1)}, \quad 0 \le k \le 2\pi\\\\
\displaystyle{2 -\frac{k}{2\pi} \ln\left(\frac{k/\pi+1}{k/\pi -1}\right)}, \quad k > 2\pi,
\end{array}\right.
\label{1DSalLog1_S}
\end{eqnarray}
\begin{eqnarray}
S(k) = \left\{
\begin{array}{lr}
{\displaystyle \frac{k}{2\pi}}, \quad 0 \le k \le 2\pi\\\\
\displaystyle{1}, \quad k > 2\pi,
\end{array}\right.
\label{1DFermionic_S}
\end{eqnarray}
and
\begin{eqnarray}
S(k) = \left\{
\begin{array}{lr}
{\displaystyle \frac{k}{4\pi}
-\frac{k}{8\pi}\ln \Big|1-k/(2\pi)\Big|}, \quad 0 \le k \le 4\pi\\\\ \displaystyle{1}, \quad k > 4\pi.
\end{array}\right.
\label{1DSalLog2_S}
\end{eqnarray}
These ensembles correspond to the following values of the inverse temperature $\beta=(k_B T)^{-1}$:
 $\beta=1$ (COE), $\beta=2$ (CUE), and $\beta=4$ (CSE).  In all cases, we see that the structure factor $S(k)$ tends to 
zero linearly in $k$ in the limit $k \to 0$ and hence, according to (\ref{power}) and (\ref{sigma-asy}), are hyperuniform of class II \cite{To18a}.
The case $\beta=2$ has been generalized to higher dimensions (detailed below) and found to possess identical distribution with a spin-polarized fermionic system \cite{torquato2008point}.

The corresponding pair correlations of the COE, CUE and CSE are respectively
\begin{equation}
\begin{split}
g_2(r)=&1-\frac{\sin^2(\pi r)}{(\pi r)^2} \\ &+\frac{\Big(\pi r \cos(\pi r)-\sin(\pi r)\Big)\Big(2\, \mbox{Si}(\pi r)-\pi\Big)}{2(\pi r)^2} ,\
\end{split}
\label{coe-g2}
\end{equation}
\begin{eqnarray}
g_2(r)=1-\frac{\sin^2(\pi r)}{(\pi r)^2},
\label{cue-g2}
\end{eqnarray}
and
\begin{equation}
\begin{split}
g_2(r)=&1-\frac{\sin^2(2\pi r)}{(2\pi r)^2}\\ &+\frac{\Big(2\pi r \cos(2\pi r)-\sin(2\pi r)\Big)
\,\mbox{Si}(2\pi r)}{4(\pi r)^2}, 
\end{split}
\label{cse-g2}
\end{equation}

We now apply our algorithm by  targeting these three structure factors for systems with $N=400$.
The target analytical forms for the  structure factor
for $\beta=1$, $\beta=2$, and $\beta=4$ and the corresponding simulation data
are plotted in Figs.~ \ref{1DSalLog1}, ~\ref{1DFermi}, ~and~ \ref{1DSalLog2},
respectively. We include in these figures the corresponding pair correlation functions, both the analytical forms
and the simulation data, as obtained by sampling the generated configurations. From all of these figures, we see that our targeting algorithm enables us
to realize these ensembles with high accuracy, validating its utility and applicability.
For the $\beta=2$ case, we also carried out the simulation results for a much larger system with $N=20,000$.
We see from Fig.~\ref{1DFermi} that the results for $N=400$ are indistinguishable from those for $N=20000$.

By a theorem of Henderson \cite{henderson1974uniqueness}, a pair potential that gives rise to a given pair correlation function is unique up to a constant shift, although this cannot apply at $T=0$ \cite{footnote,Co07,Co19}.
The fact that our methodology yields ensembles of configurations
with targeted pair statistics (whenever realizable)  that are determined by effective pair potentials means
that those interactions in the case of Dyson's one-dimensional COE, CUE and CSE must exactly be given by the two-dimensional
Coulombic potential.

\begin{figure}
\begin{center}
\includegraphics[width=0.45\textwidth]{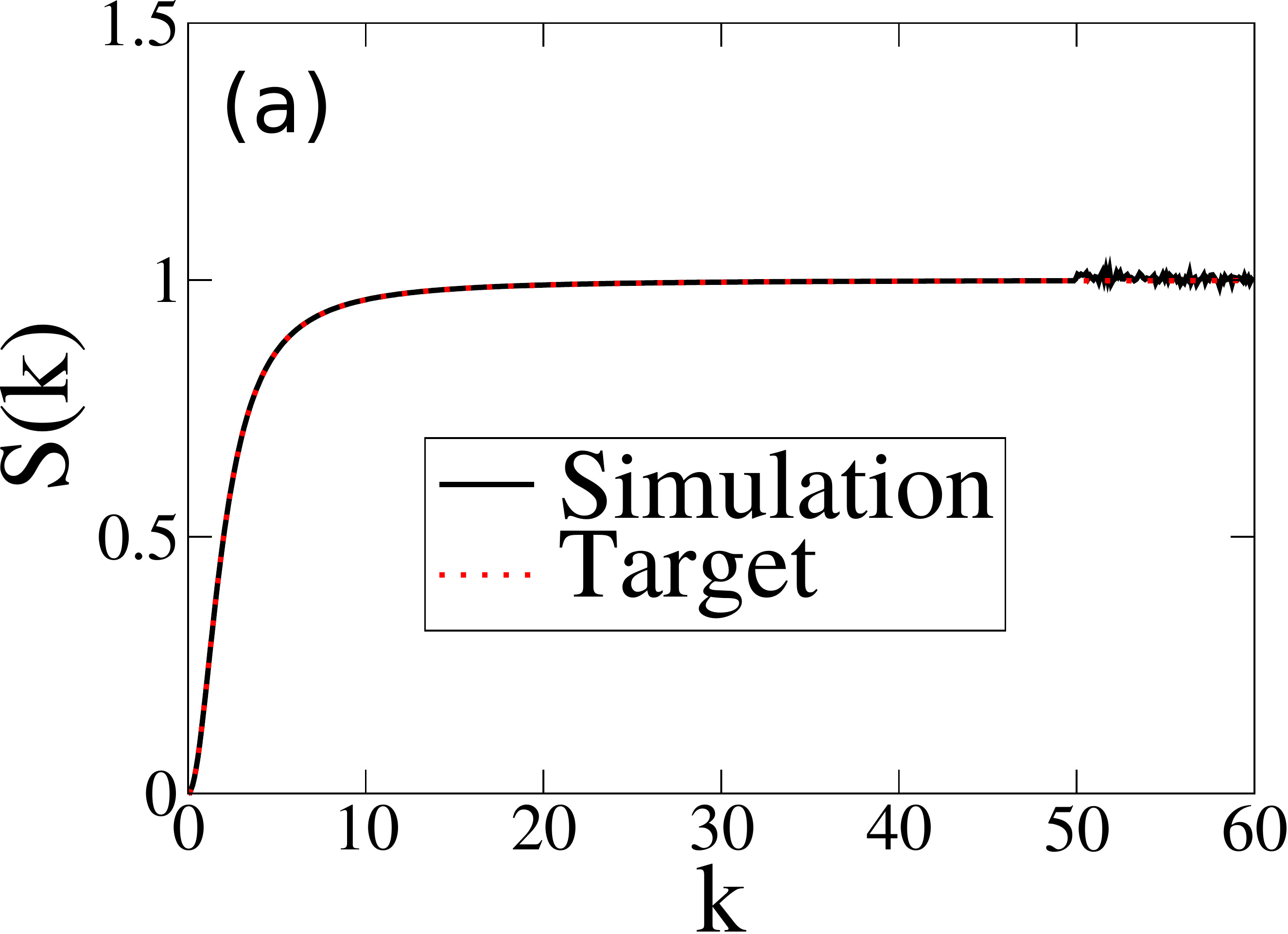}
\includegraphics[width=0.45\textwidth]{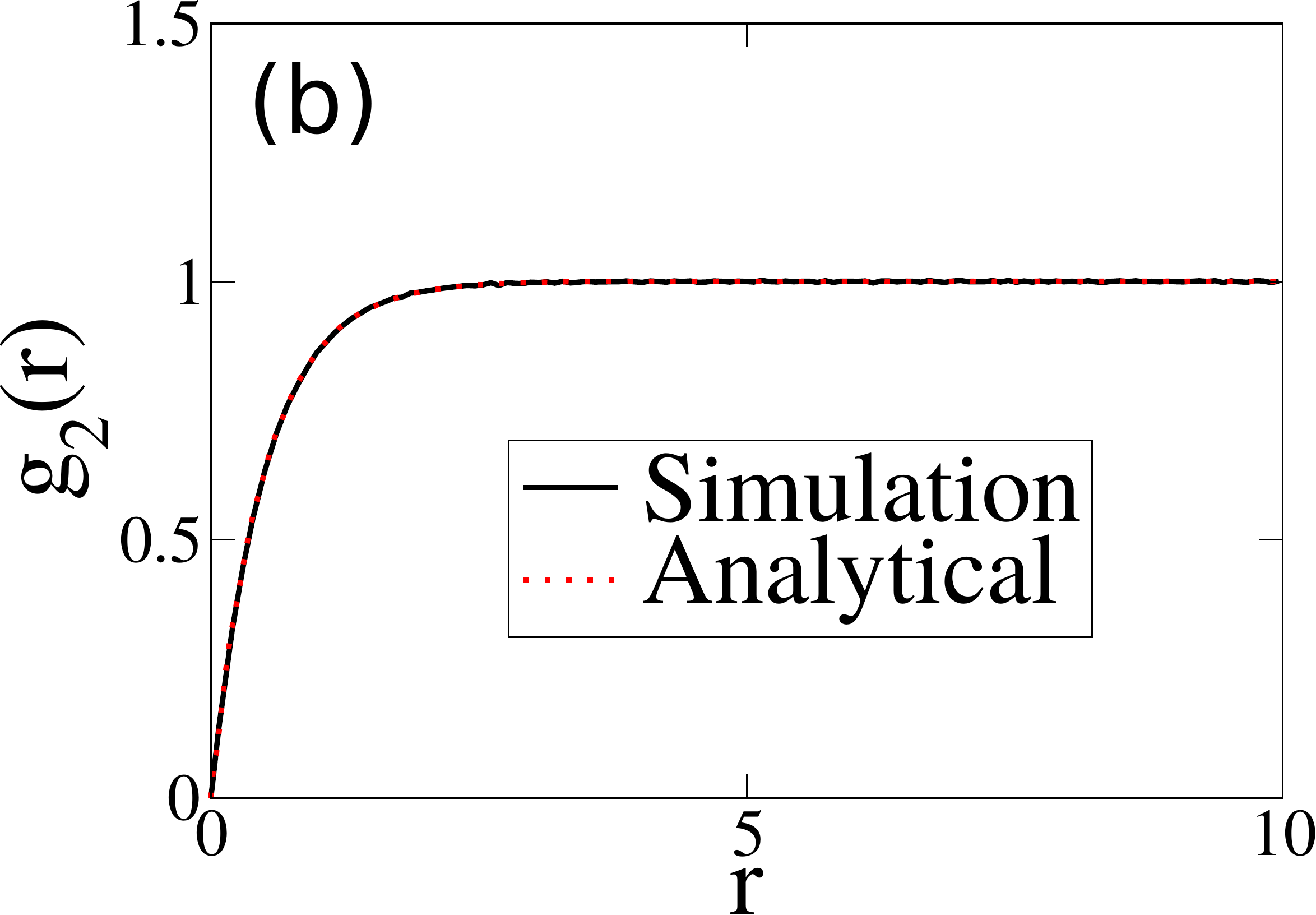}
\end{center}
\caption{(a): The structure factor obtained by sampling ensembles of 1D configurations in which
the  target function $S_0(k)$ is taken to be  Eq.~(\ref{L}) with $\lambda=2$ at $\rho=1$. (b): 
The corresponding pair correlation function sampled from simulations and the analytical formula as obtained from (\ref{h-exp}) [$g_2(r)=1-\exp(-2r)$].
Here it turns out that our usual reciprocal-space cutoff of $K=30$ is not large enough, 
and so we use $K=50$ instead.
}
\label{1DLorentzian}
\end{figure}

\subsection{One-dimensional Lorentzian target}
\label{sec:lorentzian}
Costin and Lebowitz \cite{costin2004construction} showed that there exists a one-dimensional
determinantal point process at unit density in which the total correlation function is the following
simple exponential function:
\begin{equation}
h(r) =- \exp(-\lambda r),
\label{h-exp}
\end{equation}
where $\lambda \ge 2$. This corresponds to a structure factor with a Lorentzian
form:
\begin{equation}
S(k)= \frac{\lambda(\lambda-2)+k^2}{k^2+\lambda^2}.
\label{L}
\end{equation}
 This result implies that the system is hyperuniform at the ``borderline" case of $\lambda=2$, since the structure factor $S(k)=k^2/(4+k^2)$ tends to 
zero quadratically in $k$ in the limit $k \to 0$  and hence, according to (\ref{power}) and (\ref{sigma-asy}), are hyperuniform of class I \cite{To18a}.
We have targeted the $\lambda=2$ target, 
and successfully realize it. The results for the pair correlation function and the structure factor
are presented in Fig.~\ref{1DLorentzian}.

\subsection{Fermi-sphere  targets in higher dimensions}

\begin{figure}
\begin{center}
\includegraphics[width=0.45\textwidth]{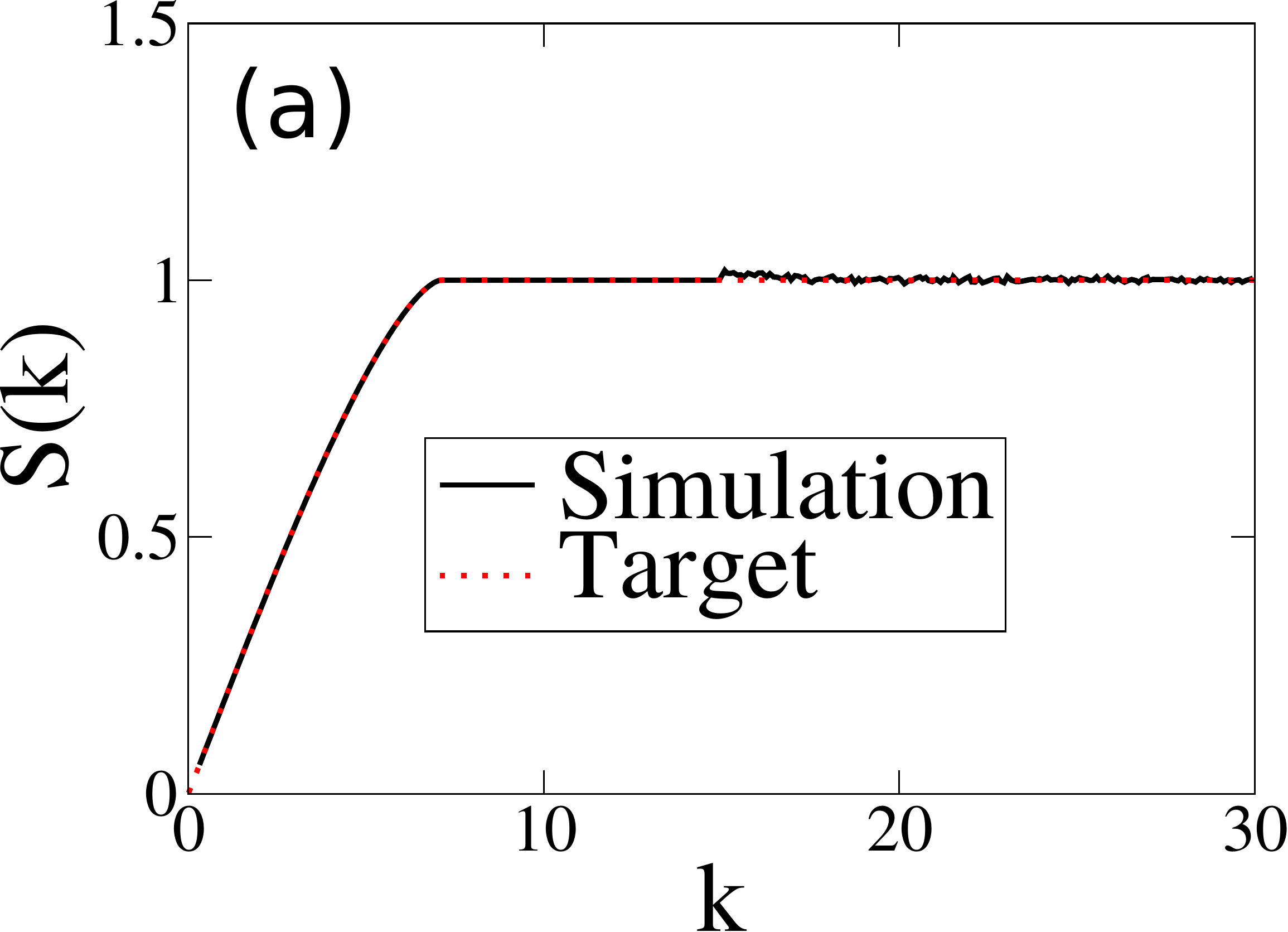}
\includegraphics[width=0.45\textwidth]{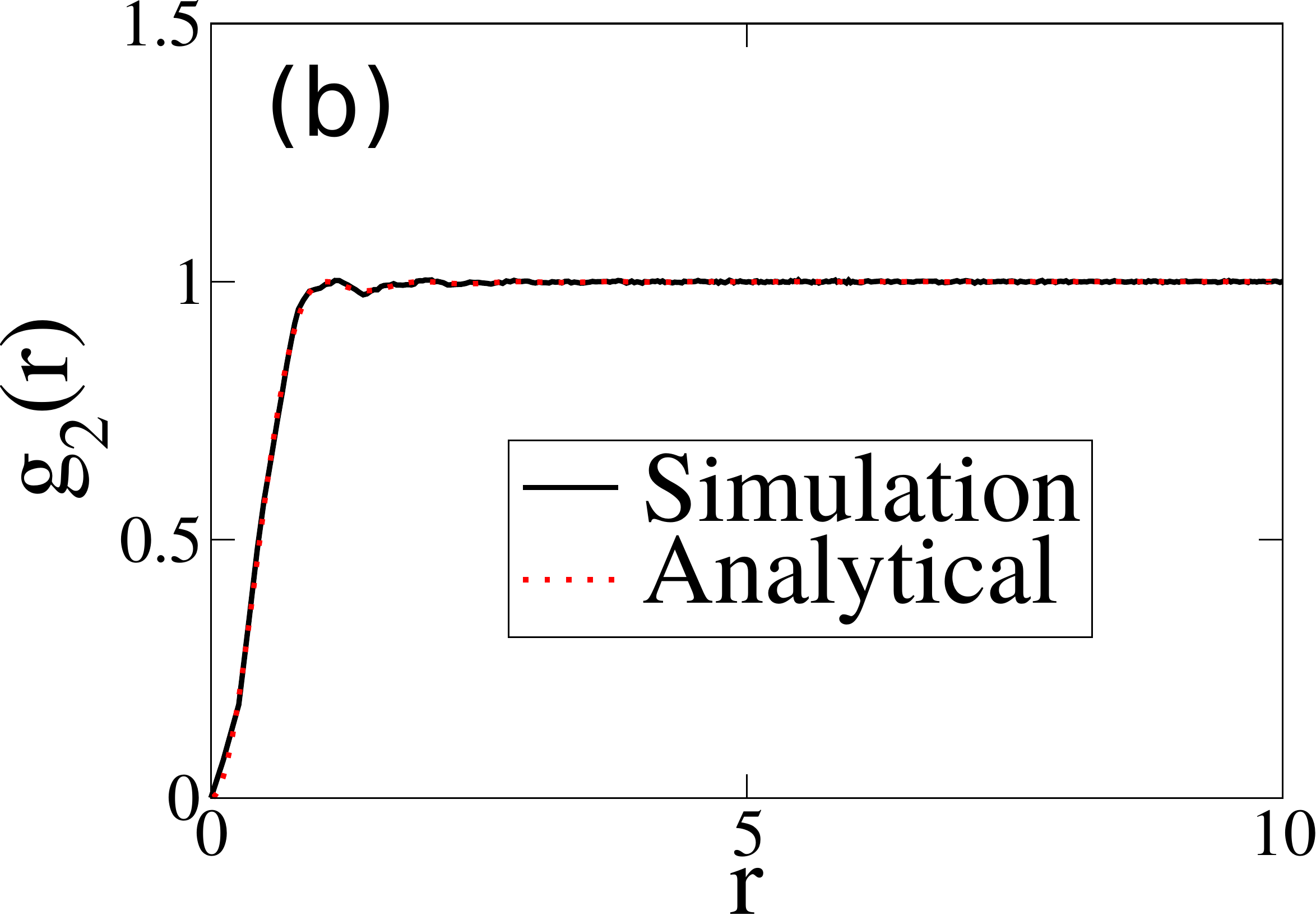}
\end{center}
\caption{(a): The structure factor obtained by sampling ensembles of 2D configurations in which
the  target function $S_0(k)$ is taken to be  Eq.~(\ref{Fermionic_S}) at $\rho=1$. (b): 
The corresponding pair correlation function sampled from simulations and the analytical formula (\ref{g2-d-2}).}
\label{2DFermionic}
\end{figure}
\begin{figure}
\begin{center}

\includegraphics[width=0.45\textwidth]{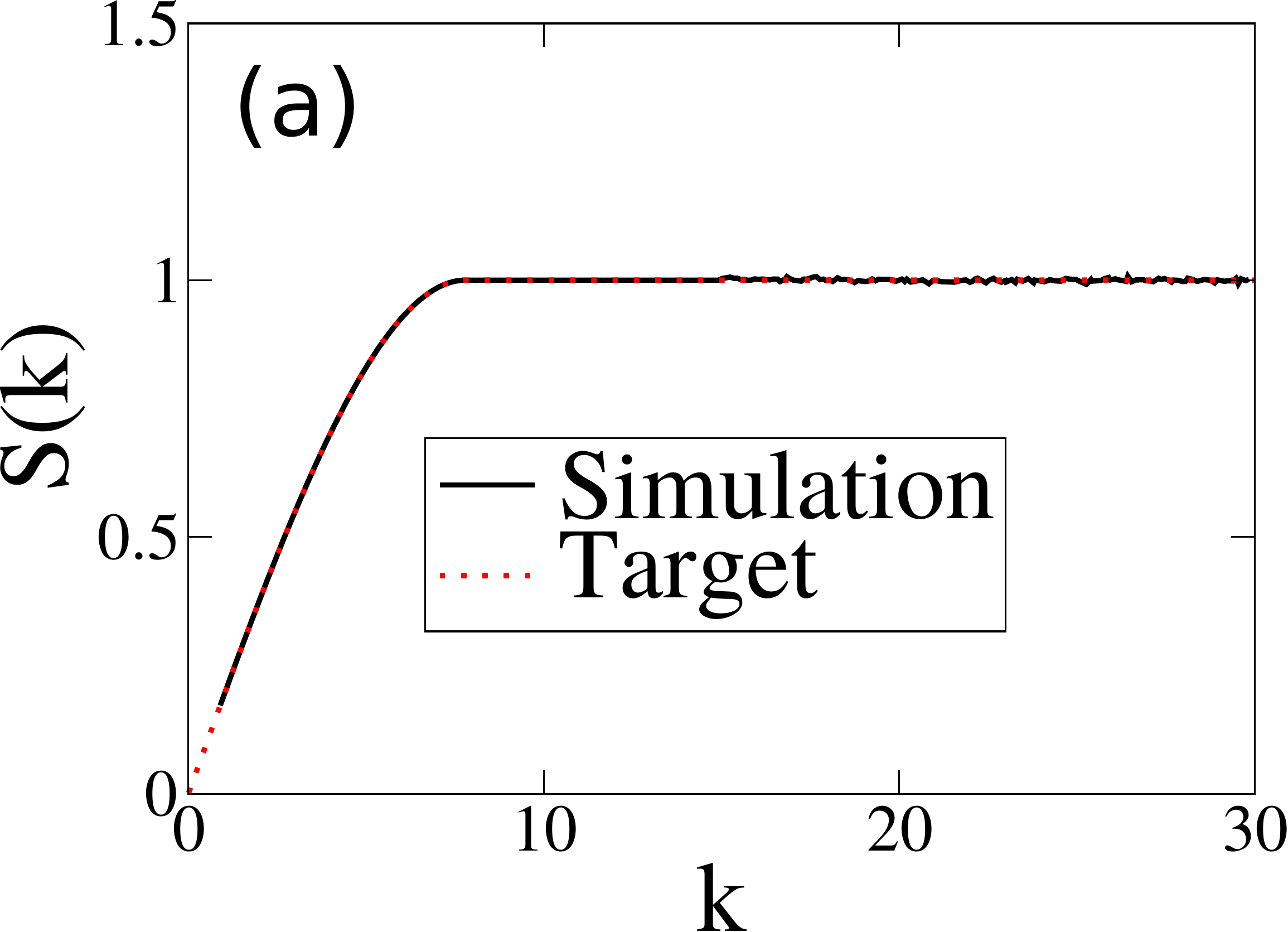}
\includegraphics[width=0.45\textwidth]{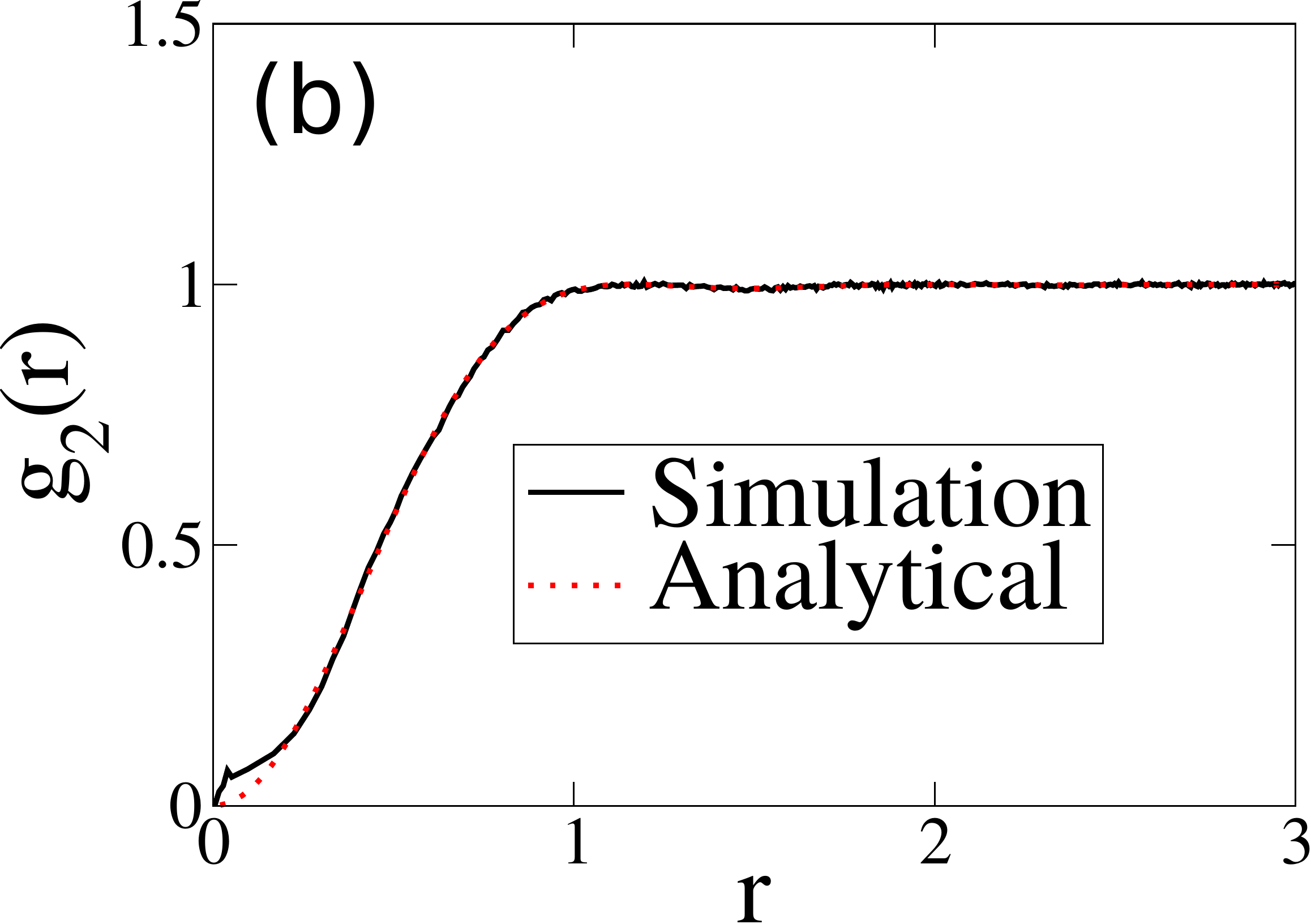}
\end{center}
\caption{(a): The structure factor obtained by sampling ensembles of 3D configurations in which
the  target function $S_0(k)$ is taken to be  Eq.~(\ref{Fermionic_S}) at $\rho=1$. (b): 
The corresponding pair correlation function sampled from simulations and the analytical formula (\ref{g2-d-2}).}
\label{3DFermionic}
\end{figure}

\begin{figure}
\begin{center}
\includegraphics[width=0.45\textwidth]{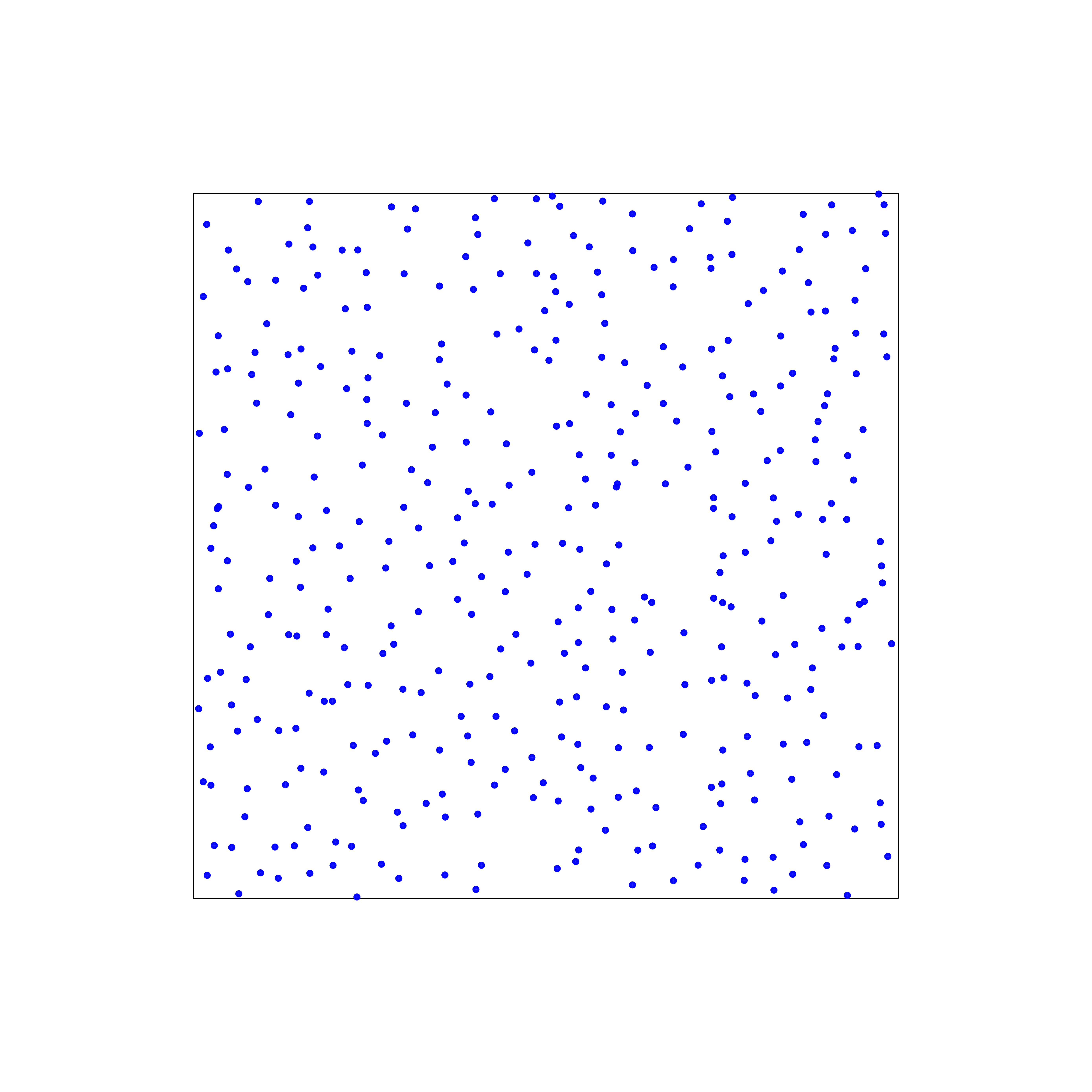}
\end{center}
\caption{A two-dimensional, 400-particle ``fermi-sphere'' configuration drawn from ensembles in which the target function $S_0(k)$ is taken to be Eq.~(\ref{Fermionic_S}) at $\rho=1$.}
\label{configurations}
\end{figure}

The 1D CUE point process with a structure factor given by  Eq.~(\ref{1DFermionic_S}) has been generalized
to so-called ``Fermi-sphere" point processes in $d$-dimensional Euclidean space $\mathbb{R}^d$  \cite{torquato2008point}. Specifically, such disordered hyperuniform 
point processes correspond to the spatial distribution of spin-polarized free fermions in $\mathbb{R}^d$,
which are special cases of determinantal processes. In particular, the structure factor in $\mathbb{R}$ at unit density
is given by
\begin{equation}
S(k)=1-\alpha(k, \kappa),
\label{Fermionic_S}
\end{equation}
where $\alpha(k, \kappa)$  is the volume common to two spherical windows of radius $\kappa$ whose centers are separated
by a distance $k$ divided by $v_1(\kappa)$, the volume of a spherical window of radius $\kappa=2\sqrt{\pi}[\Gamma(1+d/2)]^{1/d}$,
which is known analytically in any dimension \cite{torquato2006new}.  This result implies that the structure factor $S(k)$ tends to 
zero linearly in $k$ in
the limit $k \to 0$ and hence are hyperuniform of class II \cite{To18a}.
The corresponding pair correlation function of such a point process 
is given by
\begin{eqnarray}
g_2(r)= 1-2^d\Gamma(1+d/2)^2\frac{J^2_{d/2}(\kappa r)}{(\kappa r)^{d}},
\label{g2-d-2}
\end{eqnarray}
where $J_{\nu}(x)$ is the Bessel function of the first kind of order $\nu$.

We have applied our algorithm to target the structure factor (\ref{Fermionic_S}) in two and three dimensions using $K=15$. 
The results are presented in Figs.~\ref{2DFermionic}~and~\ref{3DFermionic} along with the corresponding
pair correlation functions sampled from the generated configurations as well as the analytical forms obtained
from (\ref{g2-d-2}). Consistent with the known realizability of these targets, we 
see excellent agreement between the targeted structure factors and those obtained from our ensemble-average
formulation. 
A two-dimensional configuration is shown in Fig.~\ref{configurations}.

Unlike the one-dimensional COE, CUE, and CSE
determinantal point configurations,  the interaction potential
for general determinantal point processes must contain at least
up to three-body potentials; see the Appendix of Ref.~\onlinecite{torquato2008point}.
Thus, for the 2D and 3D fermi-sphere targets as well as  the 1D Lorentzian target (Sec.~\ref{sec:lorentzian}),
we show for the first time that there exists  effective pair interactions that mimic the higher-order
$n$-body interactions corresponding to these determinantal
point processes.

\subsection{Gaussian target in two dimensions}

An example of a 2D determinantal point process that exhibits hyperuniform behavior is generated by
the Ginibre ensemble  \cite{jancovici1981exact}, which is a special case of the  two-dimensional
one-component plasma   \cite{jancovici1981exact}. A one-component plasma (OCP) is an equilibrium system of
identical point particles of charge $e$ interacting via the log Coulomb potential and immersed in a rigid, uniform background of opposite charge to ensure overall charge neutrality.
For $\beta=2$, the total correlation function for the OCP (Ginibre ensemble) in the thermodynamic limit
was found exactly  by Jancovici \cite{jancovici1981exact}:
\begin{eqnarray}
h(r) = -\exp\left(-\rho\pi r^2\right).
\label{h-OCP}
\end{eqnarray}
The corresponding structure factor is given by
\begin{equation}
S(k)=1-\exp\left(-\frac{k^2}{4\pi \rho}\right).
\label{S-OCP}
\end{equation}
 This result implies that the structure factor $S(k)$ tends to 
zero quadratically in $k$ in
the limit $k \to 0$ and hence are hyperuniform of class II \cite{To18a}.

\begin{figure}
\begin{center}

\includegraphics[width=0.45\textwidth]{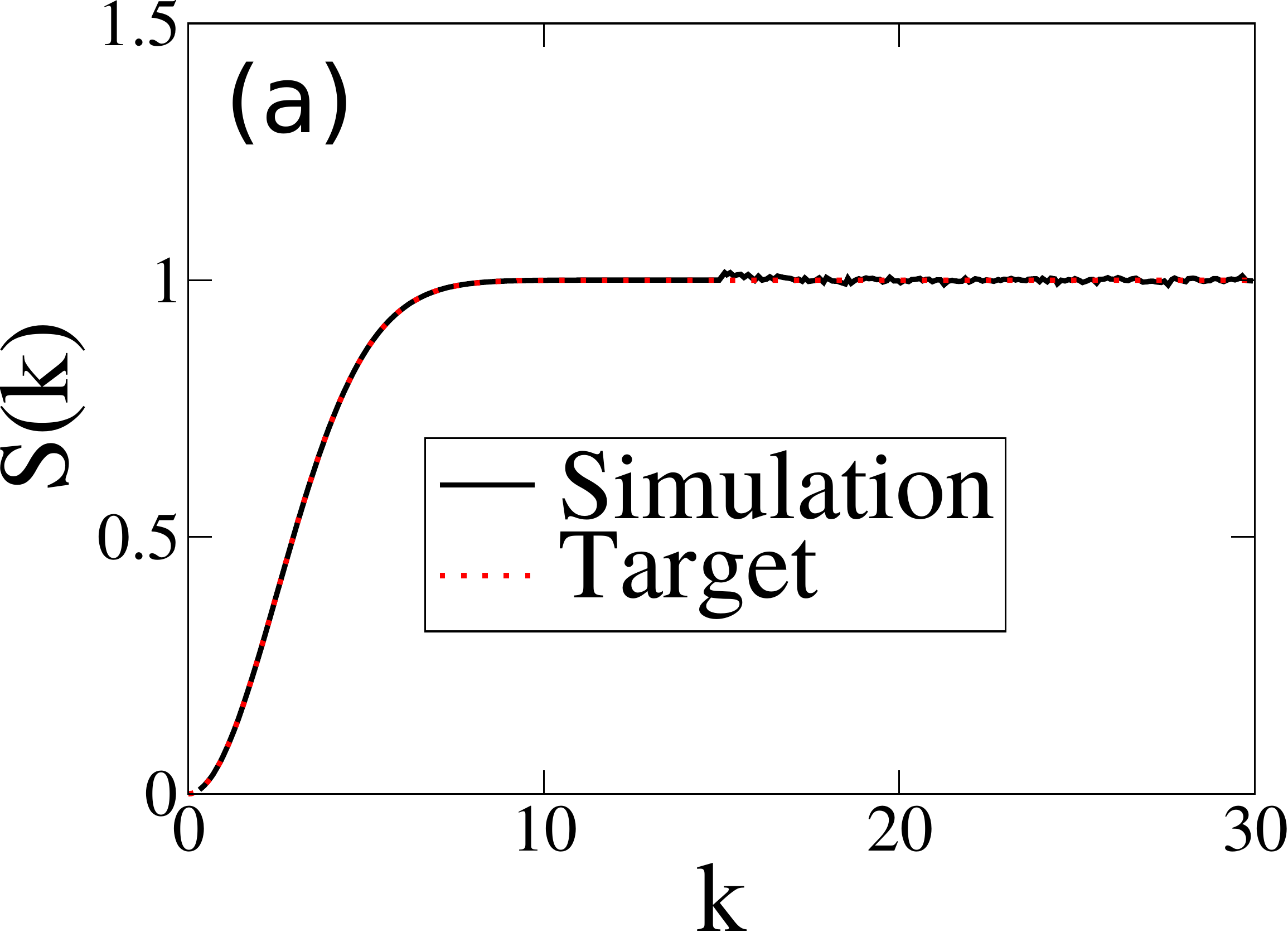}
\includegraphics[width=0.45\textwidth]{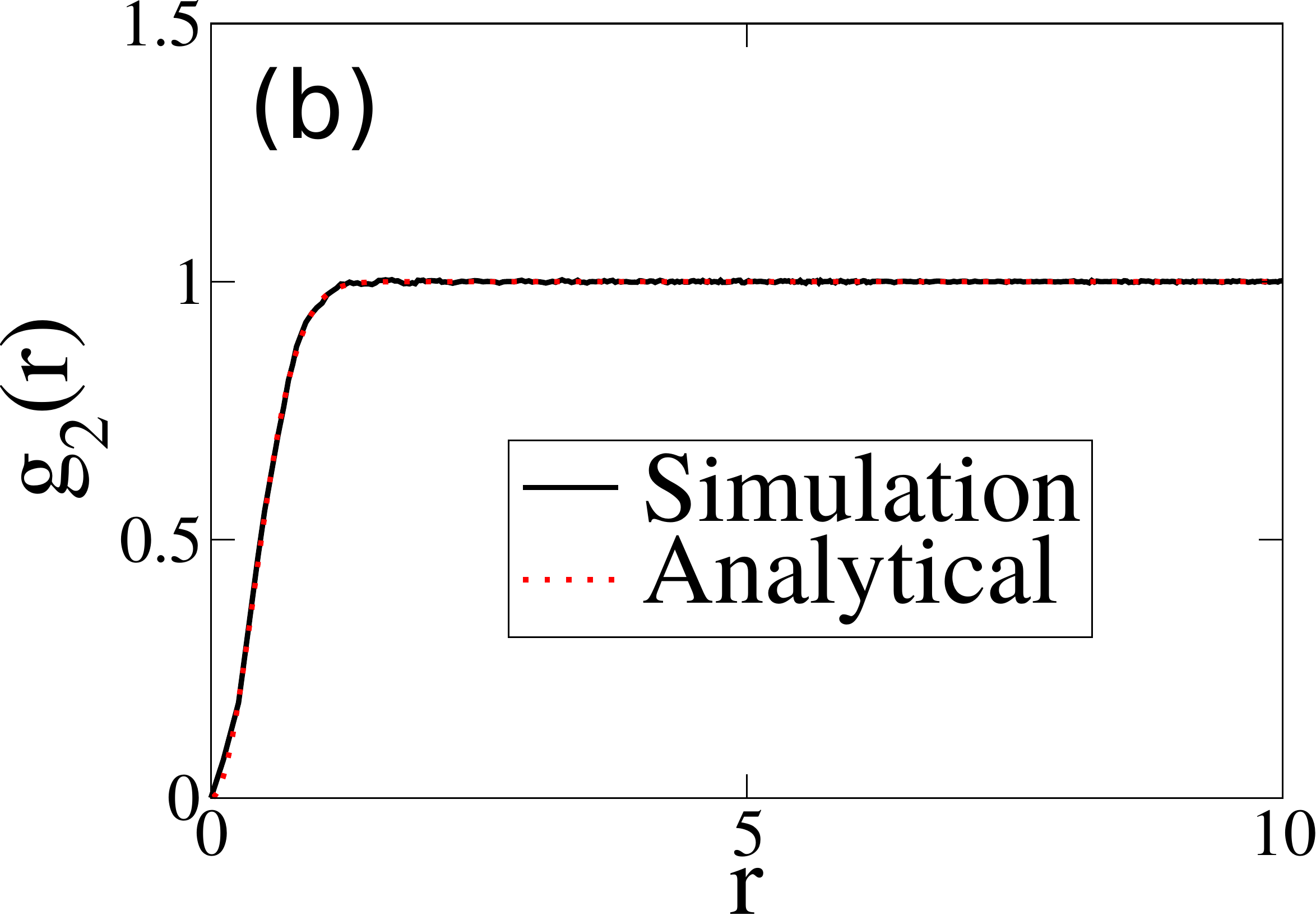}
\end{center}
\caption{(a): The structure factor obtained by sampling ensembles of 2D configurations in which
the  target function $S_0(k)$ is taken to be  Eq.~(\ref{S-OCP}) at $\rho=1$. (b): 
The corresponding pair correlation function sampled from simulations and the analytical formula (\ref{h-OCP}).}
\label{2DExpV}
\end{figure}

\begin{figure}
\begin{center}
\includegraphics[width=0.45\textwidth]{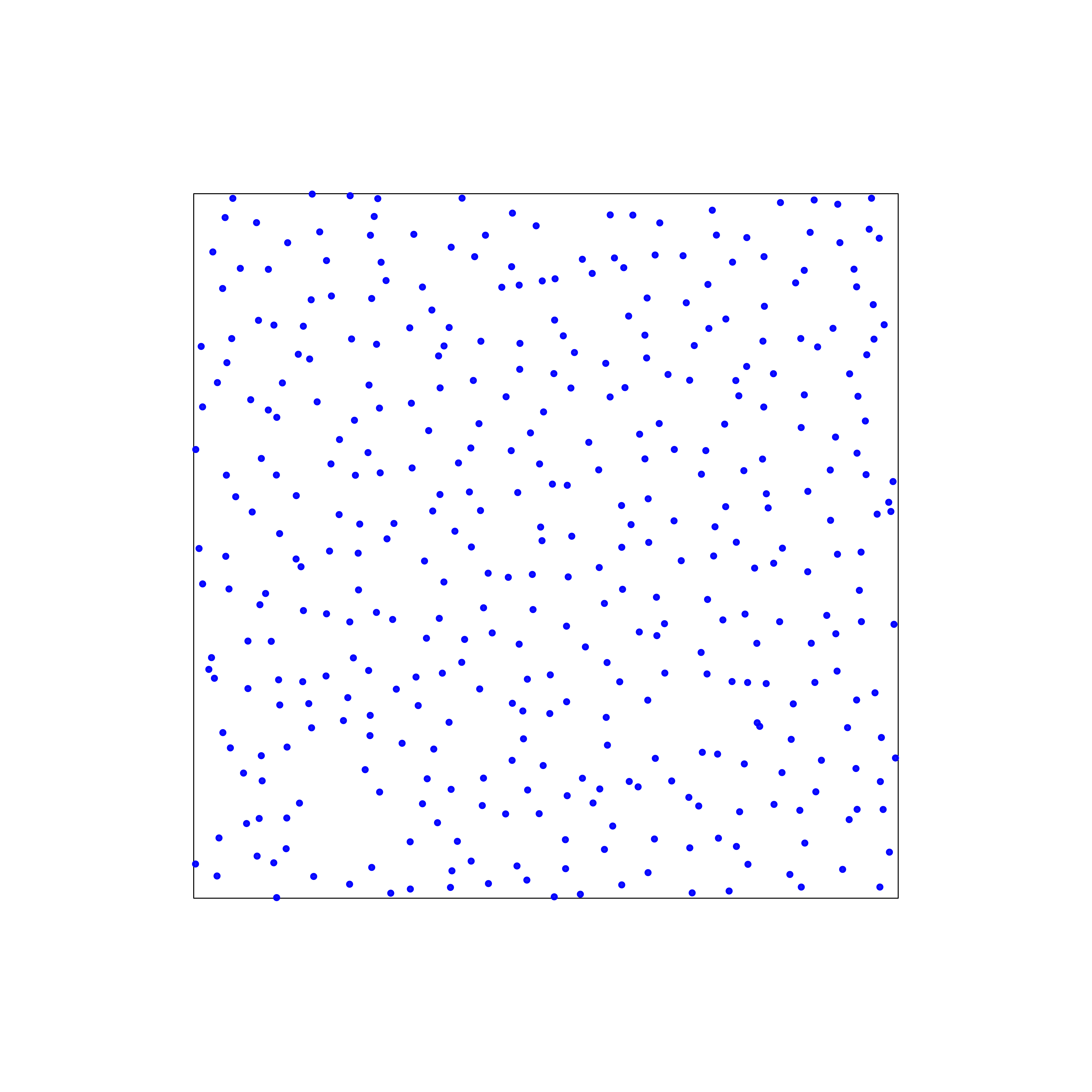}
\end{center}
\caption{A two-dimensional, 400-particle OCP configuration drawn from ensembles in which the target function $S_0(k)$ is taken to be Eq.~(\ref{S-OCP}) at $\rho=1$.}
\label{2DExpV_config}
\end{figure}

Using our method, we targeted the OCP structure factor (\ref{S-OCP}) using $K=15$.
As shown in Fig.~\ref{2DExpV}, it is seen that the algorithm is able to
realize this target with very high accuracy. The corresponding pair correlation
function obtained by sampling the resulting configurations agrees very well with 
the exact $g_2(r)$ obtained from (\ref{h-OCP}), as shown in Fig.~\ref{2DExpV}.
One configuration is shown in Fig.~\ref{2DExpV_config}.
It it noteworthy that we had previously employed a completely different algorithm to generate these configurations as well as other determinantal point processes \cite{scardicchio2009statistical}, but the maximum attainable system sizes 
were substantially much smaller ($N\approx 100$) in that study.

\section{Another proof of concept: Targeting a known unrealizable $S(k)$}
\label{impossible}

A severe test of our algorithm and another proof of concept would be its application  to  hypothetical functional forms
for pair statistics that meet the explicitly known necessary realizability
conditions  (\ref{g2Definition})-(\ref{yamada}), i.e., nonnegativity conditions
on $g_2(r)$ and $S(k)$ as well as the Yamada condition, but is 
known not be realizable. Such examples are rare. One particular
two-dimensional example was identified by Torquato and Stillinger \cite{torquato2006new} in which
the point configuration  would putatively correspond
to a packing of identical hard circular disks of unit
diameter with a pair correlation function given by
\begin{equation}
g_2(r)=\Theta(r-\sigma)+\frac{Z}{2\pi\rho}\delta(r-1),
\label{step3}
\end{equation}
where $\Theta(x)$ is the unit step function, $\delta(r)$
is a radial Dirac delta function, $\sigma=1.2946$ and $Z=4.0148$.  
The corresponding structure is given by
\begin{equation}
S(k)=1 -\frac{ 8\phi\sigma^{2}  }{(k\sigma)} J_{1}(k\sigma)
+   Z   J_{0}(k),
\label{exp-S}
\end{equation}
where $\phi=0.74803$ is the packing fraction.
It turns out that both $g_2(r)$ and $S(k)$ are
nonnegative functions and the Yamada condition
is satisfied. However, Torquato and Stillinger \cite{torquato2006new}
observed that the test function (\ref{step3}) 
 cannot correspond to a  packing because it violates
local geometric constraints specified by a distance $\sigma$ and 
average contact number (per particle) $Z$.
Specifically, for $Z= 4.0148$, there must be particles
that are in contact with at least five others.
But no arrangement of the five exists that is consistent with the assumed
pair correlation function (step plus delta function with a gap from 1 to
1.2946).

\begin{figure}
\begin{center}

\includegraphics[width=0.45\textwidth]{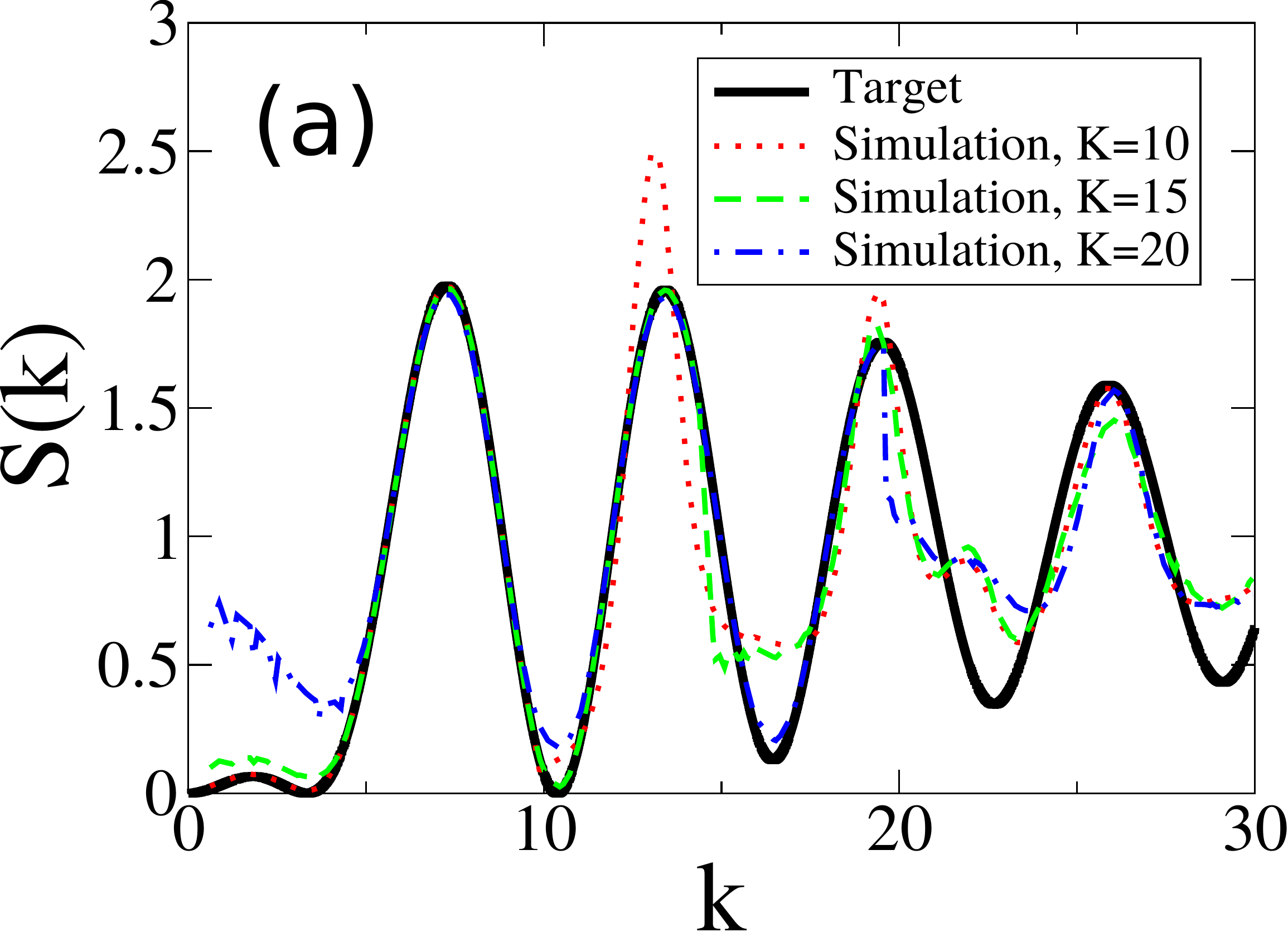}
\includegraphics[width=0.45\textwidth]{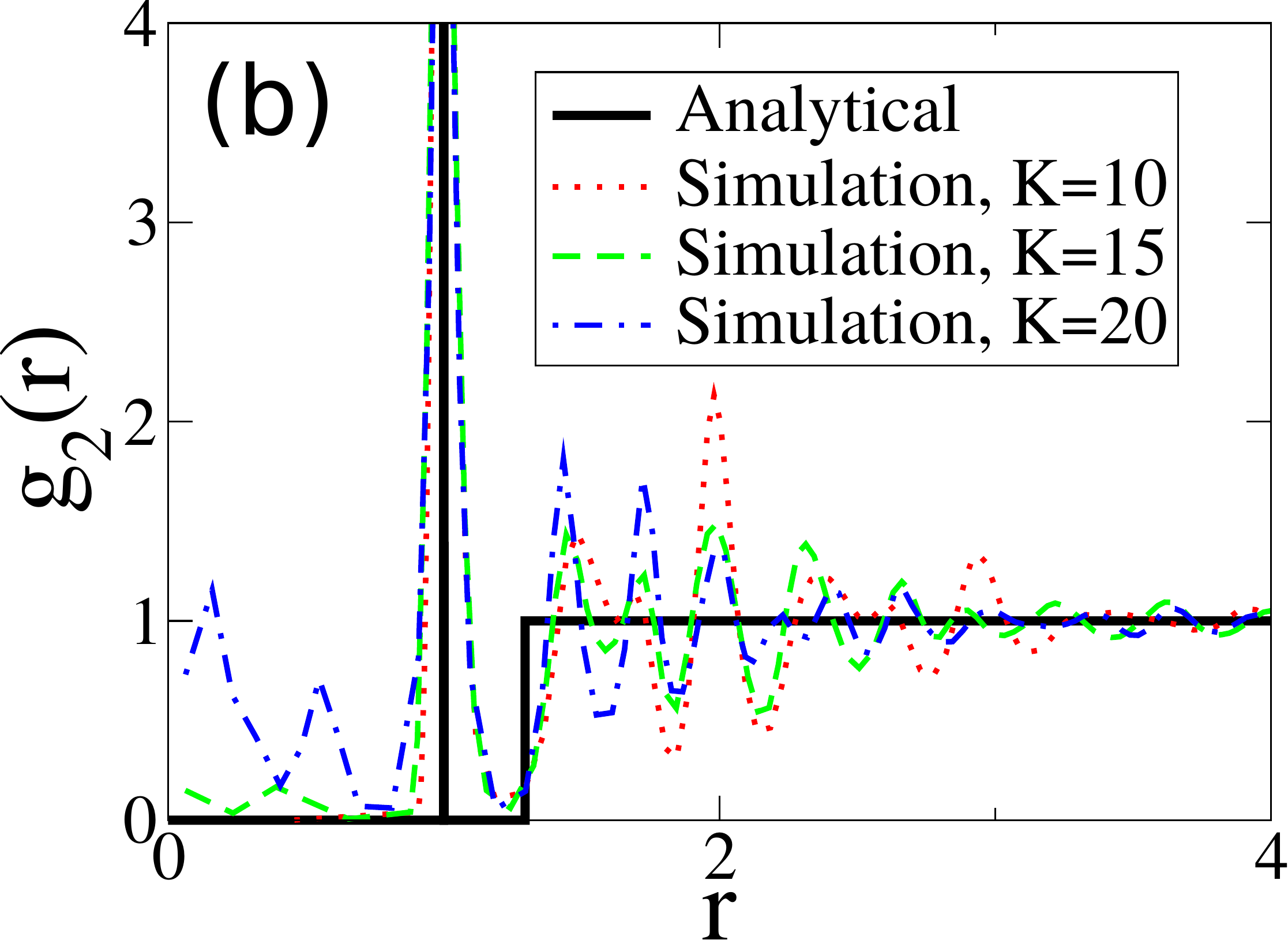}

\end{center}
\caption{(a): The structure factor obtained by sampling ensembles of 1D configurations in which
the  target function $S_0(k)$ is taken to be  Eq.~(\ref{exp-S}) at $\rho=1$. (b): 
The corresponding pair correlation function sampled from simulations and the analytical formula (\ref{step3}).}

\label{exp}
\end{figure}

We use our standard procedure described in Sec.~\ref{newAlgorithm} to target
the structure factor (\ref{exp-S}), but we change three parameters. We take
$N_c=1000$ (rather than $N_c=100$) to ensure that any failure is not due to lacking degrees of freedom
and use $N=100$ (rather than $N=400$) to compensate for the increase in simulation time caused by the previous change.
Finally, we experimented with several values of $K$ values (shown in Fig.~\ref{exp}), instead of the standard usage of $K=15$ in 2D.

We present results for three different reciprocal-space cutoff values: $K=10,15,20$.
For $K=10$, the structure factor can match the target 
inside the constrained region. Importantly, for the two larger $K$ values, the optimizer finds local minima, and the final structure factors 
(at the end of the minimization) does not match the target, even inside the constrained region. Therefore, we conclude
that the structure factor (\ref{exp-S}) is not realizable, which speaks to the power of our algorithm.

\section{Targeting hyperuniform structure factors with unknown realizability}
\label{unknown}

In this section, we apply our ensemble-average algorithm to target several different hyperuniform functional forms for structures factors across dimensions
that satisfy the explicitly known necessary conditions (\ref{g2Definition})-(\ref{yamada}), but are not known to be realizable. We show that all of these $d$-dimensional targets are indeed realizable.

Before presenting these results, it is instructive
to comment on the effect of space dimensionality on realizing
a prescribed structure factor. It is generally known
that the lower the space dimension, the more difficult it is to satisfy
realizability conditions \cite{torquato2006new}. This is consistent with 
{\it decorrelation principle} \cite{torquato2006new}, which states that unconstrained correlations
in disordered many-particle systems vanish asymptotically in high
dimensions and that the $n$-particle correlation function $g_n$  for any $n \ge 3$
can be inferred entirely from a knowledge of $\rho$ and $g_2$. This in turn
implies that the nonnegativity of $g_2(r)$ and $S(k)$ are sufficient conditions for realizability. It was
also shown that the decorrelation principle applies more generally to lattices
in high dimensions \cite{An16}.

\subsection{Gaussian Structure Factor Across Dimensions}

\begin{figure}
\begin{center}

\includegraphics[width=0.45\textwidth]{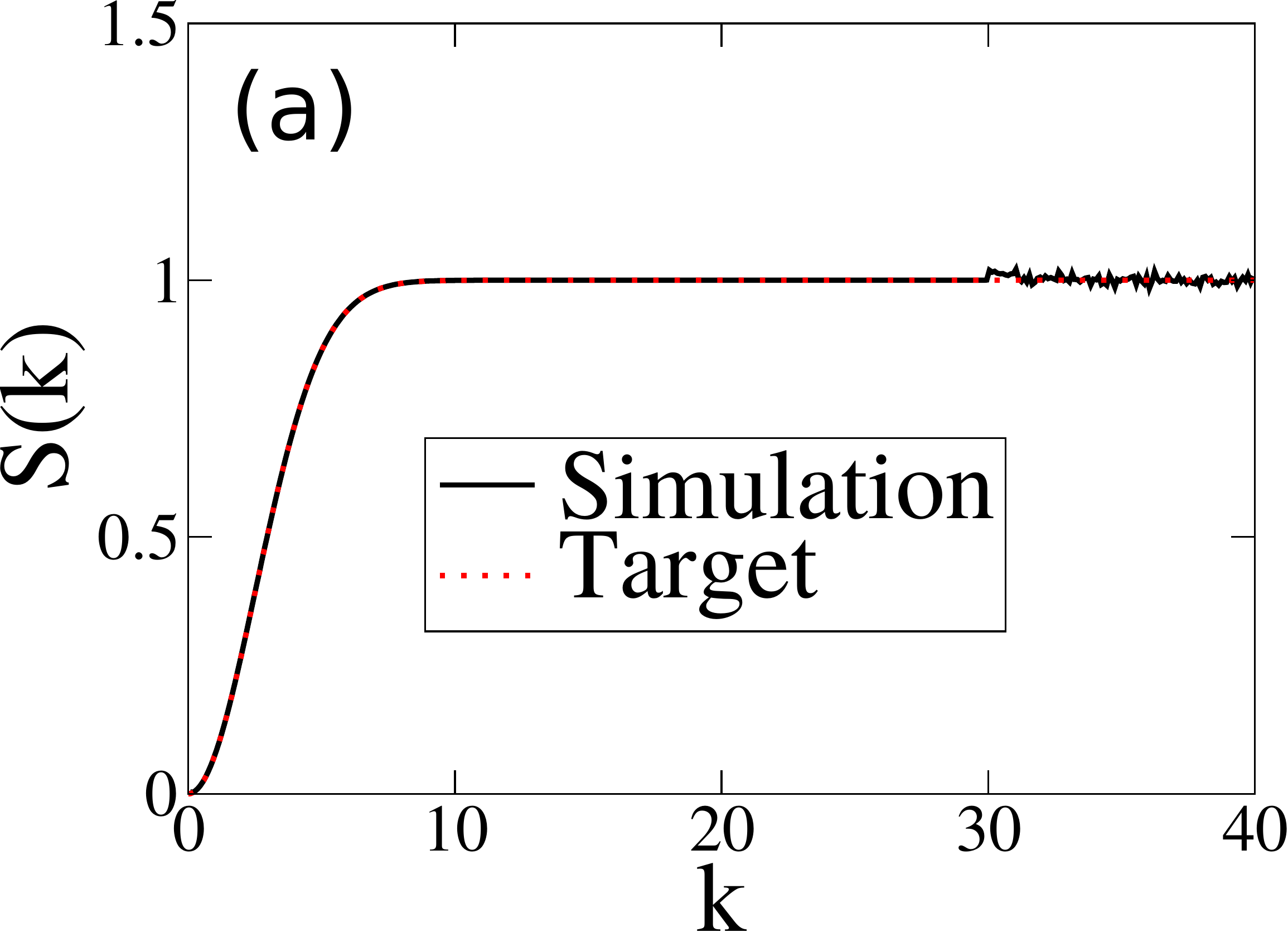}
\includegraphics[width=0.45\textwidth]{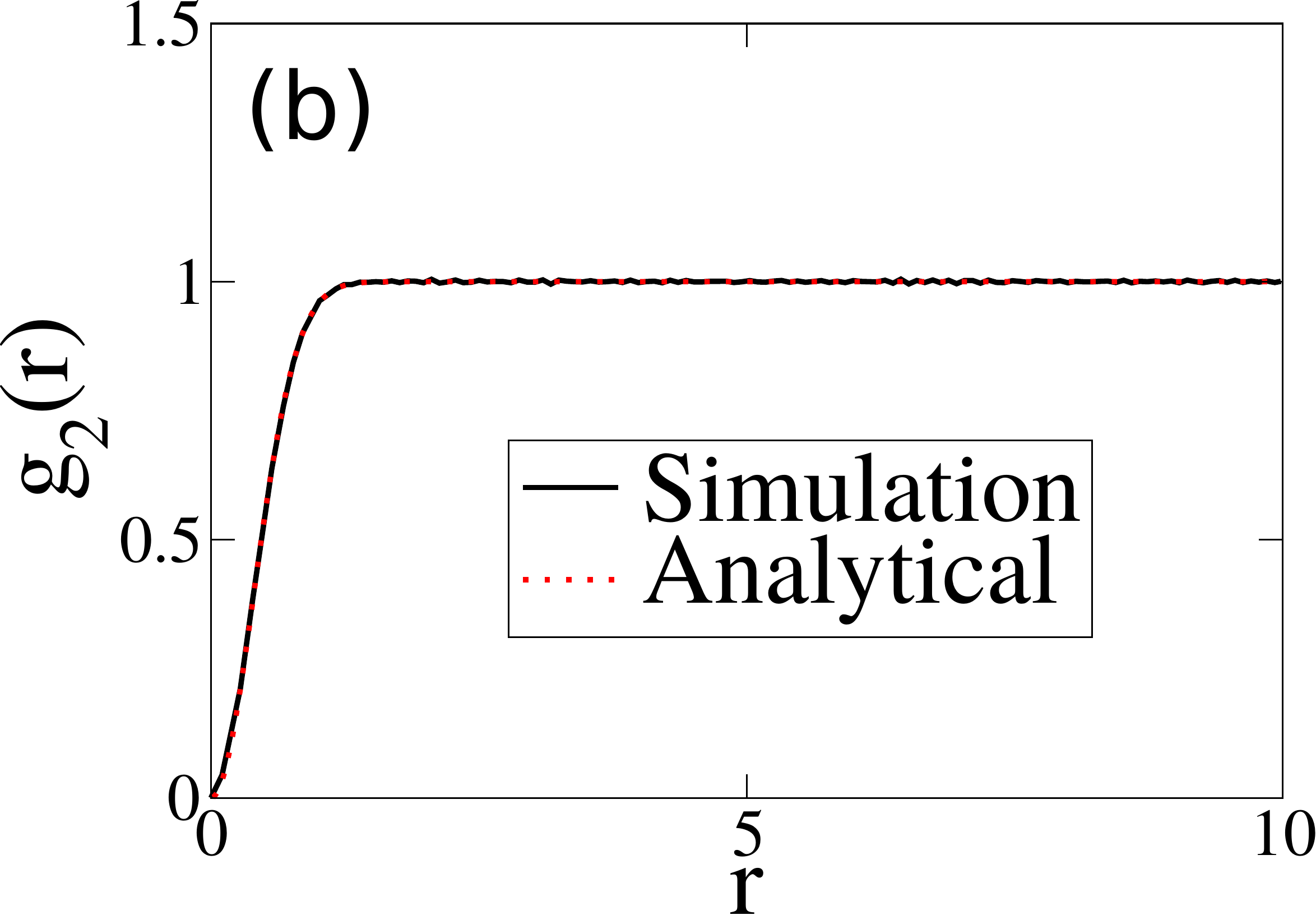}
\end{center}
\caption{(a): The structure factor obtained by sampling ensembles of 1D configurations in which
the  target function $S_0(k)$ is taken to be  Eq.~(\ref{S-d-OCP}) with $a=1/\sqrt{\pi}$ at $\rho=1$. (b): 
The corresponding pair correlation function sampled from simulations and the analytical formula (\ref{h-d-OCP}).}

\label{1DGaussian}
\end{figure}

\begin{figure}
\begin{center}

\includegraphics[width=0.45\textwidth]{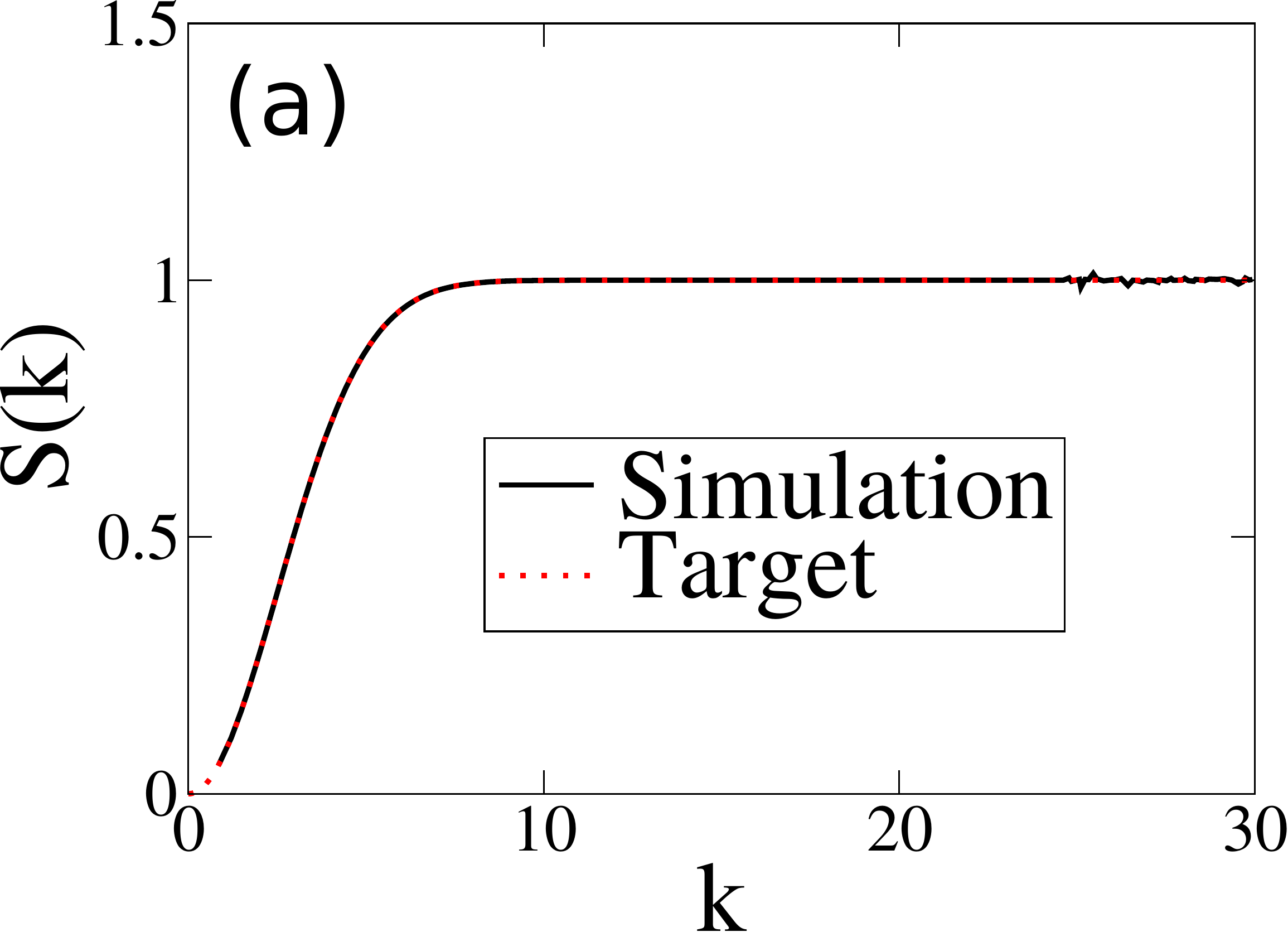}
\includegraphics[width=0.45\textwidth]{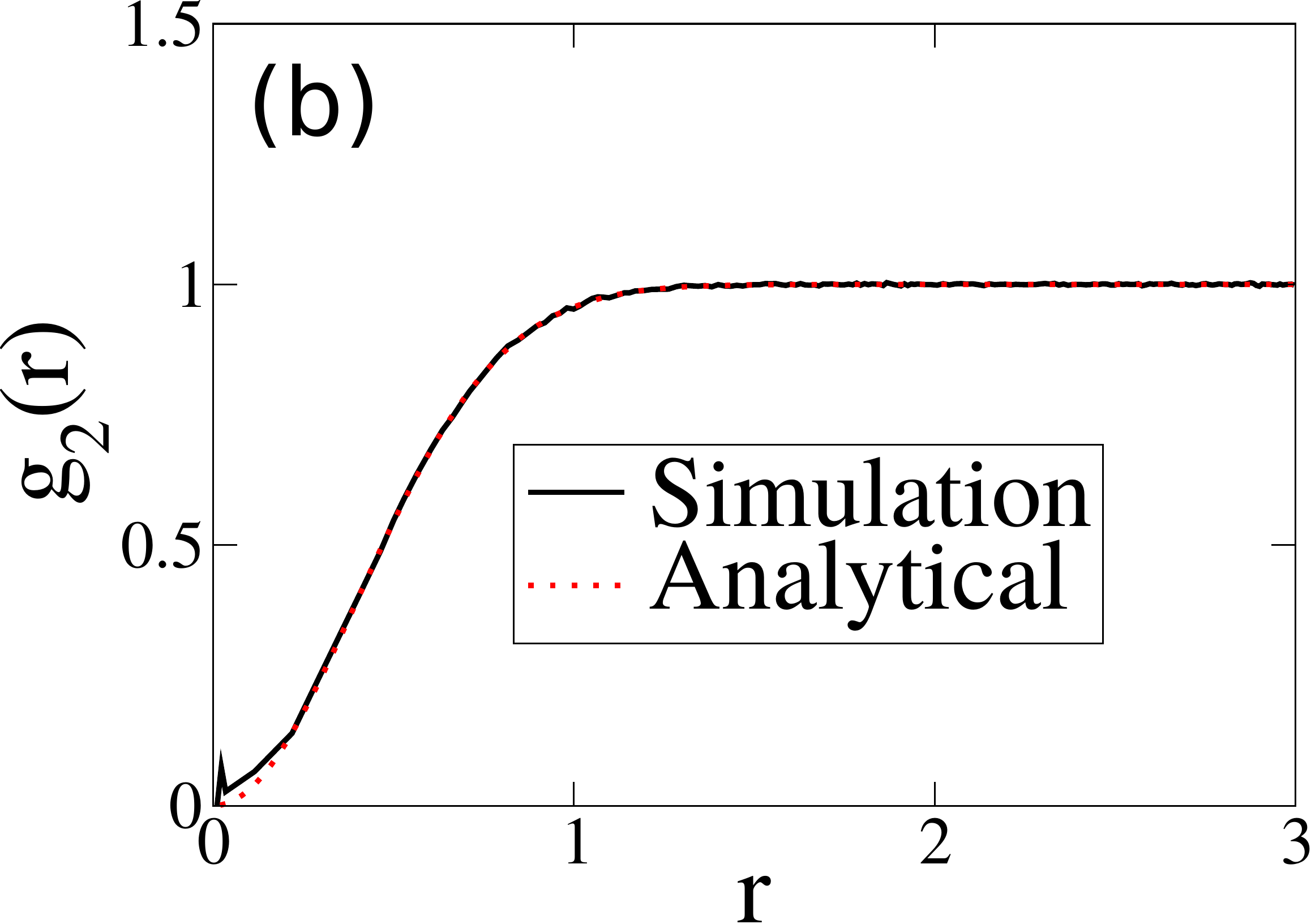}
\end{center}
\caption{(a): The structure factor obtained by sampling ensembles of 3D configurations in which
the  target function $S_0(k)$ is taken to be  Eq.~(\ref{S-d-OCP}) with $a=1/\sqrt{\pi}$ at $\rho=1$. (b): 
The corresponding pair correlation function sampled from simulations and the analytical formula (\ref{h-d-OCP}). Here it turns out that our usual reciprocal-space cutoff of $K=15$ is not large enough, 
and so we use $K=25$ instead.}
\label{3DGaussian}
\end{figure}

\begin{figure}
\begin{center}
\includegraphics[width=0.45\textwidth]{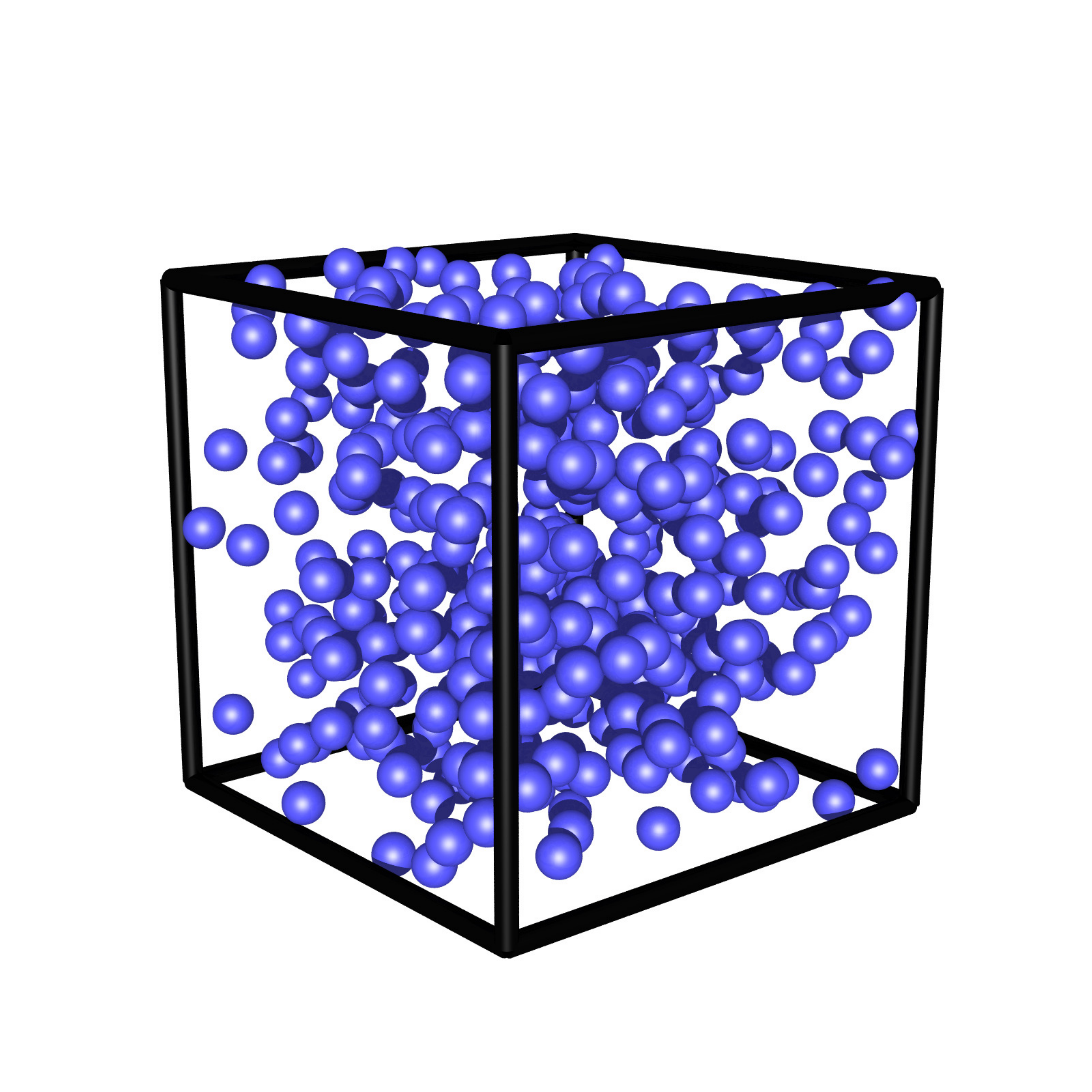}
\end{center}
\caption{A three-dimensional, 400-particle configuration drawn from ensembles in which the target function $S_0(k)$ is taken to be Eq.~(\ref{S-d-OCP}) at $\rho=1$.}
\label{3DGaussian_config}
\end{figure}

To begin, we ask whether  a total correlation function with the following
Gaussian form is realizable as a hyperuniform system  across dimensions:
\begin{equation}
h(r)=-\exp\left(-(r/a)^2\right),
\label{h-d-OCP}
\end{equation}
where $a$ is a positive constant. The corresponding structure factor is given by
\begin{equation}
S(k)=1- \rho a^d \pi^{d/2} \exp\left(-\frac{k^2}{4\pi \rho}\right),
\label{S-d-OCP}
\end{equation}
which implies that $S(k)$ tends to zero
quadratically as $k \to 0$ for all $d$. 
The hyperuniformity condition requires the unique density to be given by  $\rho=(a\sqrt{\pi})^{-d}$.
Note that the case $d=2$ is exactly the same as the two-dimensional OCP system
in which $h(r)$ and $S(k)$ are given by (\ref{h-OCP}) and (\ref{S-OCP}), respectively.

We target such structure factors  in one and three dimensions and find that
they are realizable as hyperuniform systems at unit density.
For $d=1$, we find excellent agreement between the simulated and target structure factors
are obtained, as shown in Fig. \ref{1DGaussian}.
This strongly suggests that such systems are realizable in one dimension,
the most difficult dimensionality case. Indeed, we also find 
the same excellent agreement between the simulated and target structure factors
in three dimensions, as illustrated in Fig. \ref{3DGaussian}. We conclude that such targets are
realizable as disordered hyperuniform systems of class I \cite{To18a} in any space dimension
whenever  $\rho=(a\sqrt{\pi})^{-d}$. 
A 3D configuration is shown in Fig.~\ref{3DGaussian_config}.

\subsection{$d$-dimensional generalization of the OCP pair correlation function}

\begin{figure}
\begin{center}

\includegraphics[width=0.45\textwidth]{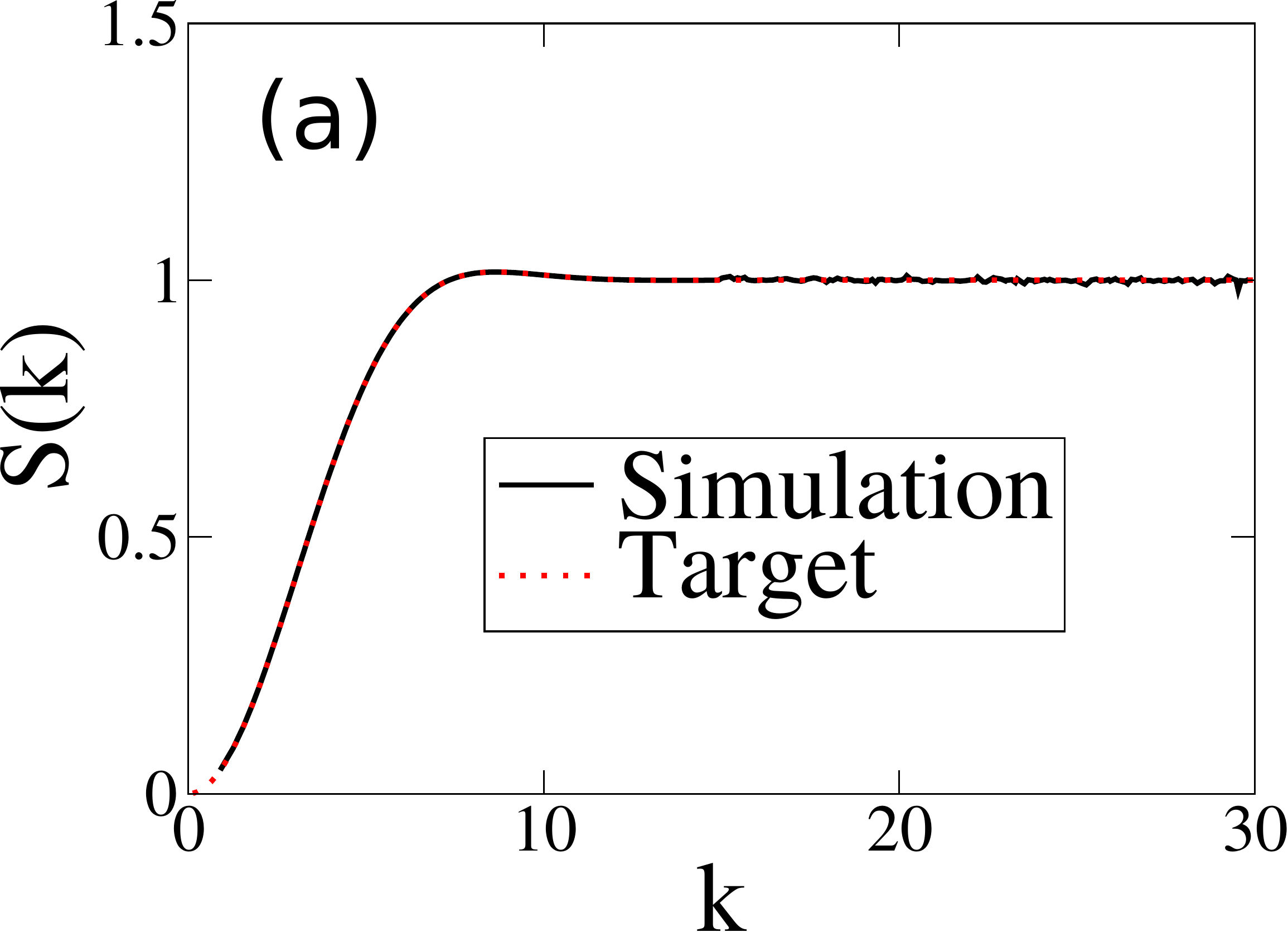}
\includegraphics[width=0.45\textwidth]{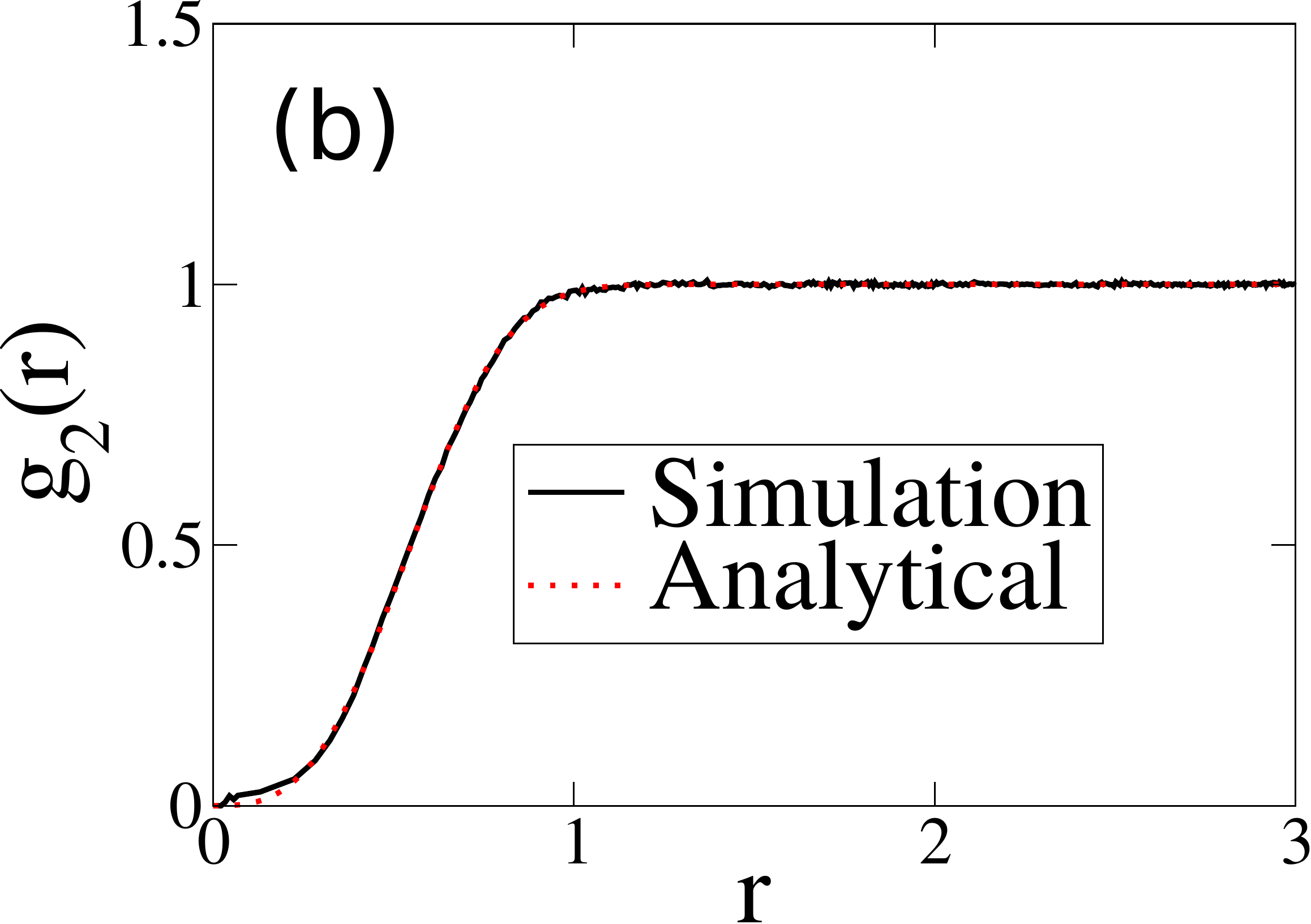}
\end{center}
\caption{(a): The structure factor obtained by sampling ensembles of 3D configurations in which
the  target function $S_0(k)$ is taken to be  Eq.~(\ref{sk-d-OCP}) at $\rho=1$. (b): 
The corresponding pair correlation function sampled from simulations and the analytical formula (\ref{d-OCP}).}

\label{3DHrExpV}
\end{figure}

\begin{figure}
\begin{center}
\includegraphics[width=0.45\textwidth]{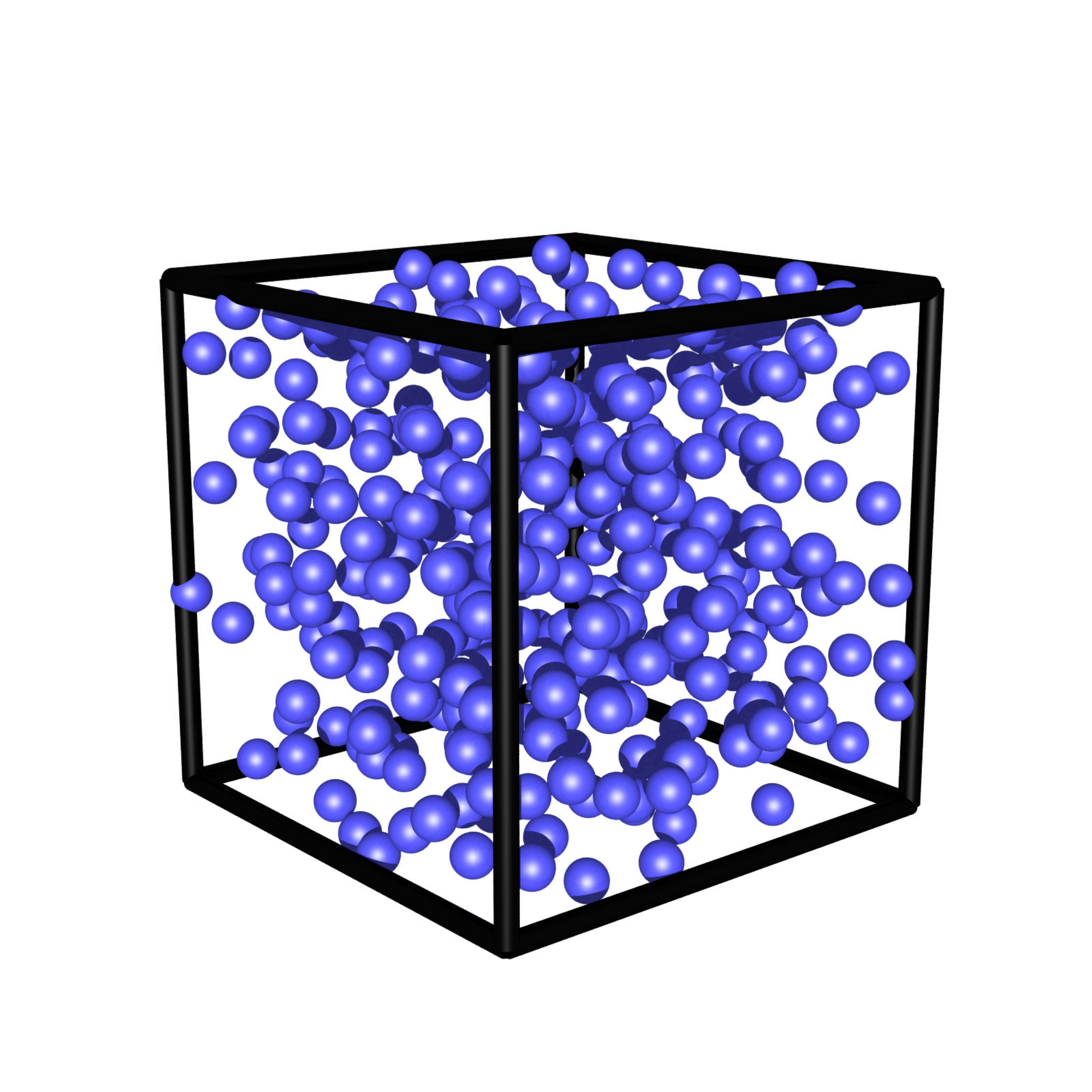}
\end{center}
\caption{A three-dimensional, 400-particle configuration drawn from ensembles in which the target function $S_0(k)$ is taken to be Eq.~(\ref{sk-d-OCP}) at $\rho=1$.}
\label{3DHrExpV_config}
\end{figure}

Consider the following $d$-dimensional generalization
of the total correlation function of the OCP:
\begin{equation}
h(r)=-\exp[-\rho v_1(r)],
\label{d-OCP}
\end{equation} 
where $v_1(r)$ is the volume of a sphere of radius $r$ [cf. (\ref{v1})].
It is noteworthy that such a total correlation function automatically satisfies
the hyperuniformity requirement for {\it any positive density and any $d$}, since
${\tilde h}(k=0)= \int_{\mathbb{R^d}} h(r) d{\bf r}=-1/\rho$ [cf. (\ref{eq:Sk_g2}) and (\ref{hyper})].
Note that when $d=1$, this is identical to the 
realizable total correlation function (\ref{h-exp}) with $\lambda=2 \rho$. Moreover,
when $d=2$, this is identical to the realizable OCP function (\ref{h-OCP}).

It is not known whether configurations corresponding to (\ref{d-OCP})
for $d \ge 3$ are realizable. We target the structure factor in this case in three dimensions: 
\begin{equation}
\begin{split}
S(k)=&1-\frac{1}{1080\rho^{4/3}\pi^{2/3}\Gamma(2/3)} \bigg \{ \\
&2^{1/3}3^{5/6}\pi^{1/3} {}_0F_3\left[-;\frac{4}{3},\frac{3}{2},\frac{11}{6};a(k)\right]k^4 \\
&-30(6\rho)^{2/3}\Gamma\left(\frac{2}{3}\right)^2 {}_0F_3\left[-;\frac{2}{3},\frac{7}{6},\frac{3}{2};a(k)\right] k^2 \\
&+1080\Gamma\left(\frac{2}{3}\right)\pi^{2/3}\rho^{4/3} {}_1F_4\left[1;\frac{1}{3},\frac{2}{3},\frac{5}{6},\frac{7}{6};a(k)\right] \bigg \},
\end{split}
\label{sk-d-OCP}
\end{equation}
where $\Gamma(x)$ is the gamma function, ${}_pF_q(a_1,\cdots,a_p;b_1,\cdots,b_q;z)$ is the generalized hypergeometric function, and 
\begin{equation}
a(k)=\frac{k^6}{20736\pi^2\rho^2}.
\end{equation}

For small wavenumbers, this 3D structure factor as well as those for any other
values of $d$ goes to zero quadratically in $k$ as $k$ tends
to zero; specifically,
\begin{equation}
S(k) \sim k^2 \qquad (k \to 0).
\label{quad}
\end{equation}
This means that $S(k)$ is analytic at the origin, which in turn implies
that $h(r)$ decays to zero exponentially fast or faster \cite{To18a}.
We find that this 3D structure factor is indeed realizable for $\rho=1$.
The results depicted in Fig.  \ref{3DHrExpV} show excellent agreement
between the simulated and target structure factors.
One such configuration is shown in Fig.~\ref{3DHrExpV_config}.
Since (\ref{d-OCP}) is realizable for $d=3$, it 
should be realizable in higher dimensions and hence such systems in $\mathbb{R}^d$ 
for any $d$ are hyperuniform of class I \cite{To18a}.

\subsection{Fourier dual of relation (\ref{d-OCP})}

Here we consider the Fourier dual of the function (\ref{d-OCP}) in $d$ dimensions, namely,
\begin{equation}
\rho {\tilde h}(k)=-\exp[-v_1(k)/(2\pi)^d\rho],
\label{dual-1}
\end{equation} 
which implies
\begin{equation}
S(k)=1 - \exp[-v_1(k)/(2\pi)^d\rho].
\label{dual-2}
\end{equation}
Thus, the structure factor has the
following asymptotic power-law 
behavior for any $d$:
\begin{equation}
S(k) \sim k^d \qquad (k\to 0).
\label{d-OCPSkScaling}
\end{equation}
The realizability of such structure factors, which would be hyperuniform
for any density and $d$,  has heretofore
not been studied in any space dimension, except for $d=2$, where it has the same
form as the OCP structure factor (\ref{S-OCP}). It is crucial to note that unlike the structure factor
corresponding to (\ref{d-OCP}), which is analytic at the origin [cf. (\ref{quad})], 
the structure factor (\ref{dual-2}) is nonanalytic at the origin
for any odd dimension. This attribute in odd dimensions results in pair correlation functions
that for large $r$ are controlled by  a power-law decay  $1/r^{2d}$; see Ref. \cite{To18a}
for a general analysis of such asymptotics. The corresponding total correlation functions
in the first three space dimensions are given respectively by
\begin{equation}
h(r)=\frac{-1}{(\pi\rho r)^2+1},
\label{h-1d-ocp}
\end{equation}
\begin{equation}
h(r)=-\exp(-\pi \rho r^2),
\label{h-2d-ocp}
\end{equation}
\begin{equation}
h(r)=f(r)-1,
\label{h-3d-ocp}
\end{equation}
where 
\begin{equation}
f(r)=S(k=2\pi\rho^{2/3}r)
\end{equation}
and $S(k)$ is the structure factor for the 3D generalization of OCP, given in Eq.~(\ref{sk-d-OCP}).

\begin{figure}
\begin{center}

\includegraphics[width=0.45\textwidth]{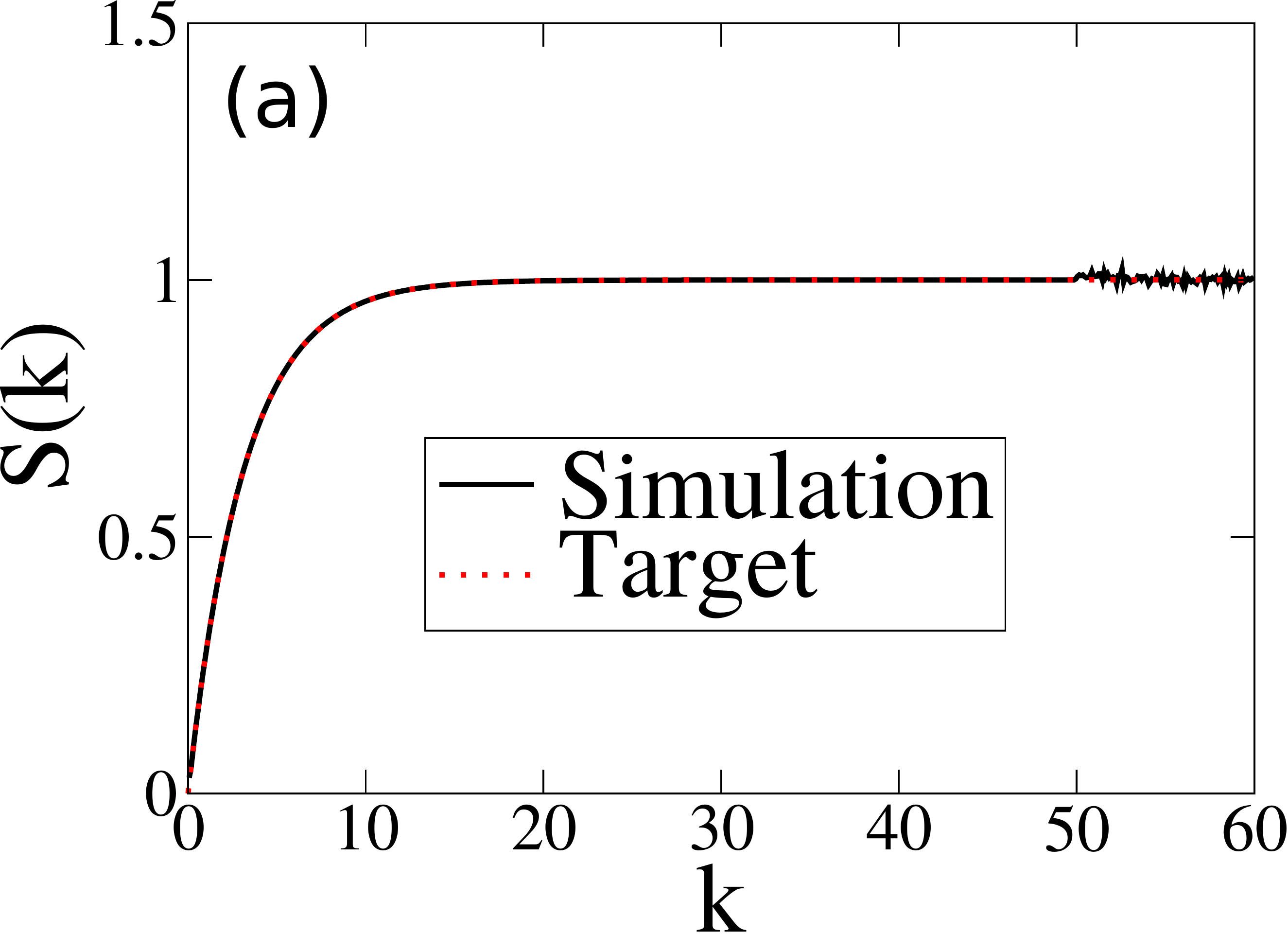}
\includegraphics[width=0.45\textwidth]{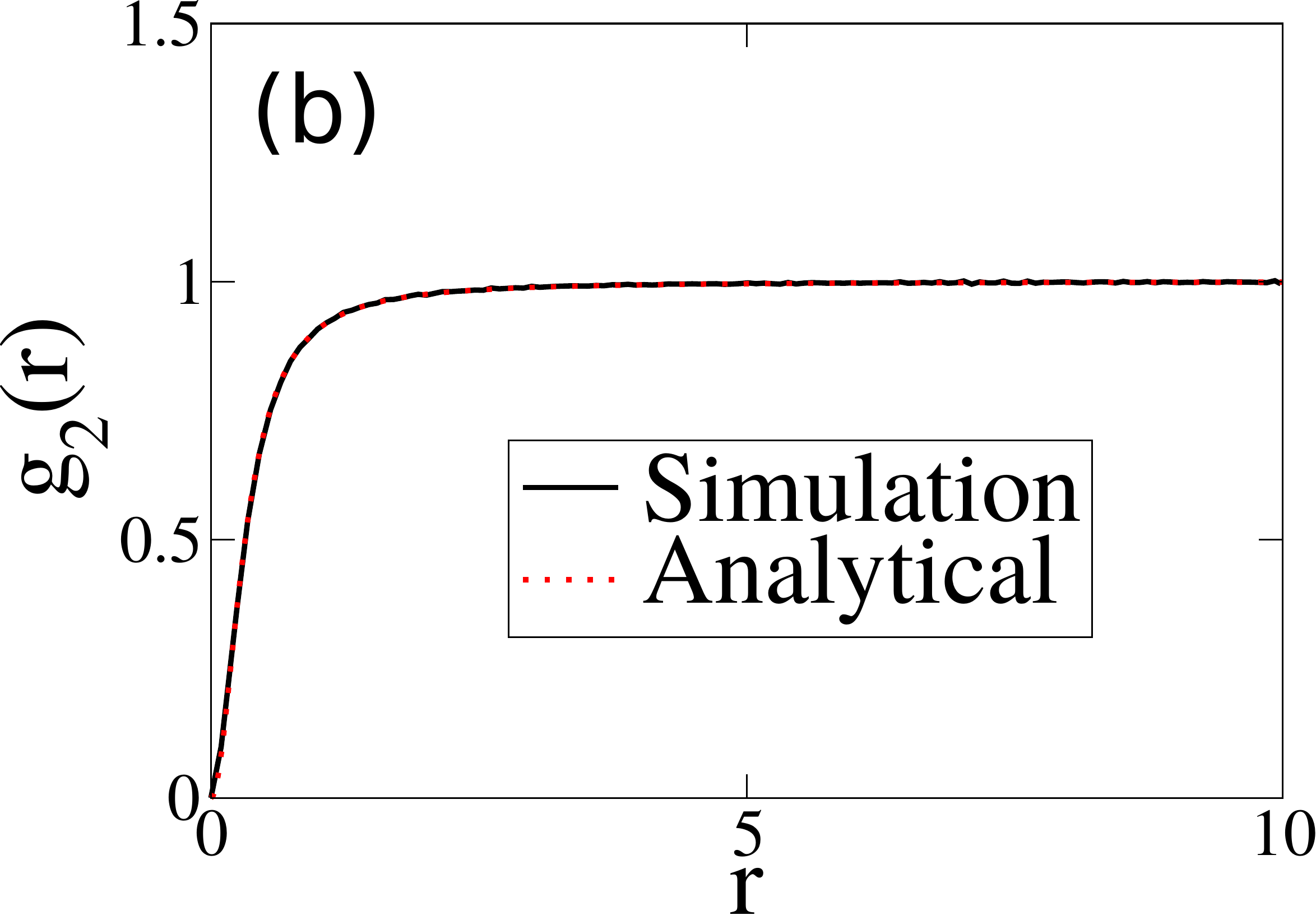}
\end{center}
\caption{(a): The structure factor obtained by sampling ensembles of 1D configurations in which
the  target function $S_0(k)$ is taken to be  Eq.~(\ref{dual-2}) at $\rho=1$. (b): 
The corresponding pair correlation function sampled from simulations and the analytical formula (\ref{h-1d-ocp}).
Here it turns out that our usual reciprocal-space cutoff of $K=30$ is not large enough, 
and so we use $K=50$ instead.}
\label{1DExpV}
\end{figure}
\begin{figure}
\begin{center}

\includegraphics[width=0.45\textwidth]{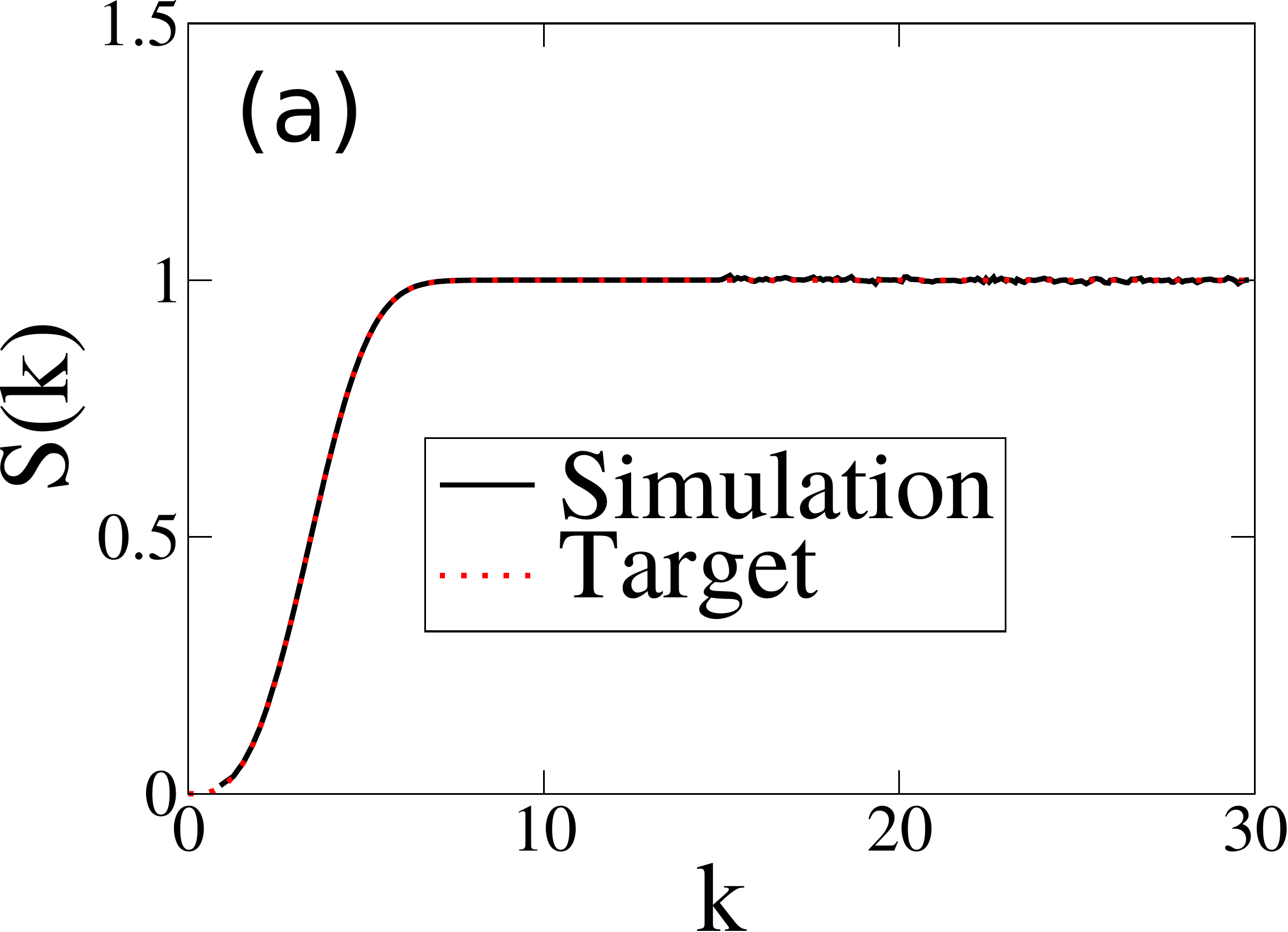}
\includegraphics[width=0.45\textwidth]{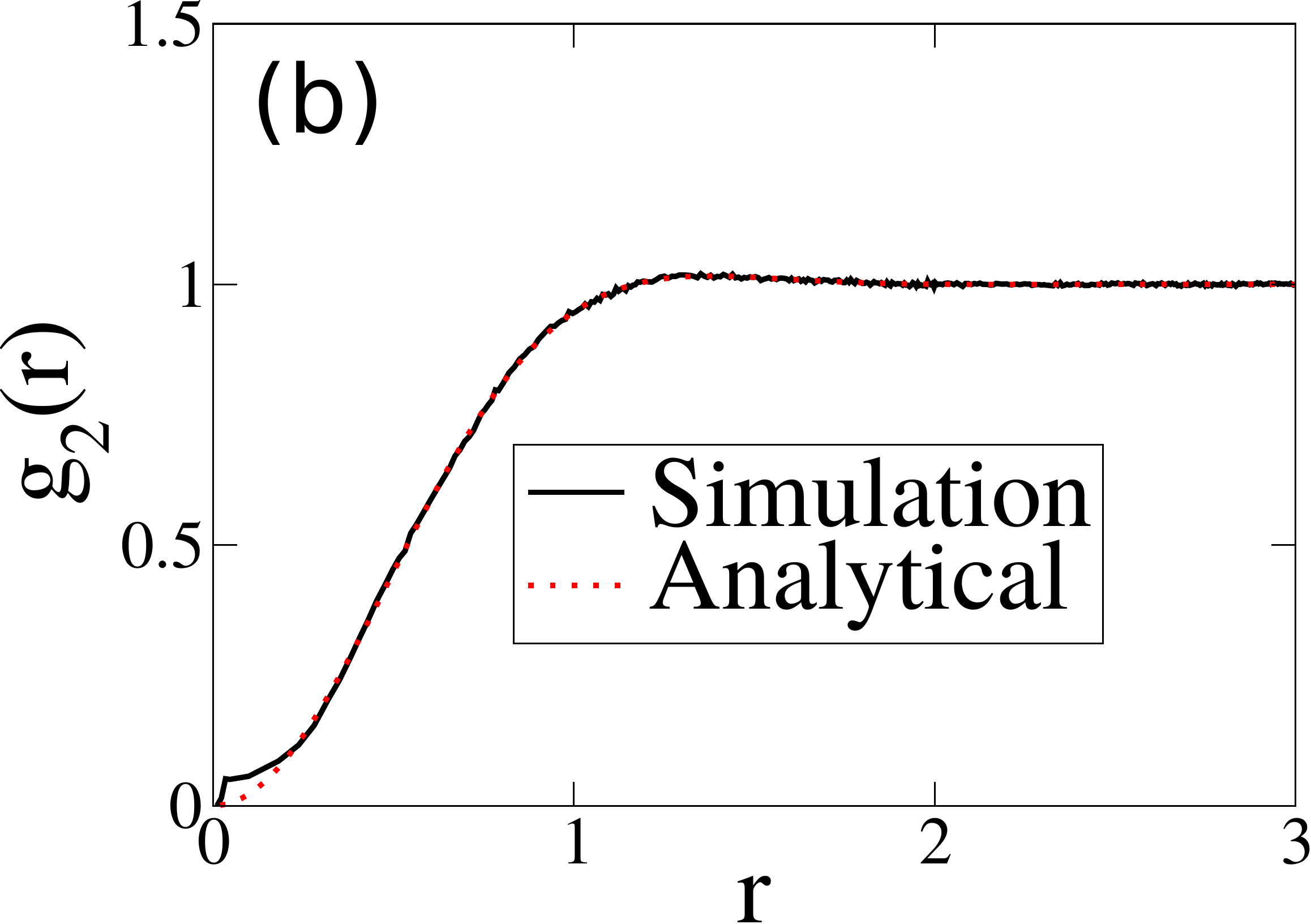}
\end{center}
\caption{(a): The structure factor obtained by sampling ensembles of 3D configurations in which
the  target function $S_0(k)$ is taken to be  Eq.~(\ref{dual-2}) at $\rho=1$. (b): 
The corresponding pair correlation function sampled from simulations and the analytical formula (\ref{h-3d-ocp}).}

\label{3DExpV}
\end{figure}

\begin{figure}
\begin{center}
\includegraphics[width=0.45\textwidth]{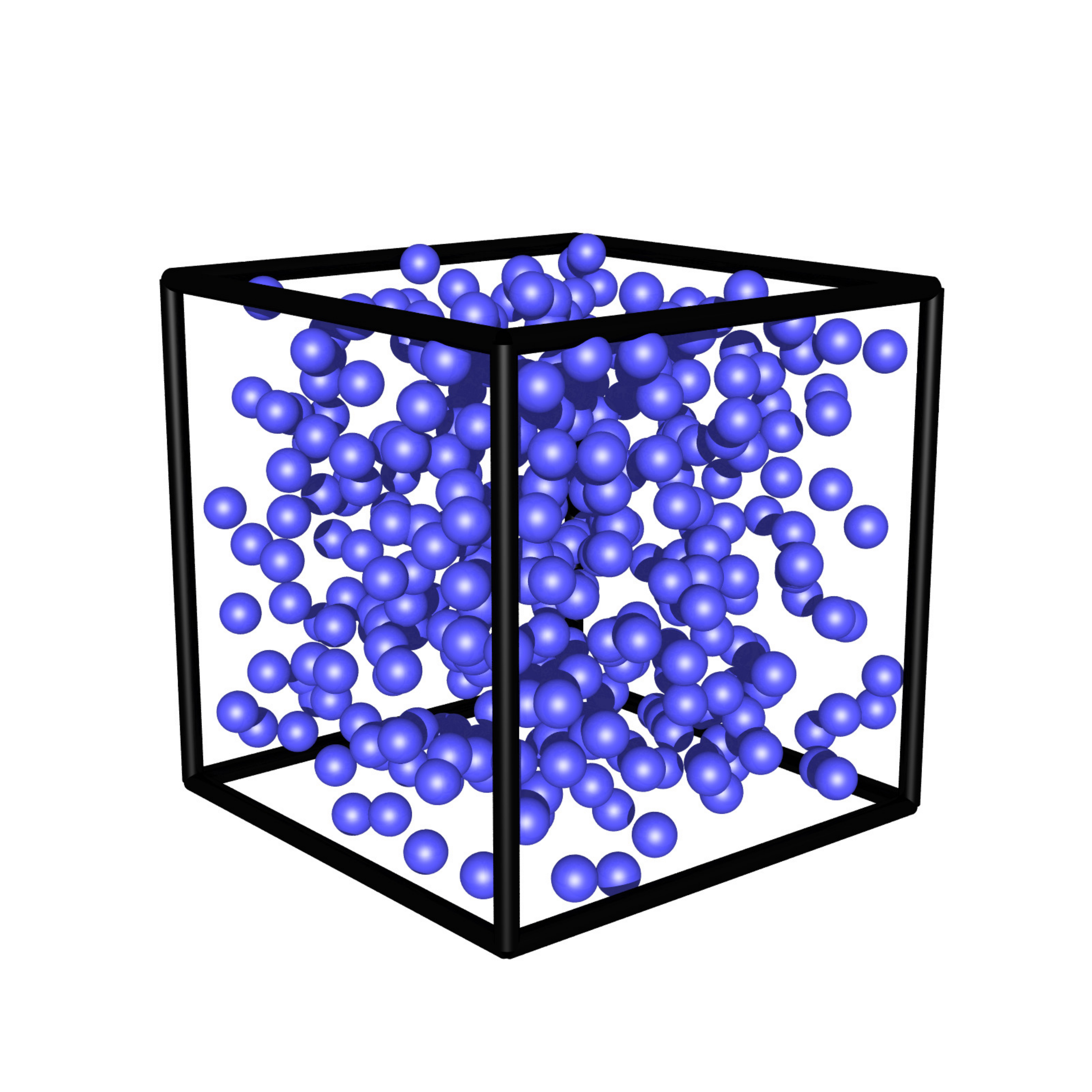}
\end{center}
\caption{A three-dimensional, 400-particle configuration drawn from ensembles in which the target function $S_0(k)$ is taken to be Eq.~(\ref{dual-2}) at $\rho=1$.}
\label{3DExpV_config}
\end{figure}

We target such structure factors  in one and three dimensions and find that
they are realizable as hyperuniform systems at unit density.
Excellent agreement between the simulated and target structure factors
are obtained, as shown in Figs. \ref{1DExpV} and \ref{3DExpV}. 
A three-dimensional configuration is shown in Fig.~\ref{3DExpV_config}.
For aforementioned reasons, this means that the function (\ref{dual-2}) is realizable for higher dimensions ($d\ge 4$) and hence for all positive dimensions.
Therefore, we see from
(\ref{power}), (\ref{sigma-asy}) and (\ref{d-OCPSkScaling}) that such systems
are hyperuniform of class II for $d=1$
and of class I for $d \ge 2$.

\section{Targeting nonhyperuniform structure factors with unknown realizability}
\label{unknown-nonhu}

In this section, we apply our ensemble-average algorithm to target two different nonhyperuniform functional forms for the structure factor in 2D and 3D. 
As before, they both satisfy the explicitly known necessary conditions (\ref{g2Definition})-(\ref{yamada}), but are not known to be realizable. We show that all of these targets are indeed realizable.

\subsection{Hyposurficial structure factors in two and three dimensions}

\begin{figure}
\begin{center}

\includegraphics[width=0.45\textwidth]{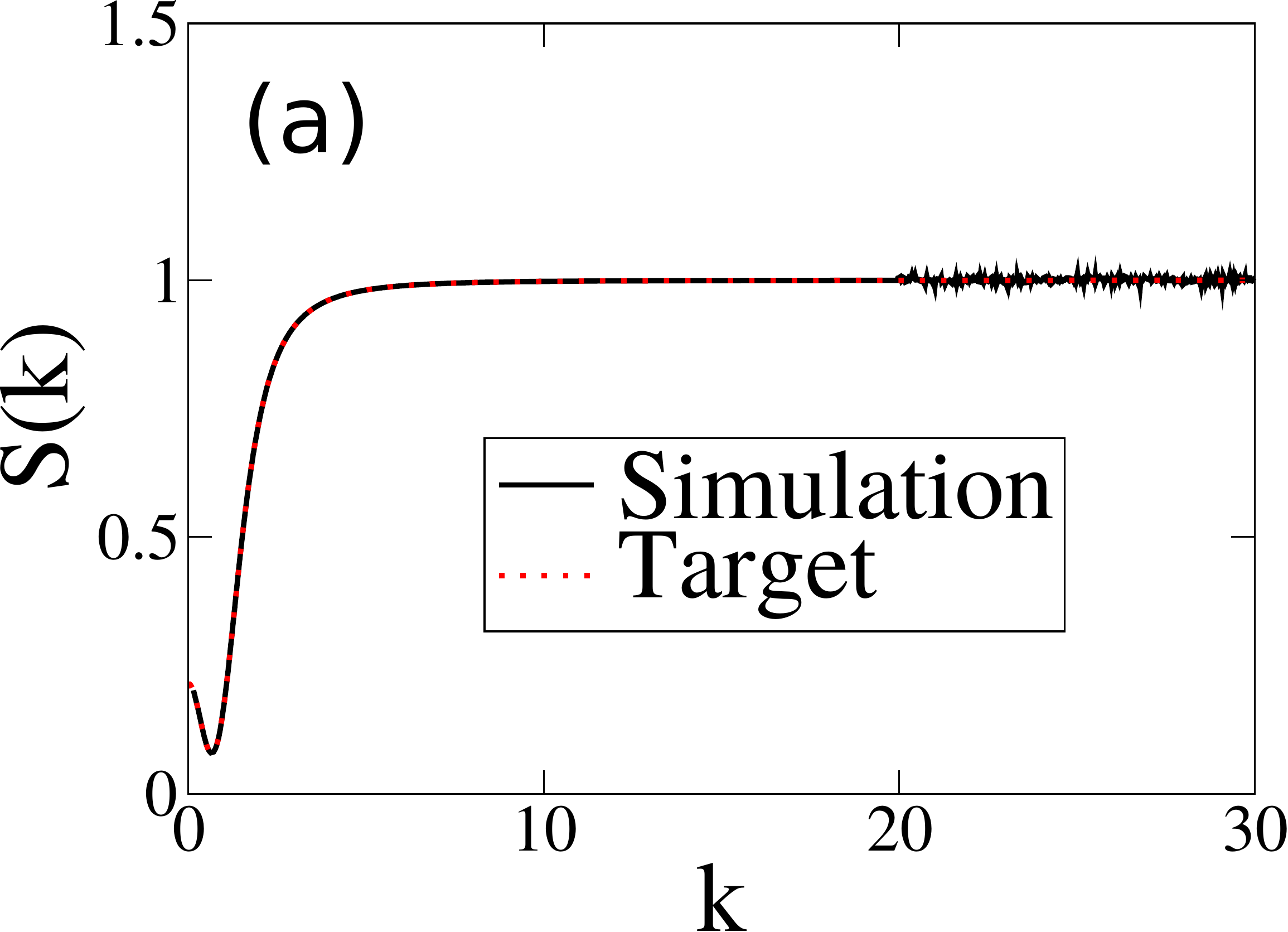}
\includegraphics[width=0.45\textwidth]{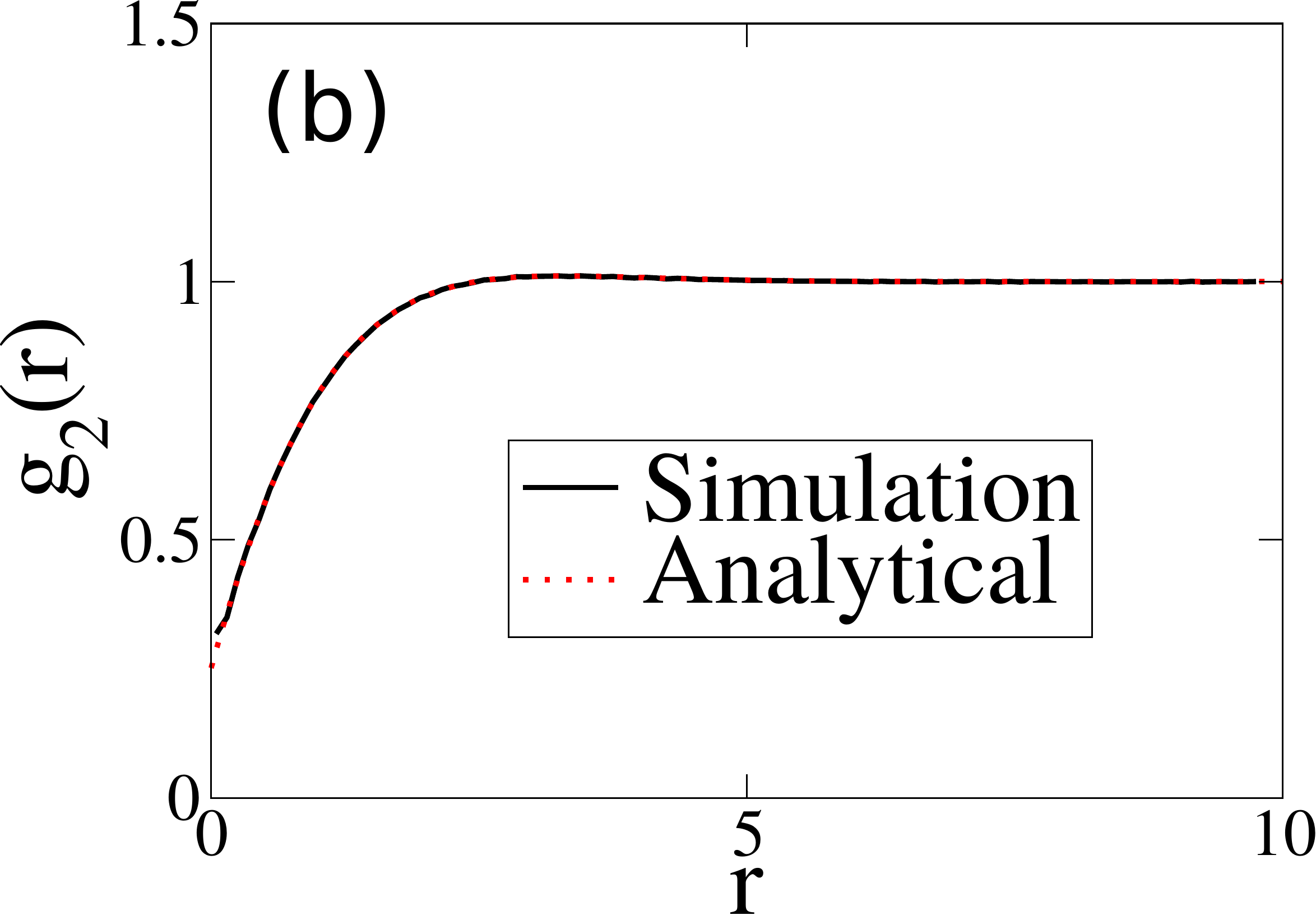}
\end{center}
\caption{(a): The structure factor obtained by sampling ensembles of 2D configurations, in which
the target function $S_0(k)$ is numerically computed from Eq. (\ref{eq:Sk_g2}) and (\ref{eq:2DHyposurficial}), at $\rho=0.5$. (b): 
The corresponding pair correlation function sampled from simulations and the analytical formula (\ref{eq:2DHyposurficial}).}

\label{2DHyposurficial}
\end{figure}

\begin{figure}
\begin{center}

\includegraphics[width=0.45\textwidth]{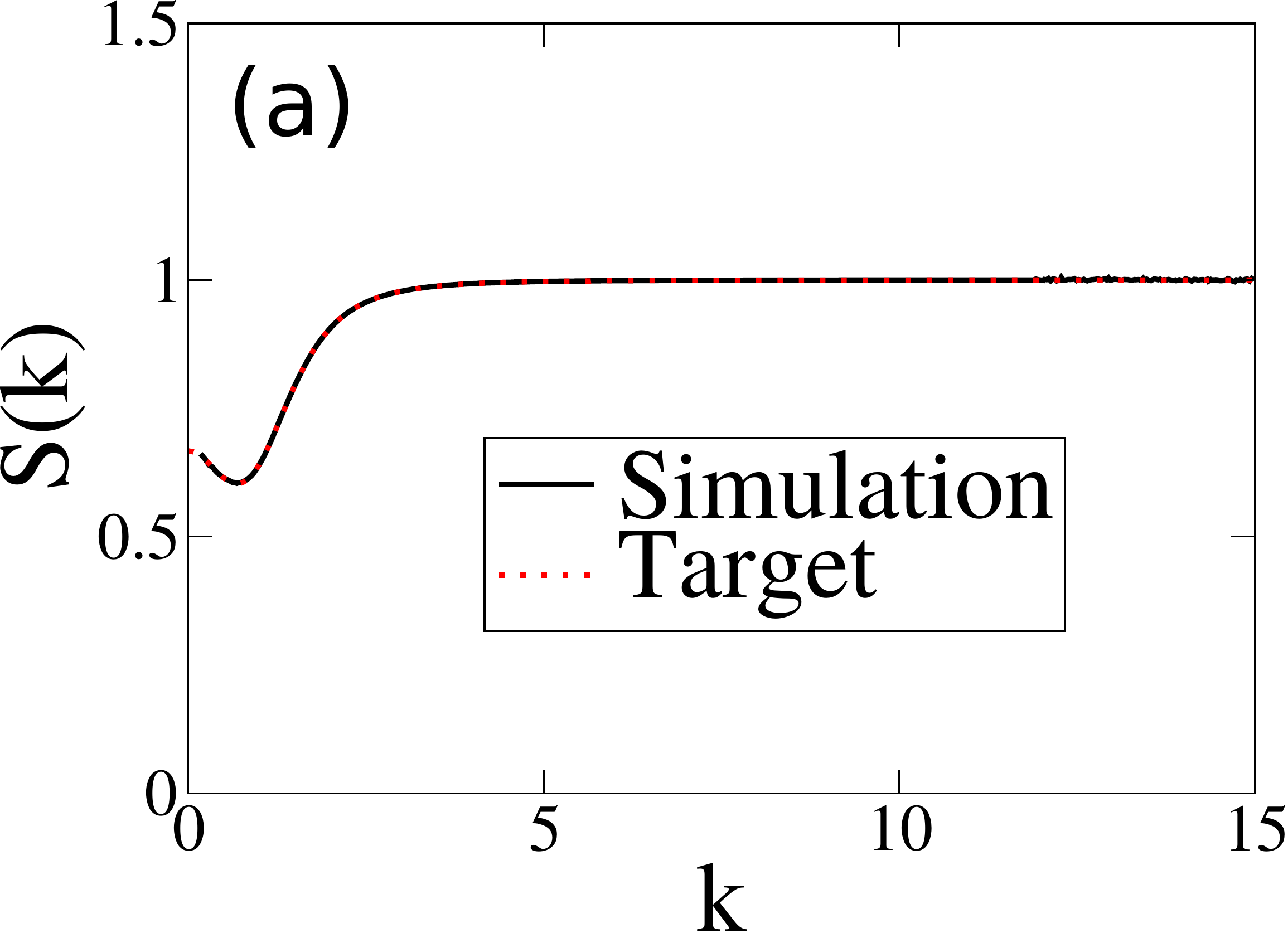}
\includegraphics[width=0.45\textwidth]{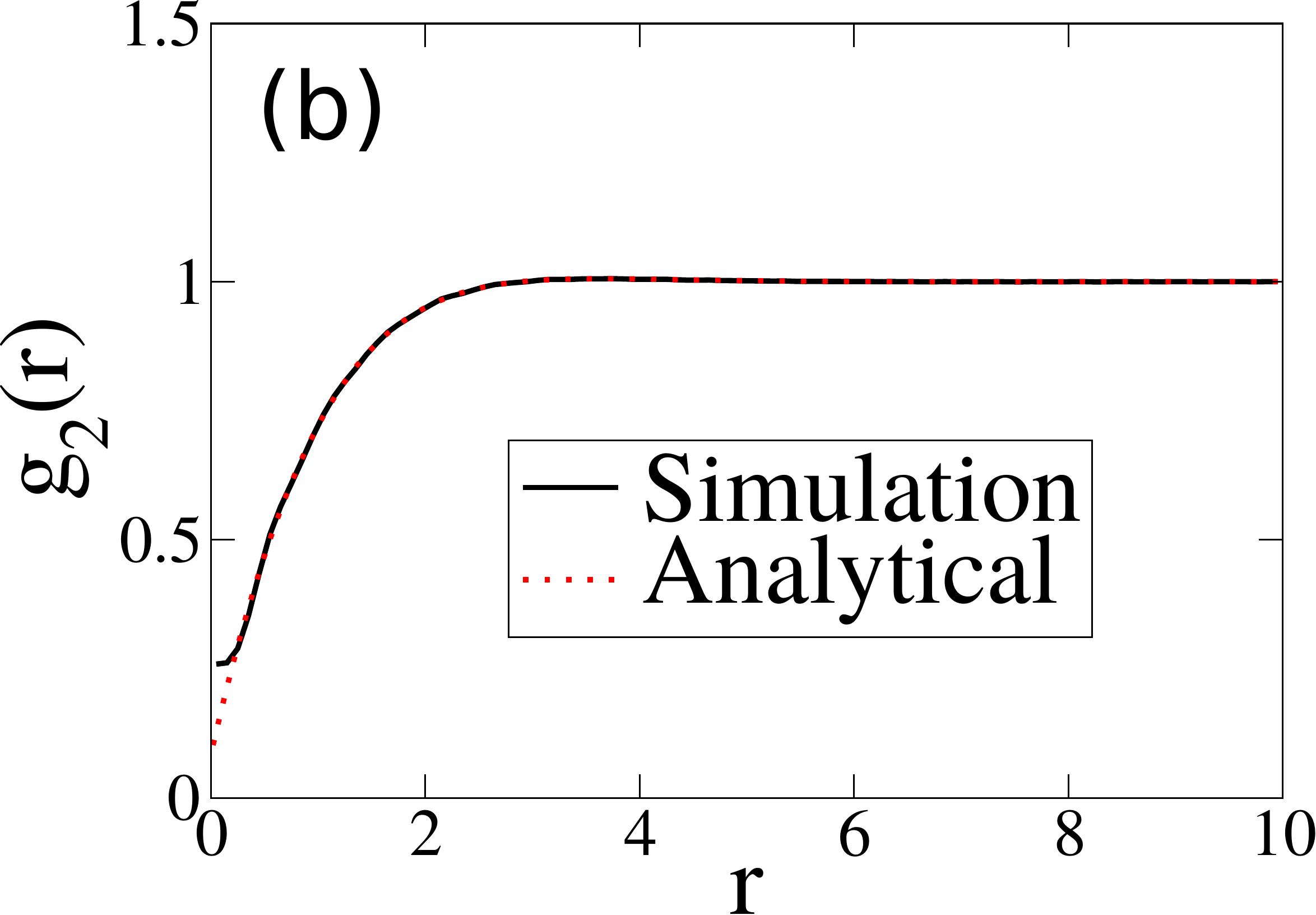}
\end{center}
\caption{(a): The structure factor obtained by sampling ensembles of 3D configurations, in which
the target function $S_0(k)$ is taken to be  Eq. (\ref{eq:3DHyposurficialSk}), at $\rho=\frac{1}{4\pi}$. (b): 
The corresponding pair correlation function sampled from simulations and the analytical formula (\ref{eq:3DHyposurficial}).}

\label{3DHyposurficial}
\end{figure}

\begin{figure}
\begin{center}
\includegraphics[width=0.45\textwidth]{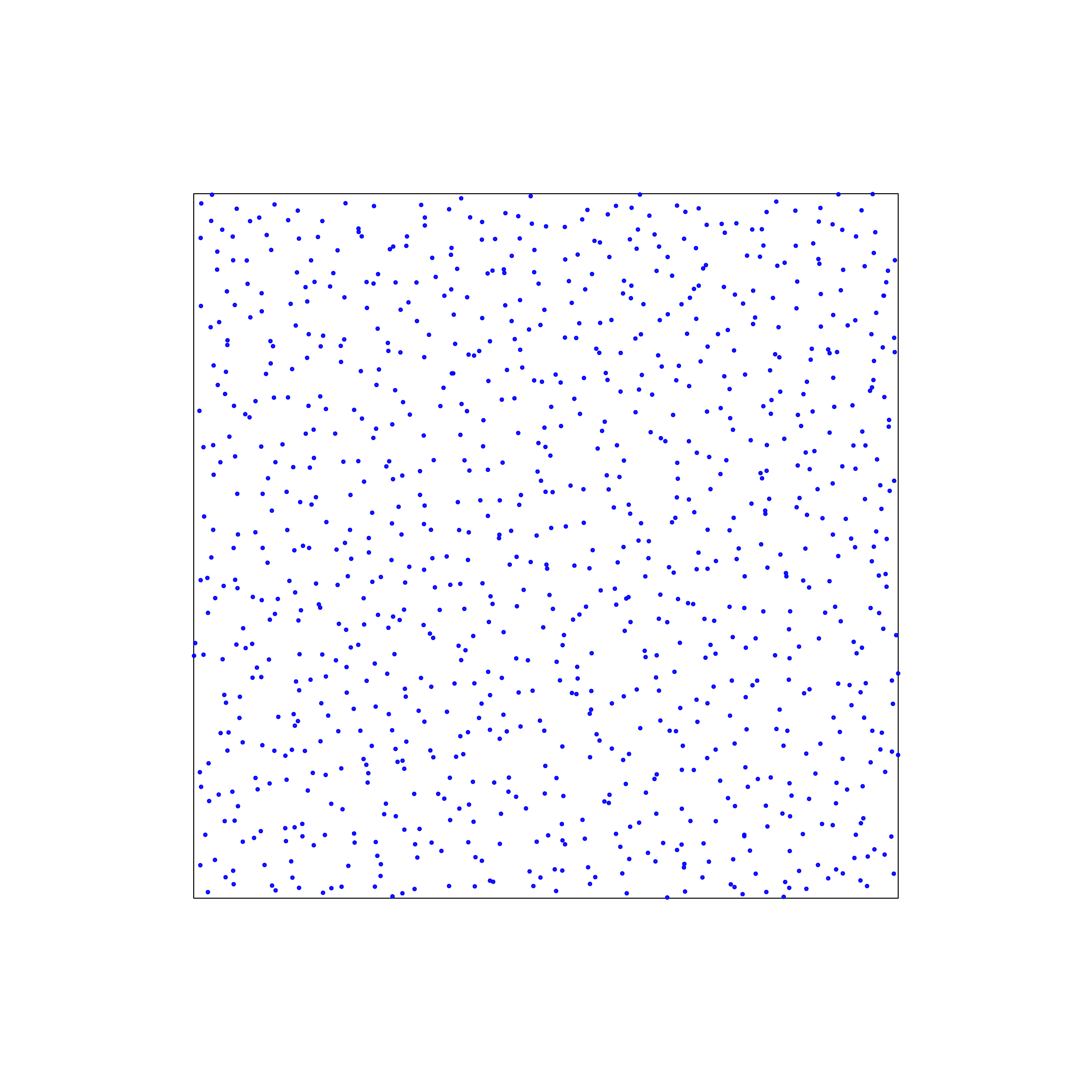}
\end{center}
\caption{A two-dimensional, 1000-particle hyposurficial configuration drawn from ensembles in which the target function $S_0(k)$ is numerically computed from Eq. (\ref{eq:Sk_g2}) and (\ref{eq:2DHyposurficial}), at $\rho=0.5$.}
\label{2DHyposurficial_config}
\end{figure}

\begin{figure}
\begin{center}
\includegraphics[width=0.45\textwidth]{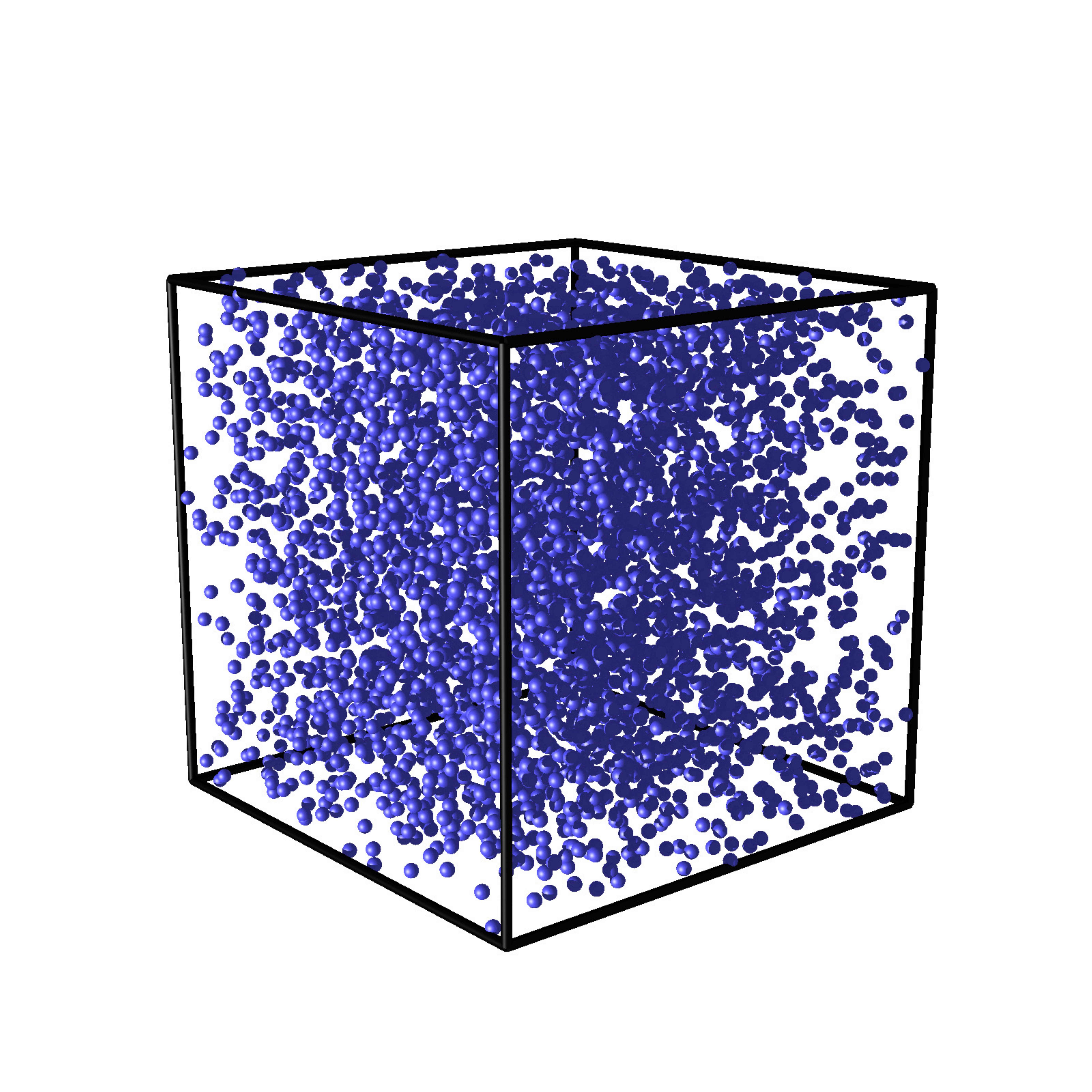}
\end{center}
\caption{A three-dimensional, 4000-particle hyposurficial configuration drawn from ensembles in which the target function $S_0(k)$ is taken to be Eq.~(\ref{eq:3DHyposurficialSk}) at $\rho=\frac{1}{4\pi}$.}
\label{3DHyposurficial_config}
\end{figure}

As we have discussed in the introduction, a hyposurficial state of matter has $A>0$ and $B=0$ in Eq. (\ref{eq:sigma}). Hyposurficiality may be considered the opposite of hyperuniformity because the latter implies $A=0$ and $B>0$. Although there is numerical evidence of $B$ vanishing at a particular pressure in a model amorphous ice \cite{martelli2017large}, a rigorous proof of the existence of hyposurficial point configurations has heretofore not been found.

Here we design and realize hyposurficial structure factors in 2D and 3D. Since $B$ is proportional to the $d$-th moment of $h(r)$ [see Eq. (\ref{eq:B})], we designed the following well-behaved hyposurficial $h(r)$ targets
\begin{equation}
h(r)=\frac{\exp(-r)}{4}-\frac{\exp(-r)\sin(r)}{r},
\label{eq:2DHyposurficial}
\end{equation}
\begin{equation}
h(r)=\frac{\exp(-r)}{4\pi}-\frac{\exp(-r)\sin(r)}{r},
\label{eq:3DHyposurficial}
\end{equation}
 in 2D and 3D respectively. We choose realizable densities $\rho=\frac{1}{2}$ in 2D and $\rho=\frac{1}{4\pi}$ in 3D. The 3D target can be analytically transformed into an $S(k)$ target
 \begin{equation}
 S(k)=\frac{6k^8+12k^6+19k^4+24k^2+16}{6(k^2+1)^2(k^2-2k+2)(k^2+2k+2)}
 \label{eq:3DHyposurficialSk}
 \end{equation}
 but the 2D $h(r)$ target has no corresponding analytical $S(k)$. We target a numerically obtained tabulated $S(k)$ instead.
 
We have successfully realized these targets with tuned parameters: $N=1000$, $N_c=200$, and $K=20$ in 2D; and $N=4000$, $N_c=1000$, and $K=12$ in 3D. We increase the system size $N$ because in both cases, $S(k)$ possesses a small kink near the origin, and we need large systems to access smaller $k$ values. Our success in realizing these targets demonstrates, for the first time, that hyposurficial point configurations indeed exist.

Reference~\onlinecite{martelli2017large} observed that hyposurficiality appears to be associated with spacial heterogeneities, and so it is interesting to see if our configurations also exhibit such characteristics. We present these configurations in Figs.~\ref{2DHyposurficial_config} and \ref{3DHyposurficial_config}, which indeed show spatial heterogeneities that are manifested by significant clustering of the particles.

\subsection{Antihyperuniform structure factors in two dimensions}

\begin{figure}
\begin{center}

\includegraphics[width=0.45\textwidth]{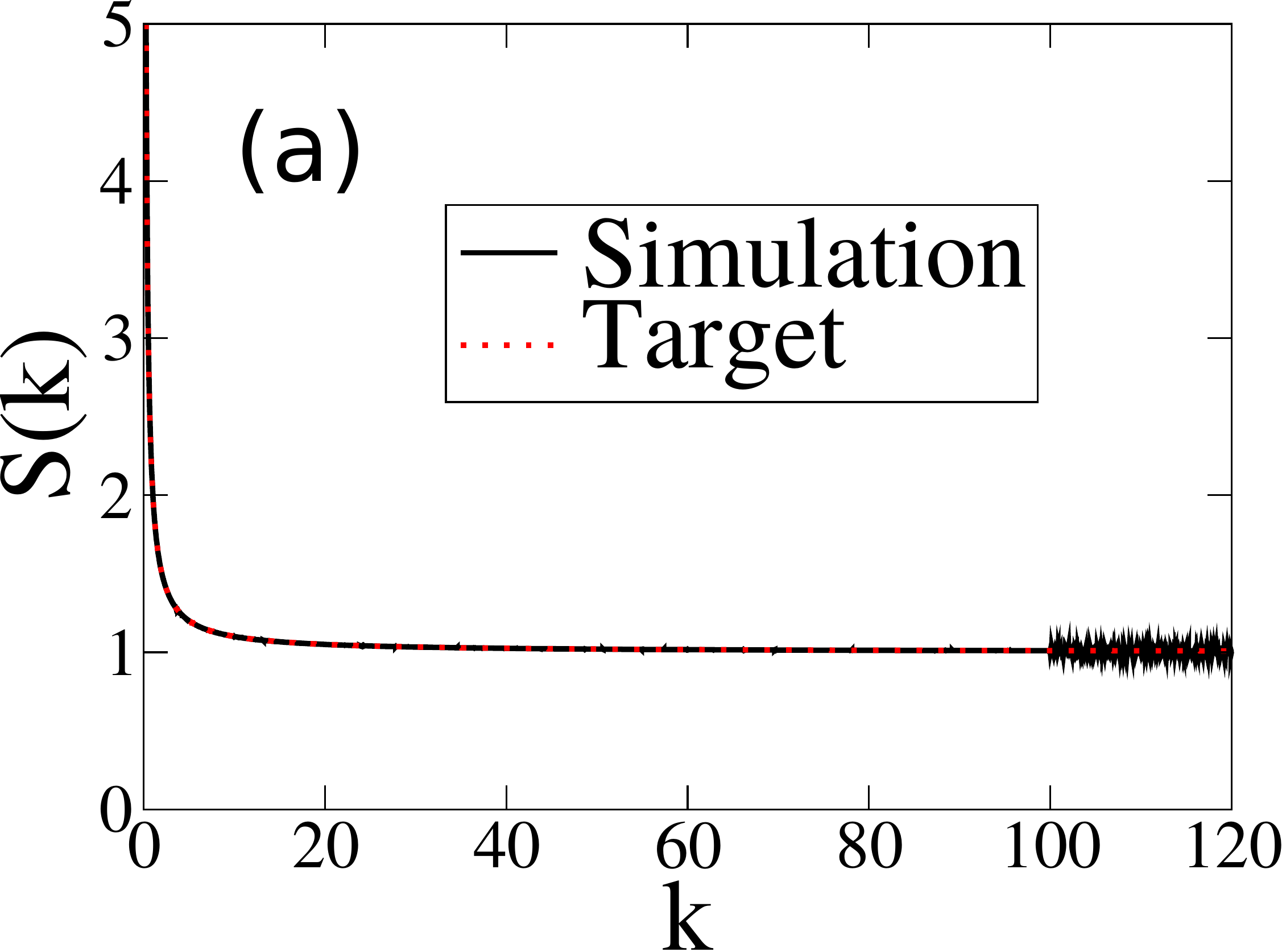}
\includegraphics[width=0.45\textwidth]{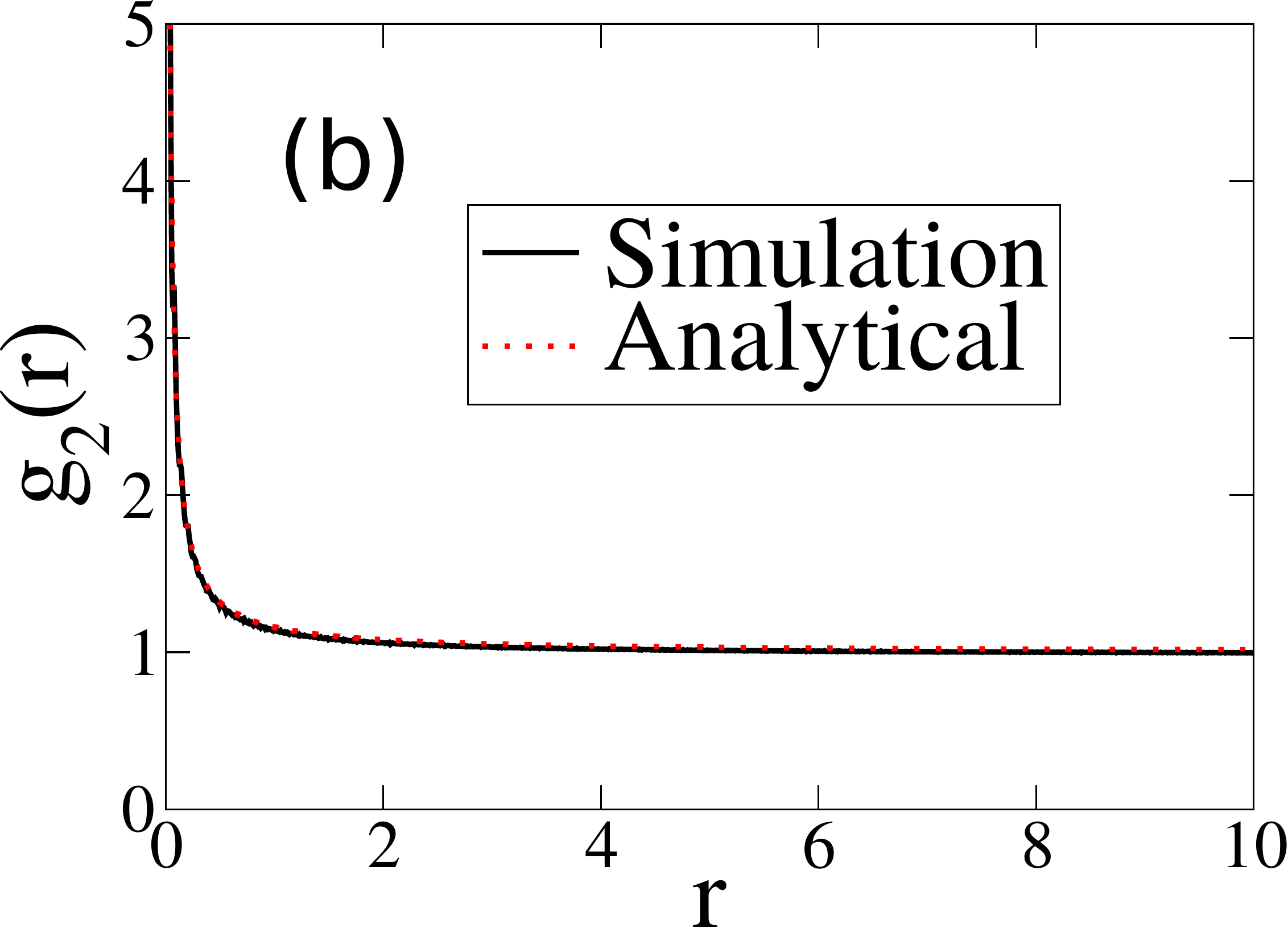}
\end{center}
\caption{(a): The structure factor obtained by sampling ensembles of 2D configurations, in which the target function $S_0(k)$ is taken to be Eq.~(\ref{eq:2DAntihyperuniform}) at $\rho=1$ and $\kappa=0$. (b): 
The corresponding pair correlation function sampled from simulations and the analytical formula (\ref{eq:2DAntihyperuniform_g2}).}

\label{2DAntihyperuniform}
\end{figure}

\begin{figure}
\begin{center}
\includegraphics[width=0.45\textwidth]{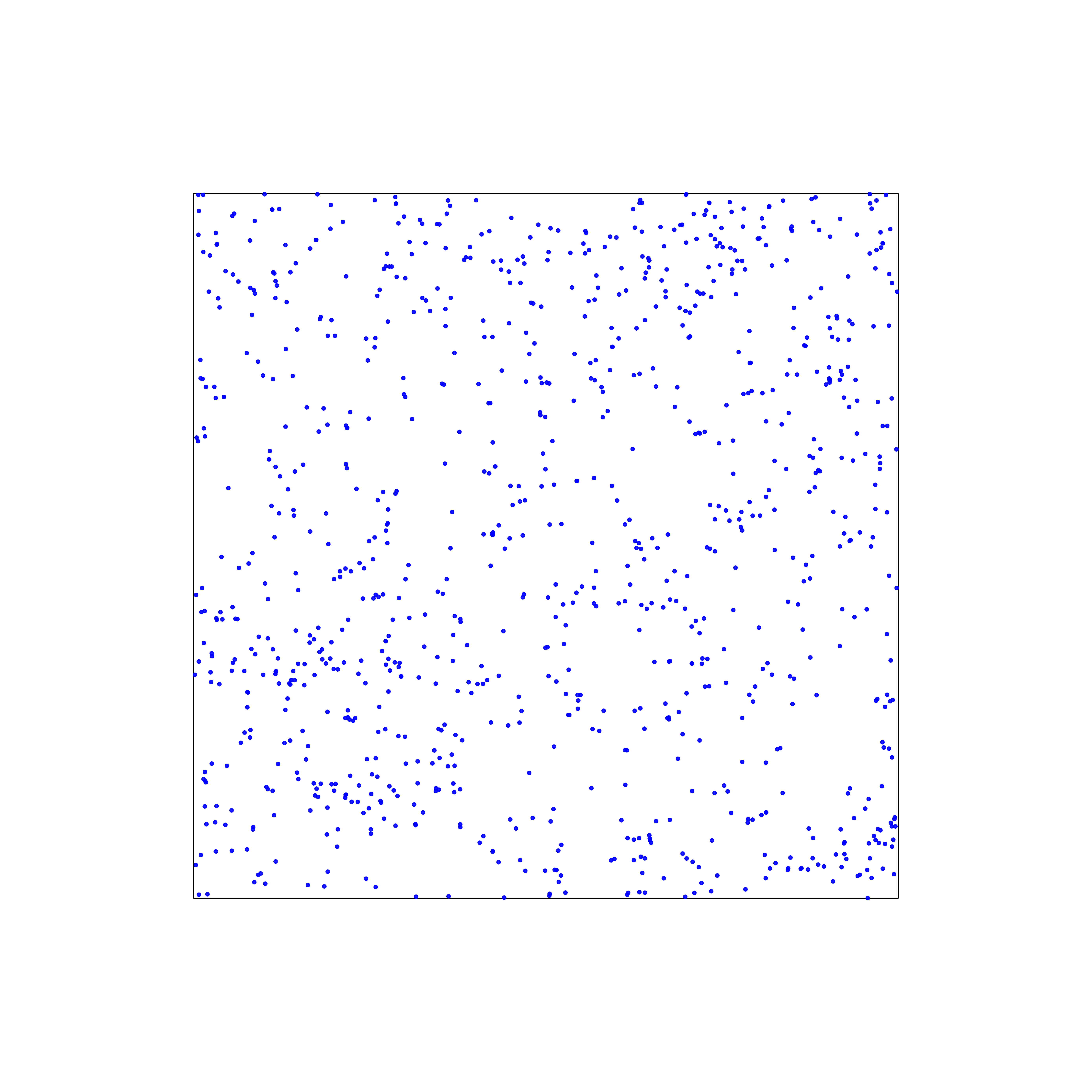}
\end{center}
\caption{A two-dimensional, 1000-particle antihyperuniform configuration drawn from ensembles in which the target function $S_0(k)$ is taken to be Eq.~(\ref{eq:2DAntihyperuniform}) at $\rho=1$ and $\kappa=0$.}
\label{2DAntihyperuniform_config}
\end{figure}

As discussed in the Introduction, an antihyperuniform configuration is one for which $S(k)$ diverges at $k=0$. Although this behavior has been observed at various critical points, it is still interesting to challenge our algorithm to generate such configurations. We designed the following target structure factor in 2D:
\begin{equation}
 S(k)=1+\frac{1}{\sqrt{k^2+\kappa^2}}.
 \label{eq:2DAntihyperuniform}
\end{equation}
Such a system, if realizable, would achieve antihyperuniformity at $\kappa=0$. 
The corresponding pair correlation function is
\begin{equation}
g_2(r)=1+\frac{\exp(-\kappa r)}{2\pi r}.
\label{eq:2DAntihyperuniform_g2}
\end{equation}
We have indeed successfully realized this target at $\rho=1$, $\kappa=0$, 0.3, and 1. We show the antihyperuniform case, $\kappa=0$, in Figs.~\ref{2DAntihyperuniform} and \ref{2DAntihyperuniform_config}. In realizing this target, we used parameters $N=1000$, $N_c=3000$, and $K=100$. We use a large value of $N$ to provide sufficient resolution to determine $S(k)$ at small $k$, and an extremely large value of $K$, since this target $S_0(k)$ decays very slowly to unity as $k$ increases. Since $K$ is large, we also need a sufficiently large value of $N_c$ to provide ample degrees of freedom. 

\section{Conjecture regarding the realizability of equilibrium and nonequilibrium configurations via effective pair interactions}
\label{sec:conjecture}

Our equilibrium ensemble-average formalism to solve the realizability problem in $\mathbb{R}^d$
raises a profound fundamental theoretical question: Can any realizable $g_2({\bf r})$ or $S({\bf k})$ associated with either 
an equlibriium or nonequilibrium ensemble be attained by an equilibrium systems with effective
pairwise interaction?
Currently, there is no rigorous proof that the answer to this question
is in the affirmative in the implied thermodynamic limit. However, there are
sound arguments and reasons to  conjecture that such an effective pair interaction can always
be found, perhaps under some mild conditions. First, our theoretical formalism strongly
supports this conjecture, since it exploits the fact that an equilibrium ensemble of configurations in $\mathbb{R}^d$
offers an infinite number of degrees of freedom to attain a realizable $g_2({\bf r})$ or, equivalently, $S({\bf k})$
in the thermodynamic limit with an associated pair potential $v({\bf r})$ at positive temperatures.
While our procedure is not suited for ground states ($T=0$), such targeted structures
are even easier to achieve by pair interactions in light of the high degeneracy
of pair potentials consistent with a ground-state structure \cite{footnote}.
Second, for finite-sized systems of particles that are restricted to
lie on lattice sites in $\mathbb{R}^d$,
it has been proved, under rather general conditions, that any realizable
$g_2({\bf r})$ can be achieved by a pair potential $v({\bf r})$ at positive temperatures \cite{kuna2011necessary}. Third,
the  success of our algorithm
in all of the known realizable targets cases with {\it nonadditive} interactions supports this conjecture, even if the
simulations were necessarily carried out on finite systems under periodic boundary conditions. 

\begin{figure}
\begin{center}

\includegraphics[width=0.45\textwidth]{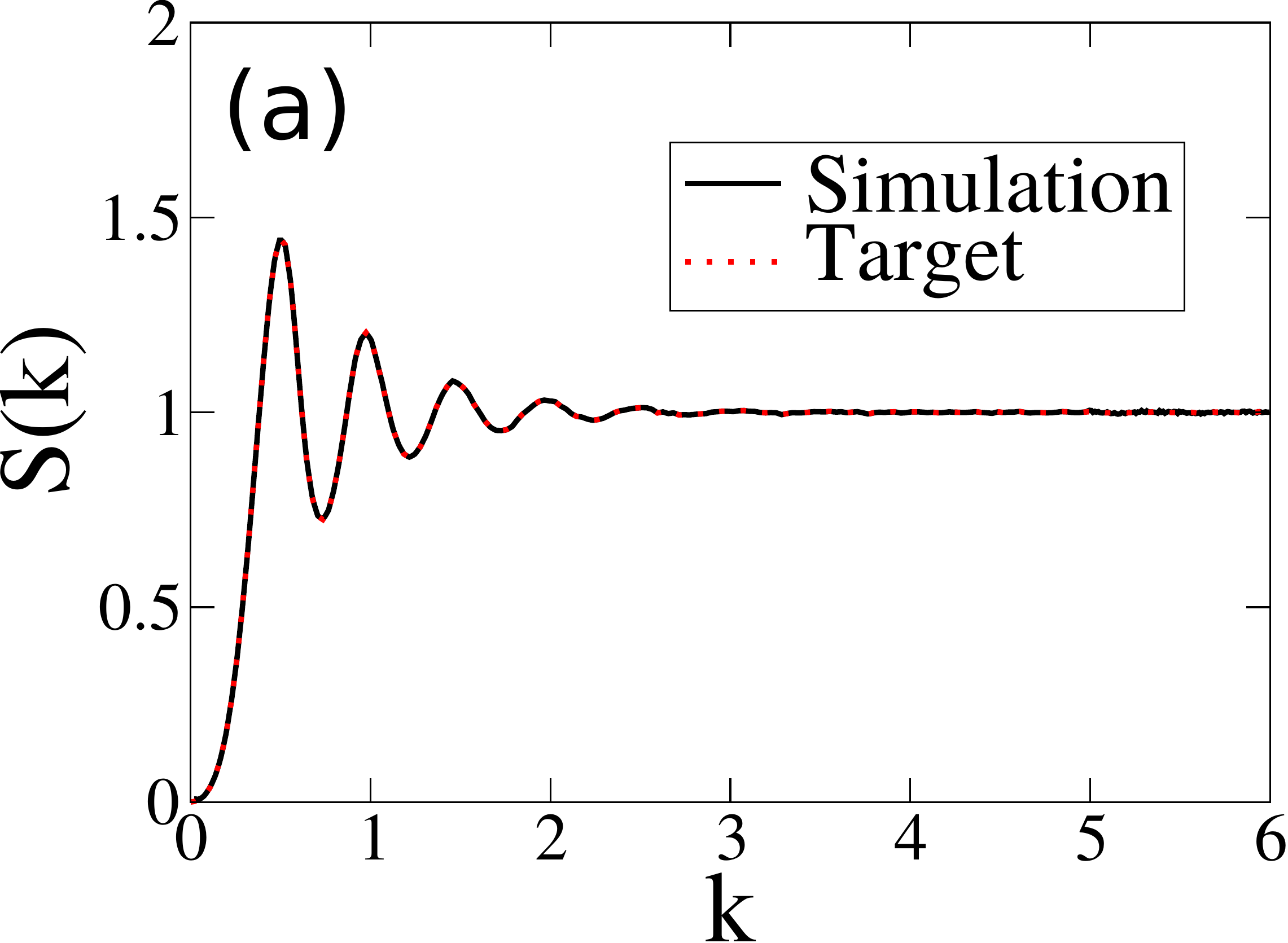}
\includegraphics[width=0.45\textwidth]{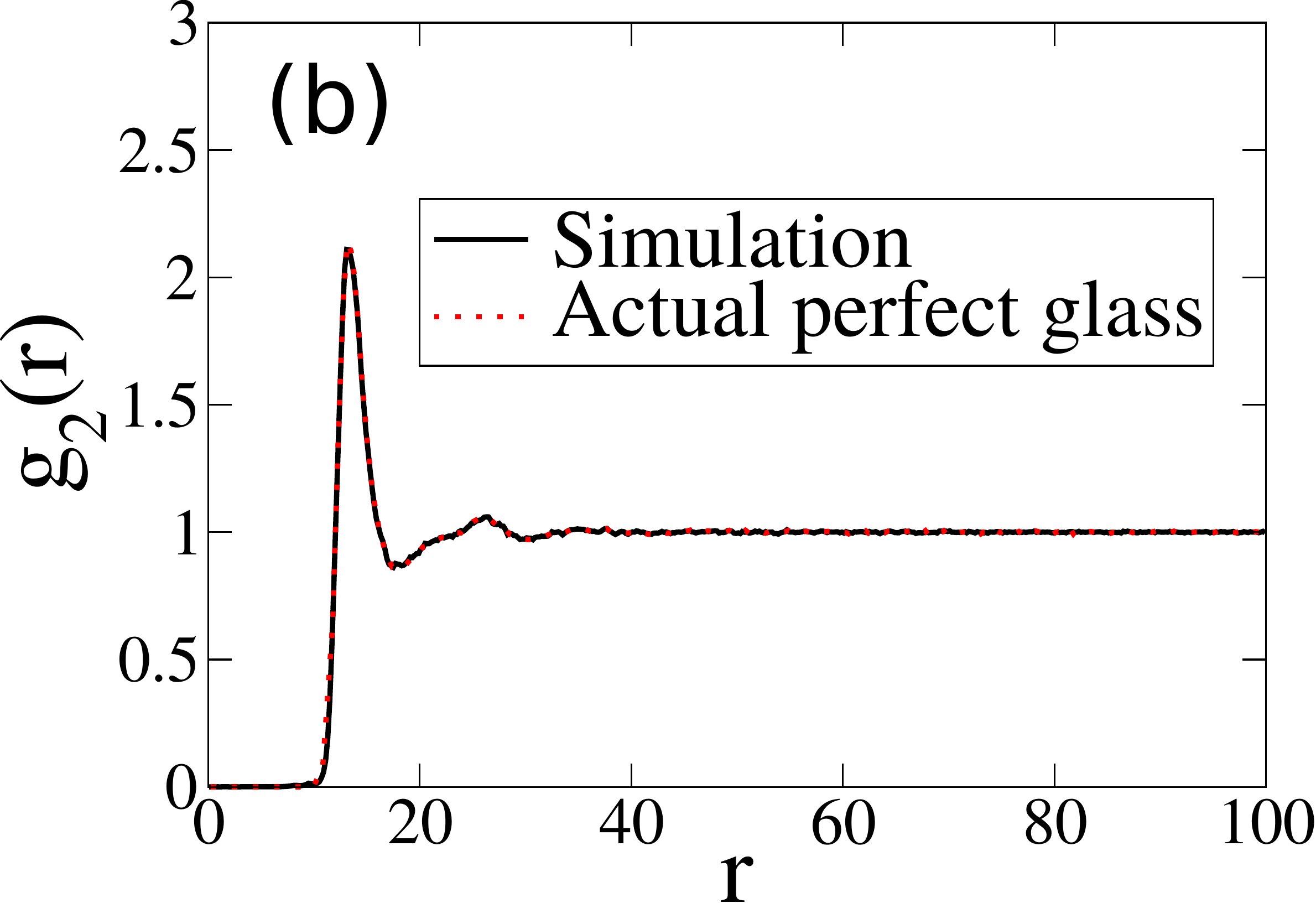}
\end{center}
\caption{(a): The structure factor obtained by sampling ensembles of 2D configurations in which the target function $S_0(k)$ is taken to be equal to the numerically measured $S(k)$ of a perfect-glass system at $\rho=0.00390625$. (b): 
The corresponding pair correlation function sampled from targeted configurations, compared with that of an actual perfect glass.}

\label{2DPerfectGlass}
\end{figure}

\begin{figure}
\begin{center}
\includegraphics[width=0.45\textwidth]{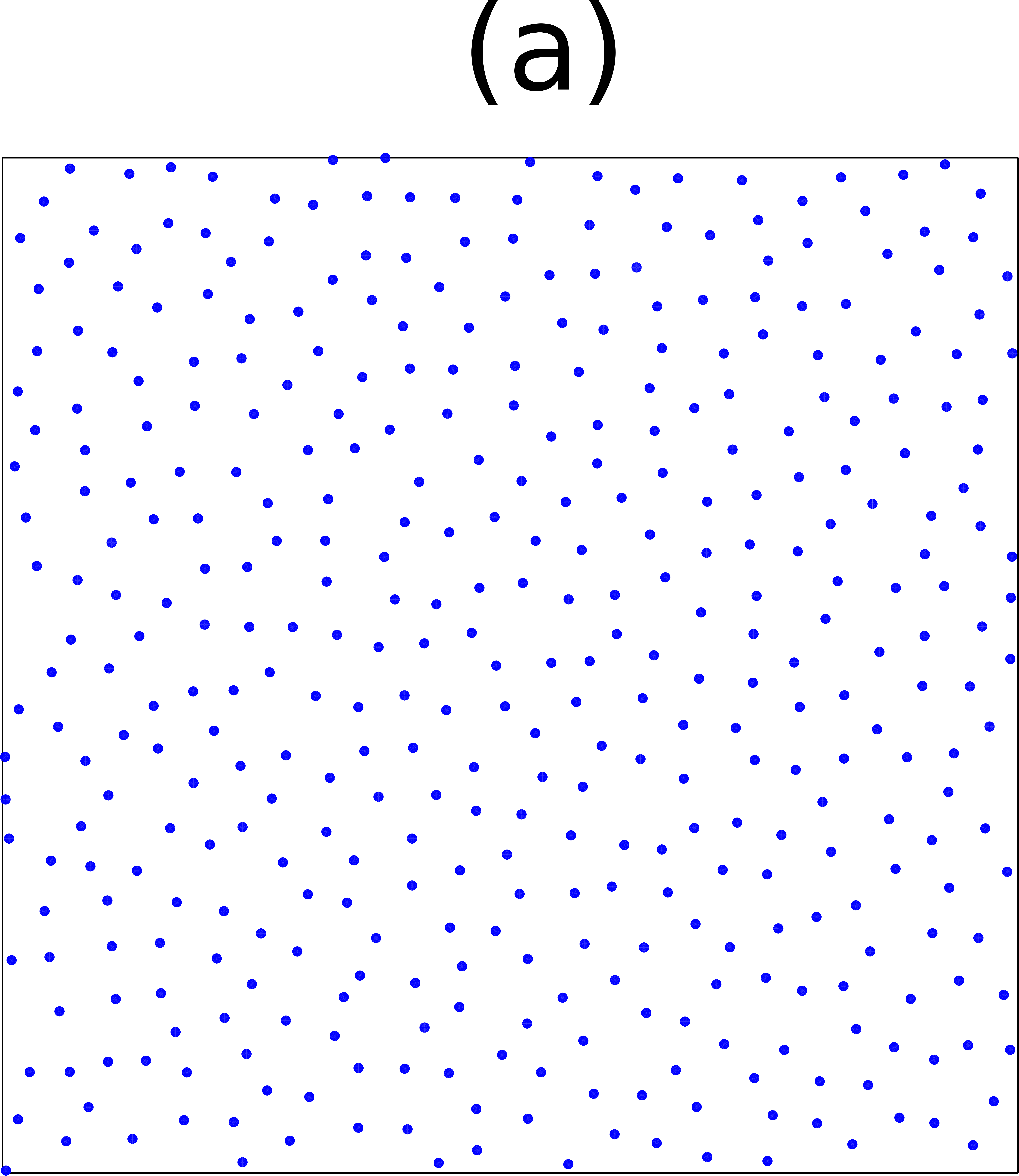}
\includegraphics[width=0.45\textwidth]{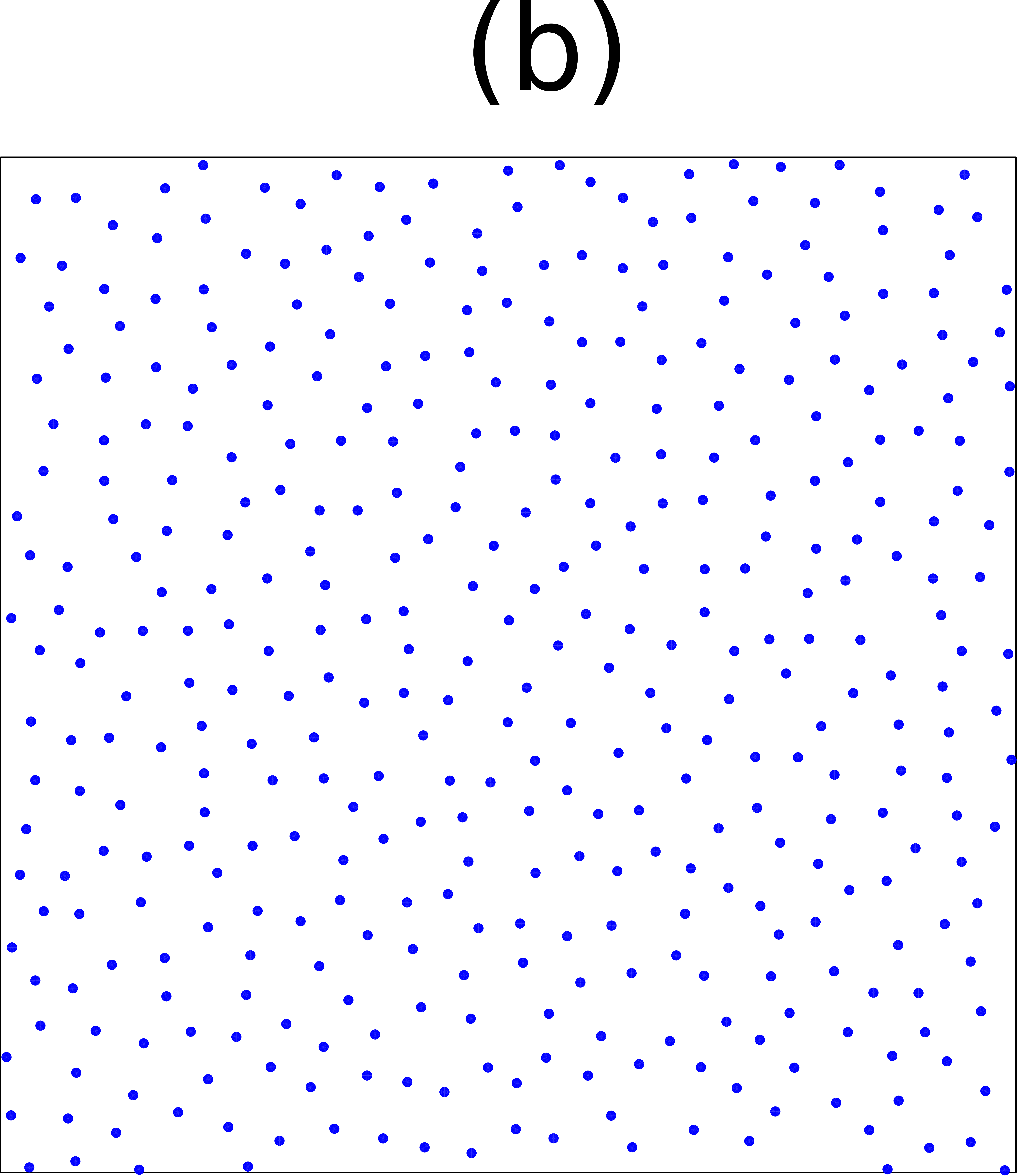}
\end{center}
\caption{(a): A two-dimensional configuration of 400 particles drawn from ensembles in which the target function $S_0(k)$ is taken to be equal to the numerically measured $S(k)$ of a perfect-glass system at $\rho=0.00390625$. (b): An actual perfect-glass configuration with the same pair statistics $g_2(r)$ and $S(k)$.}
\label{2DPerfectGlass_config}
\end{figure}

As a highly stringent test of our affirmative answer, we now target nonequilibrium ``perfect-glass'' structure factors \cite{zh16a}, which are glassy, nonequilibrium state of a many-particle system interacting with two-, three-, and four-body potentials. 
Counterintuitively, the classical ground state of this many-body interaction is unique and disordered \cite{zhang2017classical}.
By construction, it banishes crystals and quasicrystals from the ground-state manifold. We have previously investigated the quenched states from infinite temperature to zero temperature for this model with various parameter choices, and numerically computed their structure factors. Here we target the numerically-measured structure factor of a perfect glass model with parameters $d=2$, $\rho=0.00390625$, $\chi=5.10$, $\alpha=2$, and $\gamma=3$ (see Ref.~\onlinecite{zh16a} for the definition of the last three parameters). We use targeting parameters $N_c=1000$ and $K=5$. This seemingly small value of $K$ is actually relatively large considering that $\rho$ is much smaller than unity, since real-space length scale is inversely proportional to the $k$-space length scale. 
Figures~\ref{2DPerfectGlass} shows the excellent agreement between  the targeted
and simulated structure factors and
pair correlation functions. This is a remarkable suggestion that the answer to our question above is affirmative because a perfect glass is a nonequilibrium system with two-, three-, and four-body interactions while our reconstructed
system is an equilibrium state of a pair potential. 

It is interesting to compare configurations produced by the actual
perfect glass interaction and the effective pair interaction visually, as is done in Fig.~\ref{2DPerfectGlass_config}. Although these pair of configurations look strikingly similar to one another, we know that since one system involves three- and four-body interactions and the other does not, that their higher-order statistics ($g_3$, $g_4$, $\cdots$) must be different \cite{torquato2006new, jiao2010geometrical, stillinger2019structural}, even if such distinctions cannot be detected visually.

The realizability of a perfect glass
using our formalism as well as the findings reported in Secs. III-VI lead us to the following conjecture: \smallskip
\vspace{0.2in}

\noindent{\sl Given the pair correlation function $g_2({\bf r})$
of any realizable statistically homogeneous many-particle ensemble (equilibrium or not) in $\mathbb{R}^d$ at
number density $\rho$, there is an equilibrium ensemble (Gibbs measure) involving only an effective pair potential $v({\bf r})$ that gives rise to
such a $g_2({\bf r})$.}

\vspace{0.2in}
\noindent{Note that statistical homogeneity implies that thermodynamic limit
has been taken. The rigorous validity of this conjecture is an outstanding open problem.}

\section{Conclusions and Discussion}
\label{conclusions}

To address the realizability problem,  we introduced a theoretical formalism that provides a means to draw disordered particle configurations at positive temperatures from canonical ensembles with certain pairwise-additive potentials that
could correspond to disordered targeted analytical functional forms for the structure factor. This theoretical foundation
enabled  us to devise an efficient  algorithm to construct systematically canonical-ensemble particle configurations with such targeted pair statistics whenever realizable.  As a proof-of-concept, we tested this algorithm to target several different structure factor functions across dimensions that are known to be realizable and one hyperuniform target  that meets all explicitly  known
necessary realizability conditions  but is known to be nontrivially unrealizable.
Our algorithm succeeded  for all realizable targets 
and appropriately failed for the unrealizable target, demonstrating the accuracy and power of the method to numerically investigate the realizability problem.
Having established the prowess of the methodology,
we targeted  several families of
structure-factor functions that meet the known necessary realizability conditions but were heretofore
not known to be realizable, including $d$-dimensional Gaussian structure factors,
$d$-dimensional generalizations of the 2D one-component 
plasma, the $d$-dimensional  Fourier duals of the previous OCP cases, a hyposurficial target in 2D and 3D, and an antihyperuniform target in 2D.
In all of these instances, we were able to achieve the targeted structure
factors with high accuracy, suggesting that these targets are indeed truly realizable
by equilibrium many-particle systems with pair interactions at positive temperatures.
This expands our knowledge of analytical functional forms for 
$g_2(r)$ and $S(k)$ associated with disordered point configurations
across dimensions.  
When targeting hyposurficial structure factors, we confirm a previous observation that hyposurficiality is associated with spatial heterogeneities that are manifested by significant clustering of the particles \cite{martelli2017large}.
Our results, especially perfect-glass realizability, led to the conjecture that any realizable structure factor corresponding to either
an equilibrium or nonequilibrium system can be attained by an equilibrium ensemble involving only
effective pair interactions in the thermodynamic limit.

It is worth stressing that we only constrain $S(k)$ in a finite range ($0<|k|<K$) numerically for $K$ as large as feasibly possible, but do not enforce explicit constraints on $g_2(r)$. Since $g_2(r)$ for small $r$ is related to $S(k)$ for large $k$, there is no guarantee that the numerically sampled $g_2(r)$ matches its analytical counterpart at 
very small pair distances (i.e., $r < 2\pi/K$). Nevertheless, we always find impressive consistency between the simulated and analytical pair correlation functions corresponding to the target $S(k)$, which further demonstrates the success of our algorithm.

Realizable particle configurations generated with a targeted pair correlation function using our algorithm are equilibrium states of pairwise additive interactions at positive temperatures.
Such a pair potential is unique up to a constant shift \cite{henderson1974uniqueness}.
In the case of realizable hyperuniform targets with the smooth pair correlation
functions considered here,
such interactions must be long-ranged \cite{To18a}, which is a consequence of the well-known fluctuation-compressibility relation:
\begin{equation}
\lim_{k \to 0} S(k)=\rho k_B T \kappa_T,
\label{CompressibilityRelation}
\end{equation}
where $\kappa_T$ is the isothermal compressibility. Since $\lim_{k \to 0} S(k)=0$ and $\rho k_B T>0$, one must have $\kappa_T=0$. Using the analysis that relates the large-$r$ behavior 
of the direct correlation function to that of the 
pair potential  function $v(r)$ presented in Ref.~\onlinecite{To18a},  it immediately follows that the asymptotic 
behavior of $v(r)$ is Coulombic for
the Gaussian S(k) [cf. (47)] for $d \ge 3$, i.e.,
\begin{equation}
\beta v(r) \sim 1/r^{d-2}  \qquad r \to \infty,
\end{equation}
which, of course, is a long-ranged interaction.
The same asymptotic Coulombic form for the pair potential  for $d \ge 3$ arises in the $d$-dimensional generalization of the OCP pair 
correlation function (\ref{d-OCP}). It would be an interesting future research direction to find the specific functional forms of such pair interactions.
For general determinantal point processes, these would be effective pair interactions that mimic the two-body, three-body, and higher-order intrinsic interactions \cite{torquato2008point}.

Our study is also a step forward in being able to devise inverse methods \cite{torquato2009inverse, zhang2013probing} to design materials with
desirable physical properties that can be tuned by their pair statistics.
Pair statistics combined with effective pair interactions, which in principle can be obtained using inverse techniques \cite{lindquist2016inverse, jadrich2017probabilistic}, can then be used to compute all of the thermodynamic properties, such as compressibility and energy and its derivatives ({\it e.g.,} pressure or heat capacity) \cite{Han13}. 
In instances in which the bulk physical properties are primarily determined by the pair statistics, such as the frequency dependent dielectric constant \cite{Re08} and transport properties of two-phase random media \cite{To02}, our results are immediately applicable.

While applications of our ensemble-average methodology were directed
toward the realizability of target structure factor functions that putatively could
correspond to disordered hyperuniform and nonhyperuniform
({\it e.g.,} hyposurficial and anti-hyperuniform) many-particle configurations, the technique is entirely
general and hence not limited to these systems. 
In future work, we will
apply the algorithm to discover realizable families of pair correlation
functions associated with other novel configurations.
Another interesting future direction is to analytically study how the single-configuration structure factor at a particular $\mathbf k$ vector is distributed. We have proved in the present work that it is exponentially distributed when there is a single constraint, or when there are up to $d$ independent constraints. When the number of $\mathbf k$ vectors is higher than $d$, we provided strong numerical evidence that the exponential distribution still holds (see Appendix~\ref{appendix:justification}). Whether such behavior can be proved remains a fascinating open problem.

\begin{acknowledgments}
We are very grateful to JaeUk Kim, Timothy Middlemas, Zheng Ma, Tobias Kuna, and Joel Lebowitz for their helpful remarks.
S. T. acknowledges the support of the National Science Foundation under Grant No. CBET-1701843.
G. Z. acknowledges the support of the U.S. Department of Energy under Award DE-FG02-05ER46199.
\end{acknowledgments}

\appendix
\section{Numerical Tests on the Theoretical Formalism and Justification of  the Algorithm}
\label{appendix:justification}

In this Appendix, we numerically verify the major conclusions and outcomes of our theoretical canonical-ensemble 
formalism and justify our algorithm using the 1D fermionic target structure factor (\ref{1DFermionic_S}) as an example. Specifically, we show/verify: (1) to constrain the ensemble-average structure factor at a single $\mathbf k$ vector, one can alternatively perform canonical-ensemble simulations at temperature $k_BT=1$ with energy given in Eq. (\ref{eqn:tildeV}) via  the molecular dynamics algorithm; (2) to constrain $S(\mathbf k)$ at multiple wave vectors in 1D, simply performing canonical-ensemble simulations using (\ref{eqn:tildeV}) is inexact; (3) our algorithm given in Sec.~\ref{newAlgorithm} outputs configurations drawn from the canonical ensemble of a pairwise additive interaction; and (4) when employing our algorithm to reconstruct configurations in 1D, $K=30$ is a reasonable cutoff value for the constrained region.
We proved that the single-configuration structure factors $\mathcal S(\mathbf k)$ at a single constrained wave vector is exponentially distributed when there is one constraint. However, here we provide strong numerical evidence that the same distribution holds even for the multiple-constraint cases.

\begin{figure}
\begin{center}
\includegraphics[width=0.45\textwidth,clip=]{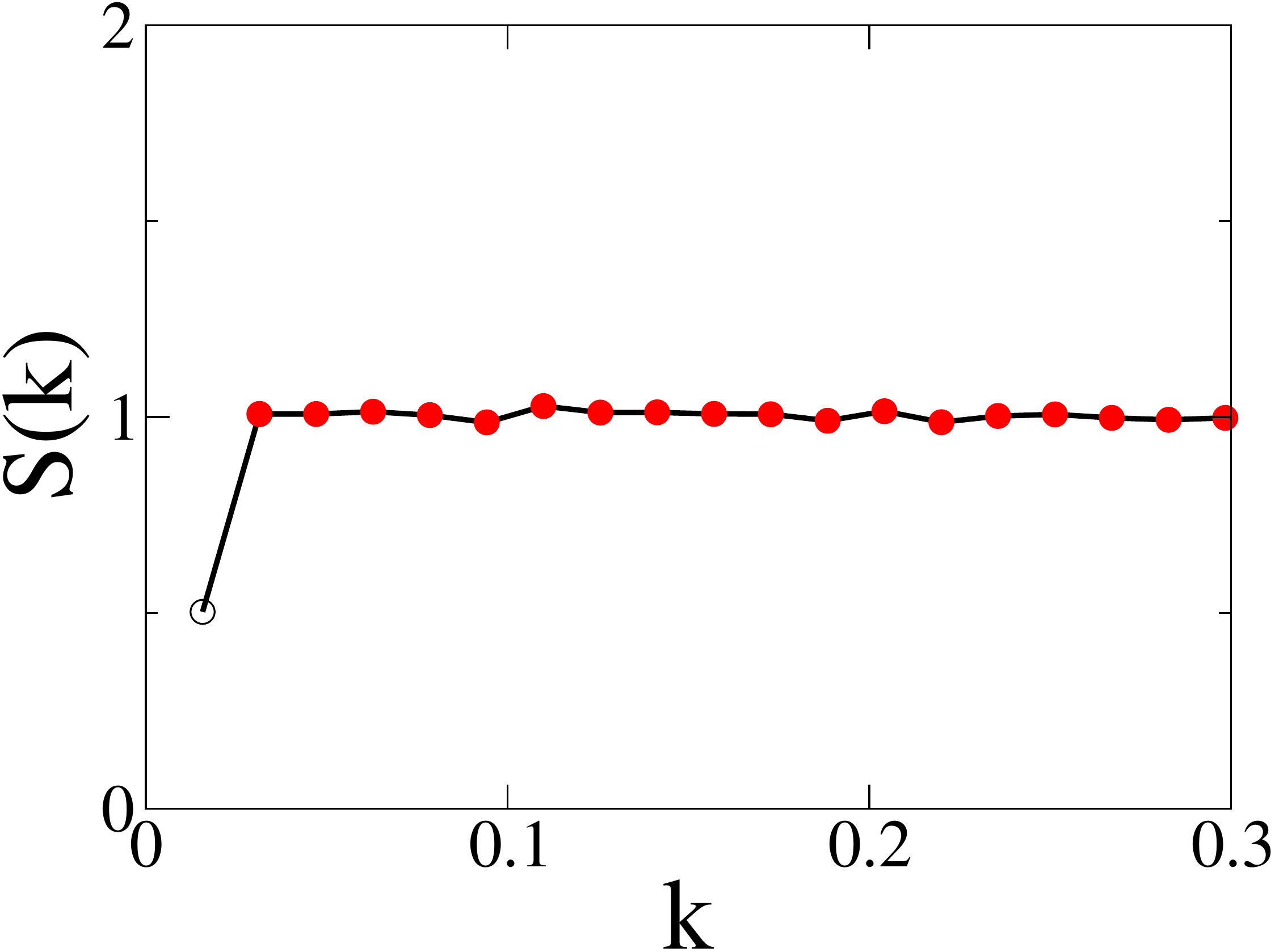}
\end{center}
\caption{Structure factor for the 1D system with energy $E=S(q) \cdot (1/S_0-1)$, where $q$ is the smallest wave vector and $S_0=0.5$, at temperature $k_BT=1$. As discussed in the text, the effect of this energy is to constrain the structure factor at $q$ to $S_0=0.5$ (unfilled circle), and leave the structure factor at other wave vectors unconstrained
(solid red circles).}
\label{StealthyMD_kmin}
\end{figure}

If we only need to constrain the structure factor at a single wave vector, then (\ref{eqn:tildeV}) is exact, and one can perform molecular-dynamics simulations [with energy defined in (\ref{eqn:canonicalEnergy1}) at temperature $k_BT=1$] to meet this constraint.
We performed such a simulation and collected 5000 snapshots. The simulation is performed on a 1D system with $N=400$ particles. We let $S_0=0.5$, and choose $q$ to be the smallest $k$ vector. The resulting structure factor is presented in Fig.~\ref{StealthyMD_kmin}. Indeed, the structure factor at $q$ averages to $S_0$.

\begin{figure}
\begin{center}
\includegraphics[width=0.45\textwidth]{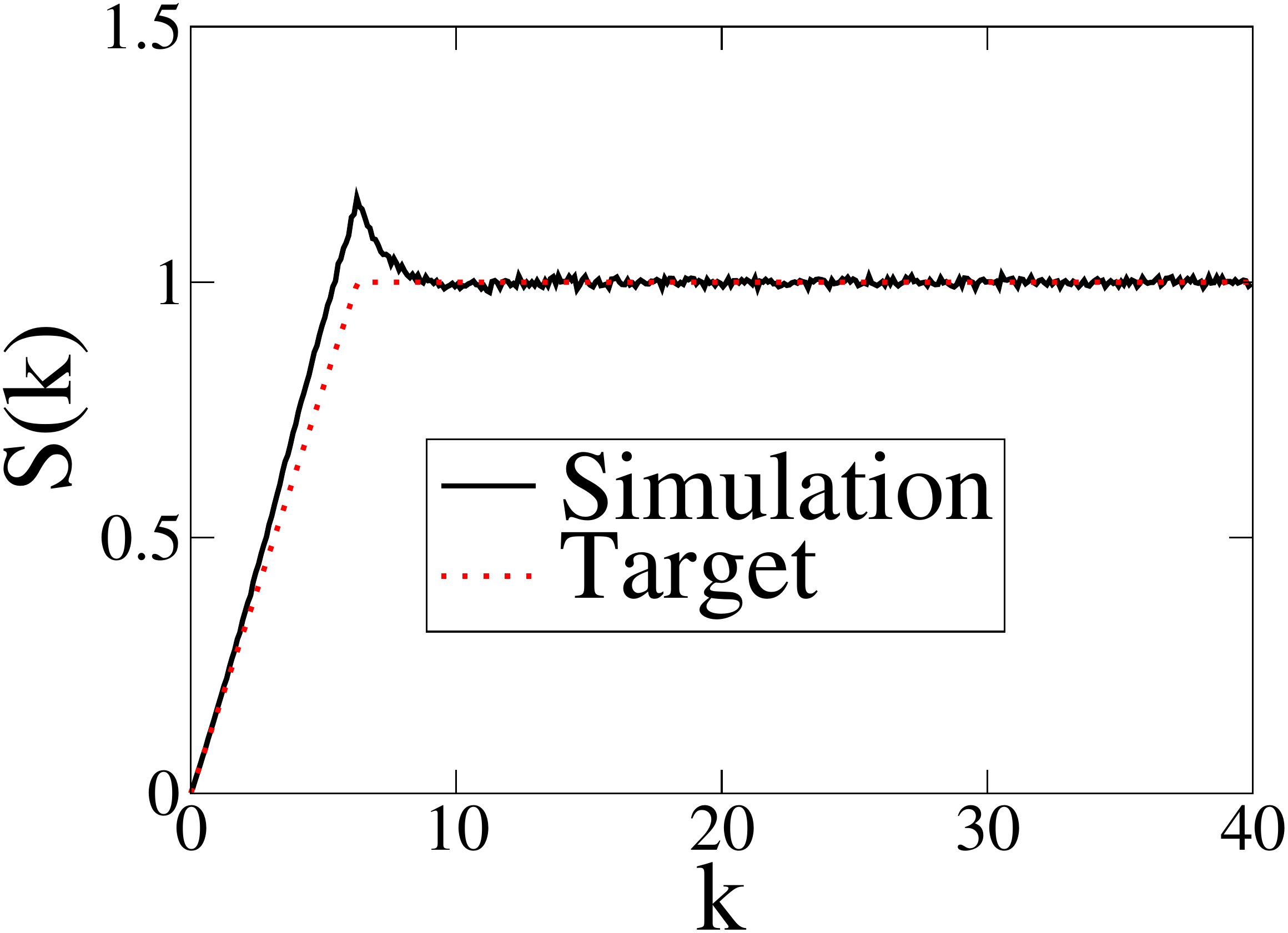}
\end{center}
\caption{Structure factor for the 1D system with energy given in Eq.~(\ref{energy_multiK}), in which $S_0(\mathbf k)$ is given in Eq.~(\ref{1DFermionic_S}), at temperature $k_BT=1$.}
\label{StealthyMD_Fermionic}
\end{figure}

We now move on to constraining the structure factor at multiple, non-independent wave vectors in order to realize the structure factor of the 1D ``fermionic'' system. Since the structure factor at multiple wave vectors are constrained, Eq. (\ref{eqn:tildeV}) is inexact due to correlations between structure-factor values at different wave vectors.
To find out how inexact it is, we again performed molecular dynamics simulations. We define the energy
\begin{equation}
E= \sum_{|\mathbf k| < 2\pi} {\tilde v}(\mathbf k) \mathcal S(\mathbf k),
\label{energy_multiK}
\end{equation}
where ${\tilde v}(\mathbf k)$ is given in (\ref{eqn:tildeV}).
Here the summation beyond $|\mathbf k| = 2\pi$ is unnecessary because for such $\mathbf k$ vectors, $S_0(\mathbf k)=1$ and ${\tilde v}(\mathbf k)$ vanishes. As previously discussed, we performed a molecular dynamics simulation with with $N=400$ particles at $k_BT=1$. The resulting structure factor, shown in Fig.~\ref{StealthyMD_Fermionic}, exhibits a noticeable deviation from its target $S_0(\mathbf k)$.

Since we were not able to derive an appropriate ${\tilde v}(\mathbf k)$ for the general case of multiple non-independent $\mathbf k$ vectors, 
we could not meet these constraints by simply performing molecular dynamics simulations.
We must therefore use the method presented in Sec.~\ref{newAlgorithm} [minimizing $\Phi$ in Eq.~(\ref{Potential_Fourier2}) for $N_c=100$ configurations simultaneously].

\begin{figure}
\begin{center}
\includegraphics[width=0.23\textwidth]{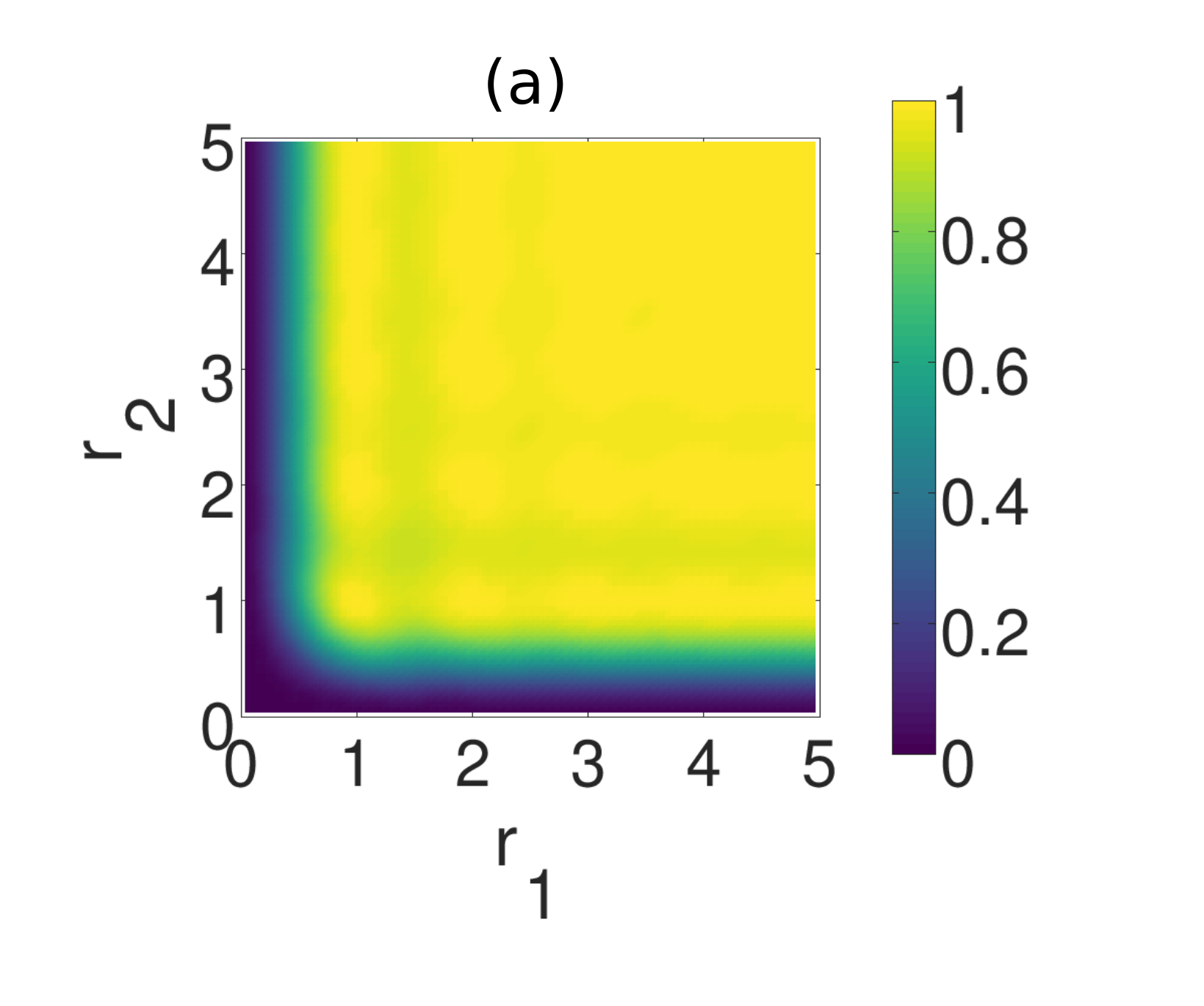}
\includegraphics[width=0.23\textwidth]{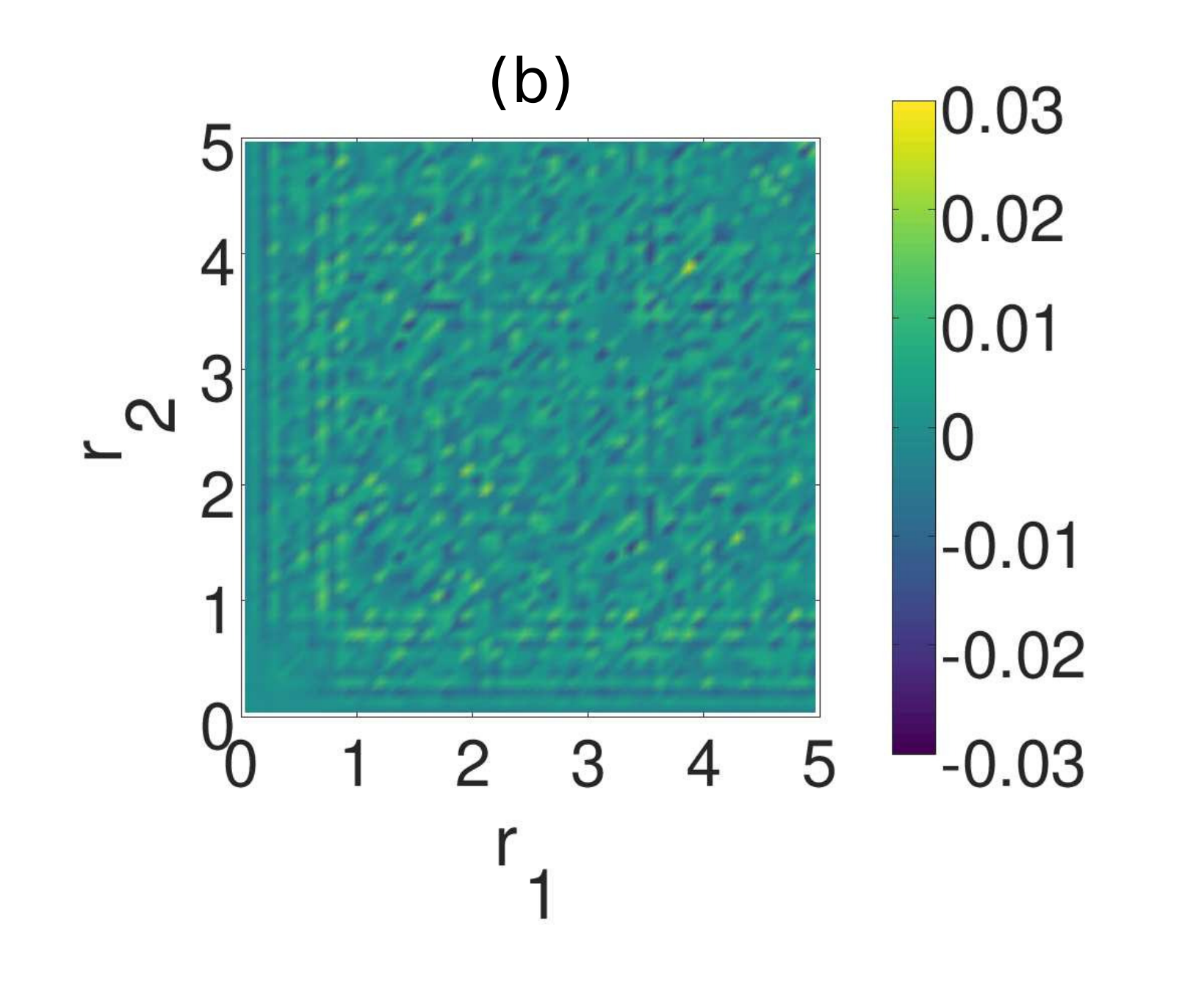}
\includegraphics[width=0.23\textwidth]{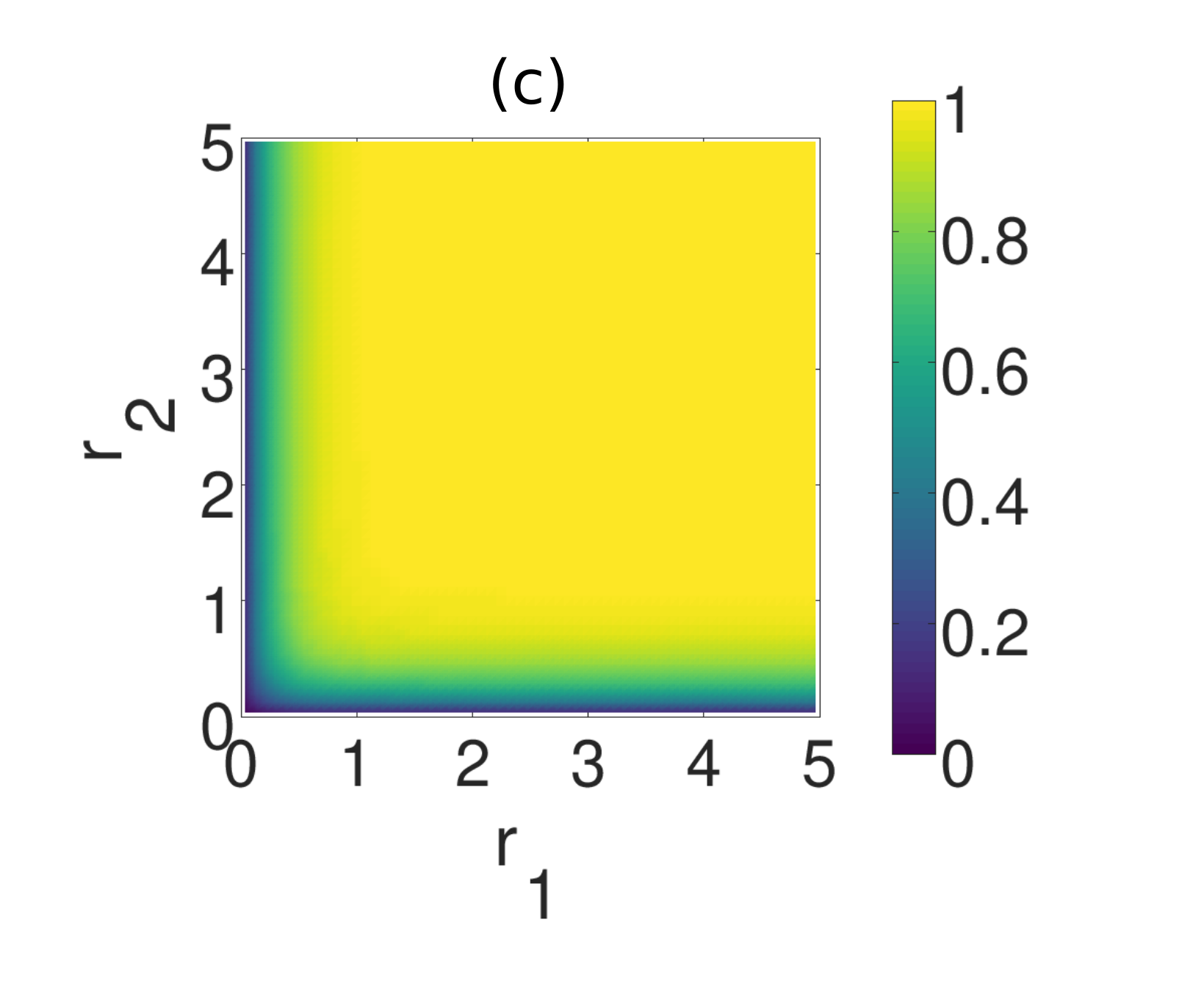}
\includegraphics[width=0.23\textwidth]{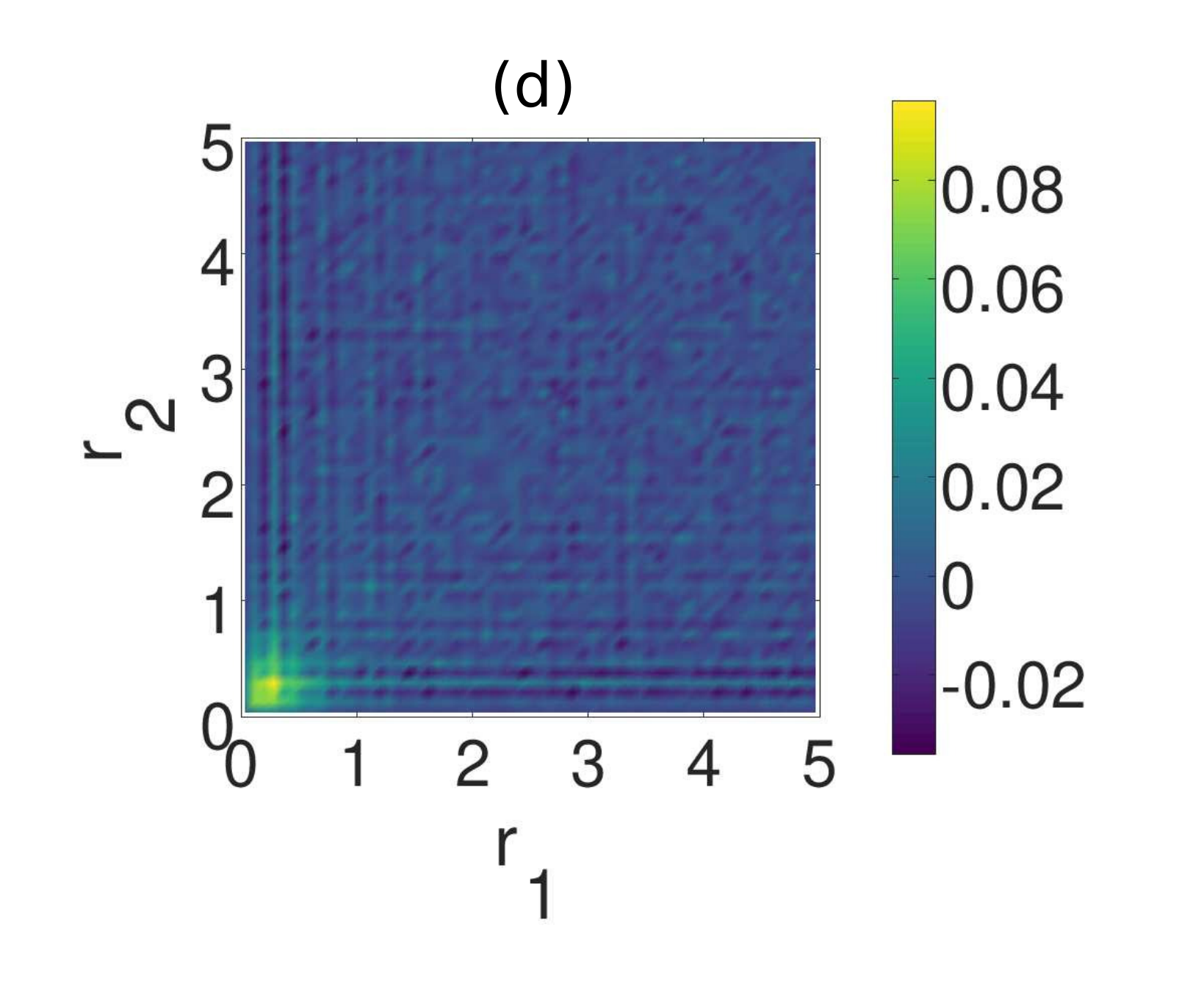}
\end{center}
\caption{(a): Analytical three-body correlation function, $g_3(r_1, r_2)$, of the 1D fermi-sphere system. (b): The difference in $g_3$ between the targeted system and the 1D fermi-sphere system, $\Delta g_3(r_1, r_2)= g_{3}^{\mbox{fermionic}}(r_1, r_2)-g_{3}^{\mbox{targeted}}(r_1, r_2)$.
(c) and (d): Same as (a) and (b), except for 1D Lorentzian target.
}
\label{1DFermionic_g3}
\end{figure}

One conclusion of  our theoretical formalism  is that our reconstructed configurations are drawn from the canonical ensemble of a pairwise additive interaction
at some positive temperature.
 We can test this conclusion because the fermionic systems in one dimension also correspond to an equilibrium state of a logarithmic pairwise-additive potential
with $\beta=2$. 
By Henderson's theorem \cite{henderson1974uniqueness}, our targeted system and the fermionic system are both equilibrium states of the same pair potential, and thus the two systems should have the same higher-order statistics ($g_3$, $g_4$, $\cdots$). To check this conclusion, we computed the three-body correlation functions of our targeted system and compare with the analytical result of the fermionic system in Fig.~\ref{1DFermionic_g3}. 
In general, $g_3$ is a function of three vectors: $\mathbf r_{12}$, $\mathbf r_{13}$, and $\mathbf r_{23}$. However, in 1D, we can set $\mathbf r_{12}=r_1$, $\mathbf r_{13}=r_2$, and $\mathbf r_{23}=r_1+r_2$, and then $g_3$ becomes a function of two scalars, $r_1$ and $r_2$.
The difference between the numerically found $g_3(r_1, r_2)$ of the targeted system and the analytical $g_3(r_1, r_2)$ of the fermi-sphere system is two orders of magnitudes smaller than unity, and appears completely random with no systematic trend, consistent with our reasoning. 

By contrast, the one-dimensional Lorentzian $S(k)$ target is also realized by a determinantal point process with an analytically known $g_3$ \cite{costin2004construction}, but it is not an equilibrium state of a pairwise additive potential \cite{torquato2008point}. Therefore, the difference between the analytic $g_3$ for the determinantal point process and the numerical $g_3$ of the reconstructed system exhibits a peak much stronger than statistical noise at around $r_1=r_2=0.3$. The statistically significant difference is consistent with our theory that the reconstructed configurations are drawn from an equilibrium state of a pair potential, and therefore must be different from determinantal point processes not realizable by pair potentials. We have similarly verified that the exact analytical expression for $g_3$ of the 2D fermionic system differs 
from the numerically determined $g_3$ of the reconstructed system with the same pair statistics, which is expected since 2D fermionic systems also involve $n$-particle interactions with $n\ge 3$ \cite{torquato2008point}.

\begin{figure}
\begin{center}
\includegraphics[width=0.45\textwidth]{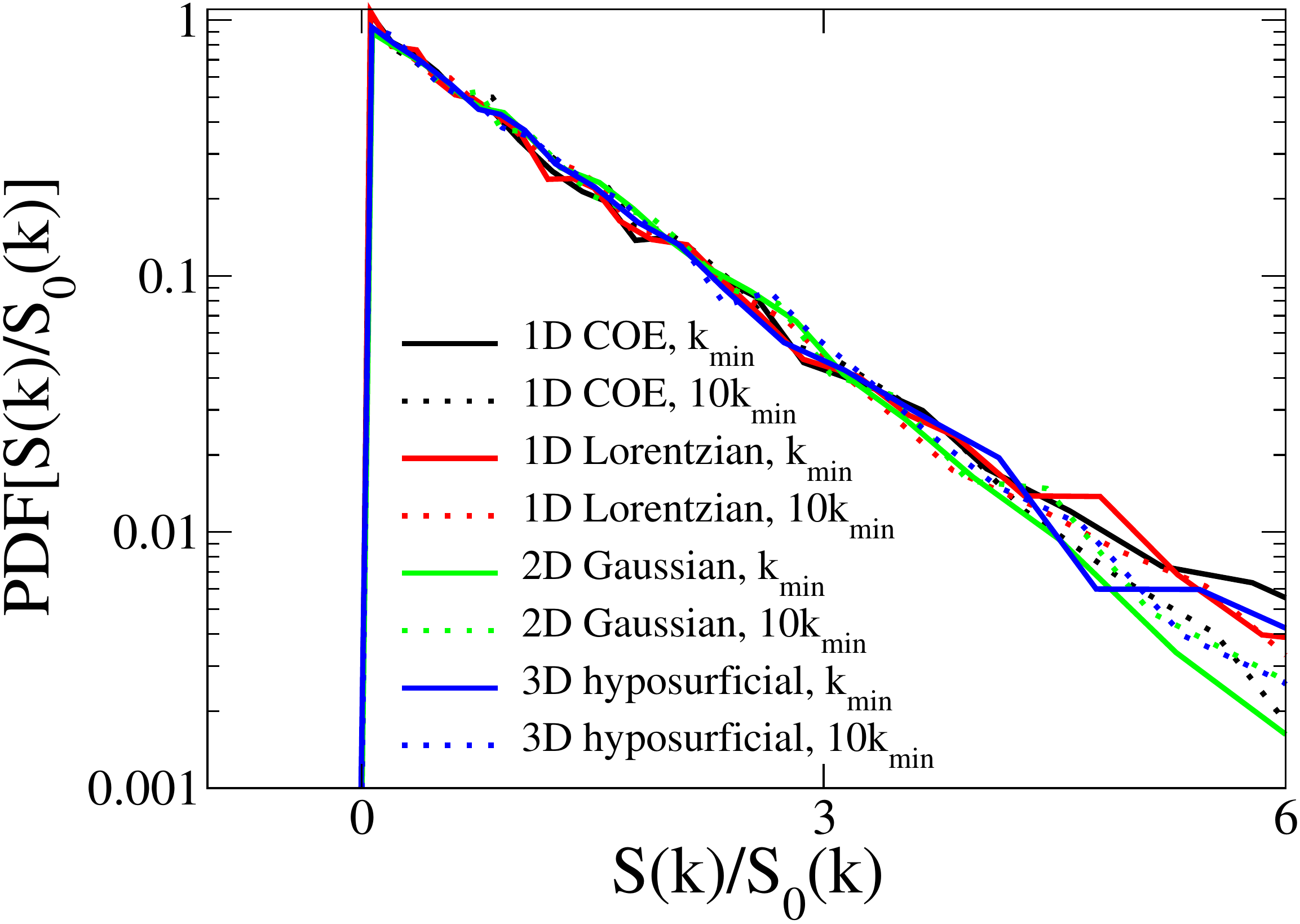}
\end{center}
\caption{Probability density function (PDF) of the structure factor, normalized by the target, at two different wave vectors for four different targets. The two wave vectors are $k_{\mbox{min}}$, the minimum wave vector in the $x$ direction, and $10k_{\mbox{min}}$. The targets are 1D COE [Eq.~(\ref{1DSalLog1_S})], 1D Lorentzian [Eq.~(\ref{L})], 2D Gaussian [Eq.~(\ref{S-OCP})], and 3D hyposurficial [Eq.~(\ref{eq:3DHyposurficialSk})]. Since the structure factor is always exponentially distributed, the normalized plot for different cases falls on a single strait line on a plot on a semi-logarithmic scale.}
\label{SkPDF}
\end{figure}

We now show numerical evidence that the structure factor at a constrained $\mathbf k$ vector is exponentially distributed. In Fig.~\ref{SkPDF}, we plot the distribution of single-configuration structure factors at two different wave vectors for four different targets. After normalizing $S(\mathbf k)$ by its mean, $S_0(\mathbf k)$, all results collapse onto a single straight line in a semi-logarithmic plot, demonstrating that $S(\mathbf k)$ is indeed exponentially distributed in all cases.

\begin{figure}
\begin{center}
\includegraphics[width=0.45\textwidth]{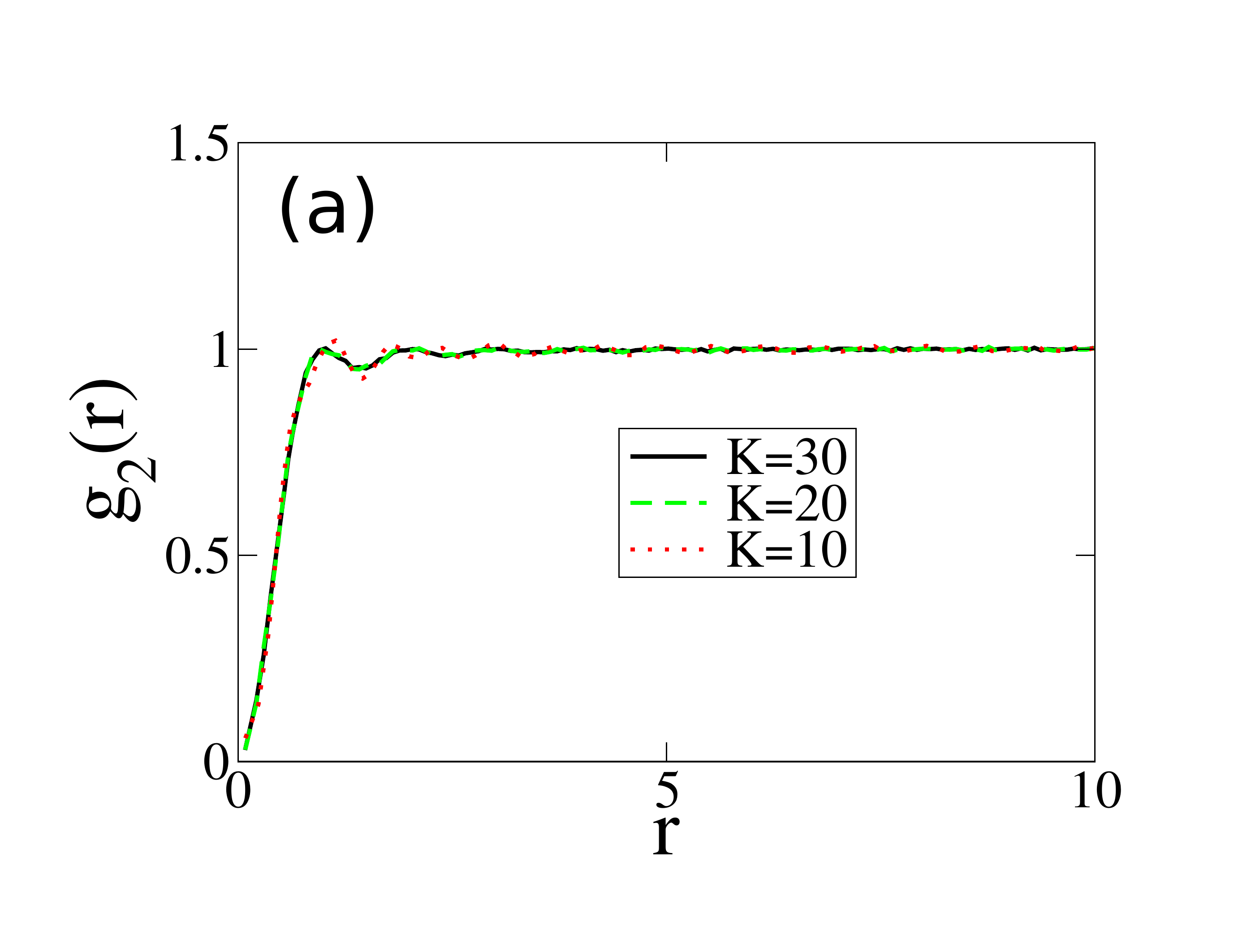}
\includegraphics[width=0.45\textwidth]{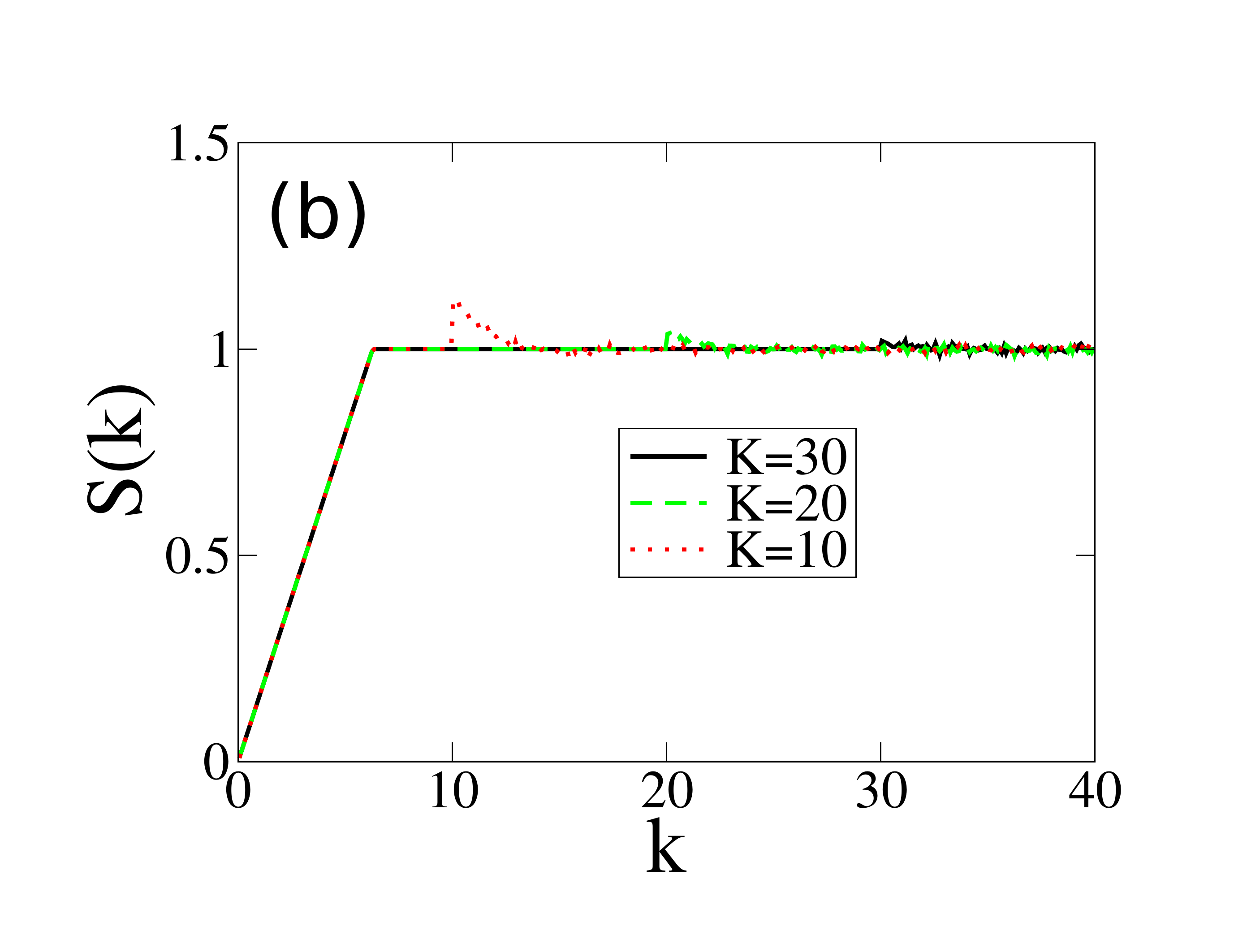}
\end{center}
\caption{Pair correlation function and structure factor obtained by targeting Eq.~(\ref{1DFermionic_S}) at various $K$'s.}
\label{1DFermionic_varK}
\end{figure}

Lastly, we explore the effect of changing the cutoff value $K$. The results are presented in Fig.~\ref{1DFermionic_varK}. With $K=10$, the structure factor develops a discontinuity at the cutoff. However, as we increase $K$ to 30, the discontinuity diminishes. We have also explored many other targets, detailed in the rest of the paper, and always find that when $K$ is sufficiently large, the structure factor is continuous at the cutoff. In summary, except for some special cases discussed in the paper, the cutoff value of $K=30$ is generally suitable for one-dimensional targets that we examined, and $K=15$ is generally suitable for two- and three-dimensional targets reported here.


\begin{thebibliography}{53}
\bibitem [{foo()}]{footnote}%
  \BibitemOpen
  \href@noop {} {}\bibinfo {note} {Henderson claimed that a modification of his
  proof for positive temperature ($T>0$) can be used to prove the same
  uniqueness theorem for a system in its ground state ($T=0)$). However, this
  cannot be generally true for a classical statistical mechanical system at
  $T=0$, since there are generally an infinite number of different pair
  potential functions consistent with a particular ground-state structure.
  Indeed, the existence of ``universally optimal" solutions for certain
  ground-state structures, as described in Refs. \cite{Co07} and \cite{Co19}, rigorously
  supports this conclusion.}\BibitemShut {Stop}%
\end{thebibliography}
\end{document}